\documentclass{aa}  

\usepackage{amsmath, amssymb, amsfonts, amscd}
\usepackage{natbib}
\usepackage{graphicx}
\usepackage{txfonts}
\usepackage[final]{hyperref}
\usepackage{mathrsfs}
\usepackage{color}
\usepackage{mathabx}
\usepackage{supertabular}
\usepackage{empheq}
\usepackage{multirow}
\usepackage{makecell}
\usepackage{longtable}
\usepackage{ragged2e}
\usepackage[switch]{lineno}

\pdfpageattr {/Group << /S /Transparency /I true /CS /DeviceRGB>>}

\hypersetup{
   colorlinks=true,
   citecolor=blue,
   linkcolor=blue,
   urlcolor=black
   }

\definecolor{Green}{RGB}{0,130,56}
\definecolor{RoyalBlue}{RGB}{65,105,225}

\newcommand{\RV}[1]{#1}
\def\define{\equiv}				
\def\scale{\, \propto \,}			
\def\scalprod{\cdot}				

\newcommand{\ddroit}{{\rm d}}				
\newcommand{\vect}[1]{\boldsymbol{#1}}		
\newcommand{\evect}[1]{\vect{{\rm e}}_{#1}}	
\newcommand{\dd}[2]{\dfrac{\partial #1}{\partial #2}}		
\newcommand{\ddn}[3]{\dfrac{\partial^{#3} #1}{\partial #2^{#3}}}		
\newcommand{\DD}[2]{\dfrac{\ddroit #1}{\ddroit #2}}			
\newcommand{\Dt}[1]{\dot{#1}}				
\newcommand{\abs}[1]{\left| #1 \right|}		
\newcommand{\matx}[1]{\boldsymbol{#1}}
\newcommand{\transp}[1]{#1^{\rm T}}		
\newcommand{\infvar}[1]{\ddroit #1}
\newcommand{\normalized}[1]{\tilde{#1}}

\def\pressure{p}
\def\ihoriz{\sigma}
\def\nab{\nabla}					
\def\nabnorm{\normalized{\nab}}
\def\grad{\nab}					
\def\lap{\nab^2}					
\def\lapn{\normalized{\nab}^2}
\newcommand{\nabn}[1]{\nab^{#1}}
\def\div{\grad \cdot}				
\def\laph{\lap_{\ihoriz}}
\def\laphn{\lapn_{\ihoriz}}
\newcommand{\nabnh}[1]{\nab^{#1}_{\ihoriz}}
\newcommand{\nabnhn}[1]{\nabnorm^{#1}_{\ihoriz}}

\newcommand{\gradx}[1]{\grad_{#1}}
\newcommand{\divx}[1]{\gradx{#1} \cdot}
\newcommand{\atx}[2]{\left. #1  \right|_{#2}}

\def\nn{n}						

\newcommand{\expo}[1]{{\rm e}^{#1}}				
\newcommand{\integ}[4]{\int_{#3}^{#4} #1 {\rm d} #2 }	
\newcommand{\normal}[1]{\tilde{#1}}					

\def\rr{r}						
\def\col{\theta}					
\def\time{t}					
\def\zz{z}
\def\xx{x}

\def\XX{X}
\def\zd{u}
\def\tmin{\time_{\rm min}}
\def\tmax{\time_{\rm max}}
\def\trun{\time_{\rm run}}
\def\tradia{\time_{\rm rad}}

\def\etheta{\evect{\col}}			

\def\ipla{{\rm p}}				
\def\istar{\star}					
\def\norb{n_\istar}				
\def\spinrate{\Omega}			

\def\Mbody{M}					
\def\Rbody{R}					
\def\Mstar{\Mbody_\istar}			
\def\Rpla{\Rbody_\ipla}			

\def\iearth{\Earth}				
\def\Rearth{\Rbody_{\iearth}}		

\def\chartime{t}				
\def\ggravi{g}					

\def\Xvect{X}					

\def\period{P}					

\def\itide{{\rm T}}
\def\rtidlock{\rr_{\itide}}

\def\ihz{{\rm HZ}}
\def\rhz{\rr_{\ihz}}

\def\isurf{{\rm s}}				
\def\iground{{\rm gr}}					
\def\iatm{{\rm a}}
\def\iinc{\rm i}
\def\iturb{\rm turb}

\def\sigmaSB{\sigma_{\rm SB}}

\newcommand{\mean}[1]{\overline{#1}}	
\def\up{\uparrow}					
\def\down{\downarrow}				
\def\wiflux{F}				
\def\Fnet{\wiflux_{-}}					
\def\Ftot{\wiflux_{+}}					
\def\Fstar{\wiflux_{\star}}				
\def\Fearth{\wiflux_{\iearth}}
\def\Tearth{\temp_{\iearth}}
\def\optdepth{\tau}					
	
\def\Fup{\wiflux_{\up}}					
\def\Fdown{\wiflux_{\down}}				

\def\Fupsw{\wiflux_{\up \sw}}

\newcommand{\Fupi}[1]{\wiflux_{\up  #1}}
\newcommand{\Fdowni}[1]{\wiflux_{\down #1}}

\def\Finc{\wiflux_{\iinc}}				
\def\Fturb{\wiflux_{\iturb}}
\def\opacity{\kappa}					
\def\psurf{\pressure_{\isurf}}
\def\imom{{\rm M}}
\def\iheat{{\rm H}}
\def\ineutral{{\rm N}}
\def\cdrag{C}
\def\Fsmom{\wiflux_{\imom}}
\def\Fsheat{\wiflux_{\iheat}}
\def\Cmom{\cdrag_{\imom}}
\def\Cheat{\cdrag_{\iheat}}
\def\Cneutral{\cdrag_{\ineutral}}
\def\fcoeffs{f}
\def\fmom{\fcoeffs_{\imom}}
\def\fheat{\fcoeffs_{\iheat}}

\def\lw{{\rm L}}						
\def\sw{{\rm S}}						
\def\wordsw{{\rm W}}
\def\klw{\opacity_{\lw}}				
\def\ksw{\opacity_{\sw}}				
\def\klwww{\klw^{\wordsw}}
\def\kswww{\ksw^{\wordsw}}
\def\propangle{\alpha} 
\def\panglw{\propangle_{\lw}}
\def\pangsw{\propangle_{\sw}}

\def\betalw{\beta_{\lw 0}}				
\def\betasw{\beta_{\sw 0}}				
\def\betascat{\beta_{0}}

\def\Asurfsw{A_{\isurf}}				

\def\ftrans{\mathcal{T}}				
\newcommand{\ftransi}[1]{\ftrans_{#1}}	
\def\gammbeta{\zeta}				

\def\optdepthlw{\optdepth_{\lw}}		
\def\optdepthsw{\optdepth_{\sw}}		
\def\gamplus{\gammbeta_{+}}
\def\gamminus{\gammbeta_{-}}
\def\gampm{\gammbeta_{\pm}}
\def\Amat{\matx{A}}
\def\Bmat{\matx{B}}
\def\Cmat{\matx{C}}
\newcommand{\Amati}[1]{\Amat_{#1}}
\newcommand{\Bmati}[1]{\Bmat_{#1}}
\newcommand{\Cmati}[1]{\Cmat_{#1}}
\newcommand{\Acoeffi}[1]{A_{#1}}
\newcommand{\Bcoeffi}[1]{B_{#1}}
\newcommand{\Ccoeffi}[1]{C_{#1}}
\newcommand{\bcoeffi}[1]{b_{#1}}
\def\etaflux{\eta}
\newcommand{\etafluxi}[1]{\etaflux_{#1}}
\newcommand{\lambi}[1]{\lambda_{#1}}
\newcommand{\mui}[1]{\mu_{#1}}
\def\radfluxvect{\vect{\wiflux}}
\newcommand{\Fvecti}[1]{\radfluxvect_{#1}}
\newcommand{\bvecti}[1]{\vect{d}_{#1}}
\newcommand{\bcompij}[2]{b_{#1}^{#2}}

\def\temperature{T}					
\def\temp{\temperature}
\def\Bblackbody{B}
\newcommand{\Bbbi}[1]{\Bblackbody_{#1}}

\def\iconv{{\rm sen}}

\def\Vconv{\vel_{\iconv}}
\def\Dynbudget{\wiflux}			

\def\meanDyn{\mean{\Dynbudget}}
\def\meanDconv{\meanDyn_{\iconv}}

\def\ilower{{\rm low}}
\def\iupper{{\rm up}}
\def\icollapse{{\rm C}}
\def\density{\rho}					
\def\iday{{\rm d}}					
\def\inight{{\rm n}}					
\def\Cd{C_{\rm D}}					
\def\Tatm{\temperature_{\iatm}}			
\def\Tsurf{\temperature_{\isurf}}			
\def\Tday{\temperature_{\iday}}			
\def\Tnight{\temperature_{\inight}}		
\def\Tnightup{\temperature_{\inight ; \iupper}}
\def\Tnightlow{\temperature_{\inight ; \ilower}}

\def\pclow{\press_{\icollapse ; \ilower}}
\def\pcup{\press_{\icollapse ; \iupper}}
\def\iturbd{{\rm TD}}
\def\inoturbd{{\rm NTD}}
\def\Tntd{\Tnight^{\iturbd}}
\def\Tnntd{\Tnight^{\inoturbd}}

\def\tadvtd{\tadv^{\iturbd}}
\def\tadvntd{\tadv^{\inoturbd}}
\def\tadvKA{\tadv^{\rm KA}}

\def\Cp{C_{\pressure}}				
\def\rhoatm{\density_{\iatm}}			

\def\tetasurf{\teta_{\isurf}}
\def\ilayone{{\rm SL}}
\def\rhofirst{\density_{\ilayone}}
\def\Vthetafirst{\vel_{\col ; \ilayone}}
\def\tetafirst{\teta_{\ilayone}}
\def\Vvecthfirst{\Vvect_{\ihoriz ; \ilayone}}
\def\zzfirst{\zz_{\ilayone}}

\def\deltasigfirst{\Delta \sig_{\ilayone}}

\def\ibulk{0}
\def\karman{\mathcal{K}}
\def\mixingL{\ell}

\def\mixingLmax{\mixingL_{0}}

\def\richardson{{\rm Ri}}
\def\Riz{\richardson}

\def\Ribulk{\richardson_{\ibulk}}
\def\zrough{\zz_{\rm r}}

\def\Rspec{R_{\rm d}}
\def\Rgp{R_{\rm GP}}

\def\Lnorm{L}
\def\Lconv{\Lnorm_{\iconv}}

\def\massemol{\mathcal{M}}
\def\icond{{\rm cond}}
\def\carbondiox{{\rm CO_2}}
\def\Tcondcarbdiox{\temp_{\icond , \carbondiox}}
\def\Mmolatm{\massemol_{\iatm}}
\def\ptriple{\press_{\rm tr}}

\def\area{A}				

\def\ihe{{\rm in}}

\def\tadv{\chartime_{\rm adv}}			
\def\trad{\chartime_{\rm rad}}			

\def\taugrlw{\optdepth_{\isurf ; \lw}}		
\def\ieq{{\rm eq}}
\def\Teq{\temp_{\ieq}}				
\def\thermalpower{Q}
\def\Qin{\thermalpower_{\ihe}}

\def\Qconv{\Qin}

\def\rcp{\kappa}

\def\itop{{\rm t}}
\def\iref{{0}}
\newcommand{\reference}[1]{#1_{\iref}}

\def\ygrid{Y}
\def\zgrid{Z}
\def\sig{\sigma}
\def\scord{s}
\def\coln{\normalized{\col}}
\def\bcourant{b}

\def\force{F}
\def\vel{v}
\def\Vvect{\vect{\vel}}
\def\Vvecth{\Vvect_{\ihoriz}}

\def\teta{\Theta}	
\def\mass{m}		
\def\masscol{\mass_{\rm col}}
\def\psurf{\press_{\isurf}}
\def\ptop{\press_{\itop}}
\def\pisurf{\mathfrak{p}}
\def\geopot{\phi}	
\def\Vz{\vel_{\zz}}				
\def\Vtheta{\vel_\col}				
\def\Vsig{\Dt{\sig}}
\def\Vflux{V}
\def\Wflux{W}
\def\Exner{E}
\def\press{\pressure}
\def\forceh{\force_{\col}}
\def\qmoist{q}
\def\Qheat{Q}

\def\etherm{e}
\def\dtqmoist{\Dt{\qmoist}}

\def\forcen{\normalized{\force}}
\def\forcehn{\forcen_{\col}}
\def\timen{\normalized{\time}}
\def\tetan{\normalized{\teta}}
\def\Exnern{\normalized{\Exner}}
\def\veln{\normalized{\vel}}
\def\Vthetan{\veln_{\col}}
\def\Vsign{\normalized{\Vsig}}
\def\massn{\normalized{\mass}}
\def\psurfn{\normalized{\press}_{\isurf}}
\def\pisurfn{\normalized{\pisurf}}
\def\pressn{\normalized{\press}}
\def\tempn{\normalized{\temp}}
\def\gpotn{\normalized{\geopot}}
\def\qmoistn{\normalized{\qmoist}}
\def\Qheatn{\normalized{\Qheat}}

\def\dtqmoistn{\normalized{\dtqmoist}}
\def\rhon{\normalized{\density}}
\def\vcov{\hat{\vel}}
\def\vcon{\check{\vel}}
\def\dtimen{\Delta \timen}

\def\cmetric{c}
\def\cy{\cmetric_{\ygrid}}
\def\cz{\cmetric_{\zgrid}}
\def\area{\mathcal{A}}
\def\arean{\normalized{\area}}

\def\tetar{\reference{\teta}}
\def\velr{\reference{\vel}}

\def\tempr{\reference{\temp}}
\def\Exnerr{\reference{\Exner}}
\def\tetar{\reference{\teta}}
\def\pressr{\reference{\press}}
\def\forcer{\reference{\force}}
\def\Qheatr{\reference{\Qheat}}
\def\qmoistr{\reference{\qmoist}}

\def\energyr{\reference{\etherm}}
\def\Vzr{\vel_{\zz 0}}
\def\dyndt{\Delta \time}
\def\rhor{\reference{\density}}
\def\dtqmoistr{\reference{\dtqmoist}}
\def\prefteta{\press_{\rm ref}}

\newcommand{\Exneri}[1]{\Exner_{#1}}
\newcommand{\tetai}[1]{\teta_{#1}}
\newcommand{\pressi}[1]{\press_{#1}}

\def\massl{\mass_{\lver}}
\def\Exnerl{\Exner_{\lver}}

\def\tetal{\teta_{\lver}}
\def\zgridl{\zgrid_{\lver}}

\def\timederop{\mathcal{M}}

\def\ntime{n}
\def\physvar{\psi}
\newcommand{\physvari}[1]{\physvar_{#1}}

\def\idiff{{\rm diff}}
\def\qdiff{q}
\def\diffusivity{K}
\def\Kdiffq{\diffusivity_{2 \qdiff}}
\def\Fdiff{\force_{\idiff}}
\def\taudiff{\chartime_{\idiff}}
\def\meandeltacol{\mean{\Delta \col}}
\def\Kdiffqn{\normalized{\diffusivity}_{2 \qdiff}}
\def\diffcoeff{\gamma} 
\def\nosc{n}
\def\Fdiffv{\force_{\idiff ; \vel}}
\def\Fdifftemp{\force_{\idiff ; \temp}}
\def\Kdiffbi{\diffusivity_4}
\def\Kdiffbin{\normalized{\diffusivity}_4}

\def\Kdiffhan{\normalized{\diffusivity}_2}
\def\emissurf{\varepsilon_{\rm \isurf}}
\def\canis{\alpha}

\def\Rossbyrad{L_{\rm Ro}}
\def\Rossbyradnorm{\normal{L}_{\rm Ro}}
\def\BVfreq{N}
\def\cwave{c_{\rm wave}}
\def\Hpress{H}

\def\WTGpar{\Lambda}

\def\isponge{{\rm SL}}
\def\kspon{k_{\isponge}}
\def\ksponmax{k_{\isponge; {\rm max}}}
\def\Vthetaspon{\vel_{\col ; \isponge}}
\def\sigspon{\sig_{\isponge}}
\def\tsponmin{\chartime_{\isponge ; {\rm min}}}

\def\Cvgr{C_{\iground}}
\def\iconduc{{\rm c}}
\def\condgr{\lambda_{\iground}}
\def\inertiagr{I_{\iground}}
\def\zzn{\normalized{\zd}}
\def\zdn{\normalized{\zd}}
\def\Fcond{\wiflux_{\iconduc}}

\newcommand{\zznk}[1]{\zzn_{#1}}
\newcommand{\subsuper}[3]{#1_{#2}^{#3}}
\newcommand{\zzkn}[2]{\subsuper{\zzn}{#1}{#2}}
\newcommand{\tempkn}[2]{\subsuper{\temp}{#1}{#2}}
\def\ntime{n}
\def\dtphys{\Delta \time_{\rm P}}
\newcommand{\dzkn}[2]{\delta \zznk{#1}{#2}}
\newcommand{\ck}[1]{c_{#1}}
\newcommand{\dk}[1]{d_{#1}}
\newcommand{\alphak}[1]{\alpha_{#1}}
\newcommand{\betak}[1]{\beta_{#1}}
\newcommand{\Deltak}[1]{\Delta_{#1}}
\def\vlevgr{N_{\iground}}
\def\Csstar{C_{\isurf}^*}
\def\Fsstar{\wiflux_{\isurf}^*}
\def\Csgr{C_{\isurf}}
\def\Fsgr{\wiflux_{\isurf}}
\def\muint{\mu}
\def\prefteta{\press_{\rm ref}}

\def\streamfunc{\Psi}
\def\euleriansf{\streamfunc}
\def\sfmax{\euleriansf_{\rm max}}
\def\sfmaxana{\euleriansf_{\rm max}^{\rm SL}}

\def\hlev{M}
\def\vlev{N}
\def\jhor{j}
\def\kver{k}
\def\lver{l}

\newcommand{\moy}[2]{\overline{#2}^{#1}}
\newcommand{\moyy}[1]{\moy{\ygrid}{#1}}
\newcommand{\moyz}[1]{\moy{\zgrid}{#1}}
\newcommand{\diff}[2]{\delta_{#1} #2}
\newcommand{\diffy}[1]{\diff{\ygrid}{#1}}
\newcommand{\diffz}[1]{\diff{\zgrid}{#1}}

\def\bccoeff{a}
\newcommand{\bcsurfi}[1]{\bccoeff_{\isurf ; #1}}
\newcommand{\bctopi}[1]{\bccoeff_{\itop ; #1}}
\def\bcsurfb{\bcompij{\isurf}{}}
\def\bctopb{\bcompij{\itop}{}}

\def\xvar{X}
\def\Kdiffx{\diffusivity_{\xvar}}
\def\Fdiffx{\wiflux_{\idiff ; \xvar}}
\def\fdiffx{\mathcal{F}_{\xvar}}

\newcommand{\xvkn}[2]{\xvar_{#1}^{#2}}
\newcommand{\Kdiffxkn}[2]{\diffusivity_{\xvar ; #1}^{#2}}
\newcommand{\rhokn}[2]{\density_{#1}^{#2}}

\newcommand{\zkn}[2]{\zz_{#1}^{#2}}
\newcommand{\masskn}[2]{\mass_{#1}^{#2}}

\def\alphaconv{\alpha}
\def\tetaadiab{\bar{\teta}}
\def\Vthetaadiab{\bar{\vel}_{\col}}
\def\mtop{\mass_{\rm top}}
\def\mbot{\mass_{\rm bot}}

\def\mixratio{\chi}

\def\xvect{\vect{x}}
\newcommand{\xvecti}[1]{\xvect_{#1}}

\def\gamat{\matx{\Gamma}}
\def\betavect{\vect{\beta}}
\newcommand{\gamati}[1]{\gamat_{#1}}
\newcommand{\betavi}[1]{\betavect_{#1}}

\def\nmat{\nn_{\rm MT}}
\def\nphys{\nn_{\rm P}}
\def\nrad{\nn_{\rm RT}}

\newcommand{\eq}[1]{Eq.~(\ref{#1})}
\newcommand{\eqs}[2]{Eqs.~(\ref{#1}) and~(\ref{#2})}
\newcommand{\eqsto}[2]{Eqs.~(\ref{#1}-\ref{#2})}

\newcommand{\append}[1]{Appendix~\ref{#1}}
\newcommand{\units}[1]{~${\rm #1}$}
\newcommand{\fig}[1]{Fig.~\ref{#1}}

\newcommand{\sect}[1]{Sect.~\ref{#1}}

\definecolor{emerald}{rgb}{0.3,0.85,0.2}
\definecolor{smcolor}{rgb}{0.7,0.3,0.0}
\definecolor{grey}{rgb}{0.5,0.5,0.5}

\definecolor{azure}{rgb}{0.0, 0.5, 1.0}

\newcommand{\rem}[1]{}

\begin{document} 

  \title{Meta-modelling the climate of dry tide locked rocky planets}
 
  \subtitle{}
  
  \author{P. Auclair-Desrotour\inst{1,2} \and R. Deitrick\inst{2} \and K. Heng\inst{2,3,4}
          }

  \institute{IMCCE, Observatoire de Paris, Université PSL, CNRS, Sorbonne Université, Université de Lille, 75014 Paris, France 
  	\and University of Bern, Center for Space and. Habitability, Gesellschaftsstrasse 6, CH-3012, Bern, Switzerland
	\and University of Warwick, Department of Physics, Astronomy \& Astrophysics Group, Coventry CV4 7AL, U. K.
	\and Ludwig Maximilian University, University Observatory Munich, Scheinerstrasse 1, Munich D-81679, Germany \\
  \email{pierre.auclair-desrotour@obspm.fr} 
  }

 \date{Received ...; accepted ...}

  \abstract
   {Rocky planets hosted by close-in extrasolar systems are likely to be tidally locked in 1:1 spin-orbit resonance, a configuration where they exhibit permanent dayside and nightside. Because of the resulting day-night temperature gradient, the climate and large-scale circulation of these planets are strongly determined by their atmospheric stability against collapse, which designates the runaway condensation of greenhouse gases on the nightside.} 
   { To better constrain the surface conditions and climatic regime of rocky extrasolar planets located in the habitable zone of their host star, it is therefore crucial to elucidate the mechanisms that govern the day-night heat redistribution.}
   {As a first attempt to bridge the gap between multiple modelling approaches ranging from simplified analytical greenhouse models to sophisticated 3-D General Circulation Models (GCM), we developed a General Circulation Meta-Model (GCMM) able to reproduce both the closed-form solutions obtained in earlier studies, the numerical solutions obtained from GCM simulations, and solutions provided by intermediate models, assuming the slow rotator approximation. We used this approach to characterise the atmospheric stability of Earth-sized rocky planets with dry atmospheres containing $\carbondiox$, and we benchmarked it against 3-D GCM simulations using \texttt{THOR} GCM.}
   {We observe that the collapse pressure below which collapse occurs can vary by ${\sim}40\%$ around the value predicted by analytical scaling laws depending on the mechanisms taken into account among radiative transfer, atmospheric dynamics, and turbulent diffusion. Particularly, we find (i) that the turbulent diffusion taking place in the dayside planetary boundary layer (PBL) globally tends to warm up the nightside surface hemisphere except in the transition zone between optically thin and optically thick regimes, (ii) that the PBL also significantly affects the day-night advection timescale, and (iii) that the slow rotator approximation holds from the moment that the normalised equatorial Rossby deformation radius is greater than~2.  }
   {}


\maketitle


\section{Introduction}

Launched recently from Kourou's spaceport in French Guiana, the \textit{James-Webb} Space Telescope \citep[JWST;][]{Deming2009} is on the point of unravelling the features of exoplanetary atmospheres at resolutions never reached before. With the current or upcoming transit searches of the TESS \citep[][]{Barclay2018} and PLATO \citep[][]{Ragazzoni2016proc} observatories, this telescope will accelerate the dynamics initiated by previous space missions by populating the continuum of extrasolar planets and constraining the properties of the detected objects. Many of these planets are rocky planets in close-in star-planet systems, especially planets orbiting brown dwarves and very-low-mass stars \citep[e.g.][]{PL2007,Raymond2007,Kopparapu2017} such as the seven Earth-sized planets hosted by the TRAPPIST-1 ultra-cool dwarf star \citep[][]{Gillon2017,Grimm2018}. Therefore it is crucial to better understand the mechanisms governing their climate, atmospheric circulation, and surface conditions. 

Tidal locking in 1:1 spin-orbit resonance is the most probable final spin state of planets in close-in star-planet systems \citep[][]{Goldreich1966}. This evolution results from the action of the gravitational tides raised by the perturbing tidal potential of the star. Because of dissipative mechanisms, the tidal response of the planet is delayed with respect to the perturber. As a consequence, the resulting tidal torque tends to drive the spin towards the configuration where the star is motionless in the frame of reference rotating with the planet. This spin state corresponds to spin-orbit synchronisation, and is reached when the spin angular frequency of the planet $\spinrate$ equalises its orbital frequency $\norb$. 

Additionally, gravitational tides act to decrease both the obliquity and the eccentricity of the planet, which is thereby driven towards the equilibrium configurations of coplanarity and circularity \citep[][]{Hut1980,Hut1981} unless it spirals towards the star until being engulfed by it if the system is very close \citep[][]{Hut1981,Levrard2009}. Asynchronous final spin states may also exist. For instance, eccentric orbits maintained by orbital resonances in a multiple-planet system lead to spin-orbit resonances of higher degrees where the planet can be trapped \citep[][]{Correia2014,ADLBM2019}. Similarly, it has been shown that significant thermal tides generated by stellar irradiation are able to prevent Venus-like planets to reach spin-orbit synchronisation by inducing a torque opposed to the solid tidal torque \citep[][]{GS1969,ID1978,Leconte2015,Auclair2019}. 

The probability for a planet to be tidally locked in 1:1 spin-orbit resonance with temperate surface conditions is determined by the interplay between two radii: the tidal lock radius $\rtidlock$ and the radius of the habitable zone $\rhz$. While the tidal lock radius indicates the size of the region where planets are likely to be tide locked in spin-orbit synchronisation, the radius of the habitable zone corresponds to the typical star-planet distance at which a planet can sustain liquid water at its surface \citep[][]{Kopparapu2013}. By assuming that the planet behaves as a black body, and by writing the stellar luminosity as a function of the stellar mass with the empirical formula given by \cite{Barnes2008}, it can be shown that $\rhz  \scale  \Mstar^{1.32}$ for $\Mstar \lesssim 1$ \citep[][]{ADH2020}, whereas the tidal lock radius scales as $\rtidlock \scale \Mstar^{1/3} $ \citep[][]{Peale1977,Kasting1993,Dobrovolskis2009,Edson2011}. Thus, the size of the habitable zone radius decays faster than the tidal lock radius with decreasing the stellar mass, which makes planets located in the habitable zone have more chance to be tide locked if they orbit low-mass stars than if they orbit Sun-like stars \citep[][]{Kasting1993}. 

For planets orbiting the low-mass M stars, tide-locking times are actually very short, and even extremely short in the case of lava planets (e.g. 55 Cancri~e, Kepler 10b), with maximum values hardly reaching a few million years. For instance, the time required for the cool planet LHS1140~b to become tide locked is about 14 million years \citep[][]{PH2019}, which is small compared with the typical ages of planetary systems. This strongly suggests that planets orbiting in the habitable zone of very-low mass stars such as TRAPPIST~1d are tide locked in the 1:1 spin-orbit resonance. Therefore, the rotation rate of these planets is well constrained, and is given by $\spinrate = \norb$. Besides, the strength of tidal forces makes the existence of orbital configurations with significant eccentricities or obliquities unlikely. Such planets can thus be reasonably supposed to be close to the stable equilibrium configurations of coplanarity and circularity. 

In this configuration, the planet exhibits permanent dayside and nightside centred on the star-planet axis. The dayside is irradiated by the incident stellar flux while the nightside is radiatively cooled, and the energy absorbed on the dayside is transported towards the nightside by mean flows, which act to decrease the horizontal temperature gradient \citep[e.g.][]{SG2002,Leconte2013,PH2019}. The differential thermal forcing induced by spin-orbit synchronisation plays a crucial role in the evolution of the planet's climate. Typically, as shown by the pioneering work of \cite{Joshi1997}, the thermal state of synchronously rotating rocky planets is determined by the interplay between the nightside surface temperature and the condensation temperature of greenhouse gases. From the moment that the condensation temperature of the gas exceeds the surface temperature, the nightside acts as a cold trap. The greenhouse gas initially present in the atmosphere condenses and forms an ice sheet on the surface, which induces a temperature decrease in return. This triggers a positive feedback that cools down the atmosphere until the gas has been fully condensed, which is called atmospheric collapse \citep[e.g.][]{Joshi1997,HK2012,Wordsworth2015}. In the opposite case, the atmosphere is said to be stable against collapse, its composition remaining unchanged. 

Over the past decade, a substantial effort has been made to characterise the climate of tide locked rocky planets using a broad range of modelling approaches from simplified analytical greenhouse models \citep[e.g.][]{HK2012,Wordsworth2015,ADH2020} to 3-D General Circulation Model (GCM) simulations \citep[e.g.][]{MS2010,Leconte2013,Carone2014,HaqqMisra2018,DP2020,Sergeev2020,Turbet2018}, including intermediate semi-analytical or numerical approaches \citep[e.g.][]{YA2014,KA2016,Song2021} that cannot be listed here in an exhaustive way. Although based on robust methodologies, most of these approaches cannot be related self-consistently to each other due to major differences in modelling choices. These discrepancies raise two questions. How to disentangle the possible causes of different predictions between two models? How to assess the epistemic value of a given model? This can be reformulated in a more concrete way as: how to consistently characterise the climate of tide locked planets from multiple modelling approaches? \RV{This major concern was formulated explicitly by \cite{Held2005}, who argued for the need of model hierarchies on which to base one's understanding in climate modelling. Such hierarchies appear as the only way to close the gap between idealised modeling and high-end simulations, as they allow for capturing the essence of each particular source of complexity.}

The aim of the present work is to tackle these questions from the angle of atmospheric stability against collapse. In the continuity of a former study on the atmospheric stability of tide locked rocky planets \citep[][]{ADH2020}, we developed a \RV{multi-dimensional model hierarchy} that we call a General Circulation Meta-Model (GCMM) in order to bridge the gap between the analytic theory of planetary climates and simulations performed with 3-D GCMs. \RV{This model hierarchy is based on a systematic bottom-up approach in the spirit of \cite{Held2005}.}

Let us specify the sense given here to meta-modelling. By meta-model, we mean that the model ought to be able to reproduce exactly the setups of both simplified greenhouse models and GCMs -- as well as the configurations in-between -- with the same intrinsic theoretical background. In that sense, such models are possible instances of the meta-model, which can generate any of them. Hence the so-defined GCMM allows for disentangling the effects of mechanisms that are either strongly coupled in standard GCMs or ignored in simplified analytic models. \RV{These effects are added or subtracted as a function of the number of degrees of freedom of the model. Increasing the number of degrees of freedom amounts to adding key sources of complexity.}

Typically, radiative 0-D models are essentially based on radiative exchanges between the planet's surface and the atmosphere \citep[e.g.][]{Wordsworth2015,ADH2020}. At the next level of complexity, 1-D models take the coupling between radiative transfer and the atmospheric structure into account \citep[e.g.][]{RC2012}. Two-column -- or 1.5-D -- models are the minimum setup to couple self-consistently the large-scale day-night overturning circulation with radiative transfer and the atmospheric structure \cite[e.g.][]{YA2014,KA2016}. This coupling is refined at the level of 2-D GCMs, which allow for calculating self-consistently the interaction between physical mechanisms (clouds, turbulent diffusion in the planetary boundary layer, or PBL, convection) and mean flows in the slow rotation regime \citep[e.g.][]{Song2021}. Finally, 3-D GCMs complete the picture by introducing Coriolis effects and non-axisymmetric flows where super-rotation can develop \citep[e.g.][]{Leconte2013,Carone2014,HaqqMisra2018,DP2020,Sergeev2020,Turbet2021}. 

\begin{table*}[h]
\centering
 \textsf{\caption[caption_table1]{\label{tab:models} Physics described by the four studied instances of the meta-model and \texttt{THOR} 3-D GCM$^{\rm a}$.}}
    \begin{tabular}{l c c c c c }
      \hline
      \hline
      \\[-0.3cm]
      \textsc{Grid and Physics} & \textsc{0-D}  & \textsc{1-D} & \textsc{1.5-D} & \textsc{2-D} & \texttt{THOR} \\
      \hline
      \\[-0.3cm]
      Grid format & $1 \times 1$ & $1 \times 50$ & $2 \times 50$ & $32 \times 50$ & 3-D grid \\
      \hline
      \\[-0.3cm]
      Radiative transfer & X & X & X & X & X \\
      Thermal structure &  & X & X & X & X  \\
      Day-night circulation &  &  & X & X & X \\
      Planetary boundary layer &  &  &  & X & X  \\
      Soil heat transfer &  &  &  & X & X  \\
      Coriolis effects & & & & & X  \\
       \hline
    \end{tabular}
     \begin{footnotesize}
      \begin{flushleft}
      \justifying
    $^{\rm a}$ All instances of the meta-model use the same parameters and theoretical background. Physical mechanisms are gradually \RV{captured} by grid formats, which are defined by numbers of horizontal $\times$ vertical grid intervals. In ascending orders of grid resolution, the main physical features described by the meta-model's instances and by \texttt{THOR} GCM are (i) the radiative exchanges between the planet's surface and the atmosphere, (ii) the vertical thermal structure of the atmosphere (i.e. temperature-pressure profile) in radiative equilibrium, (iii) the day-night large-scale circulation, (iv) the convective turbulent diffusion due to friction in the PBL, (v) the soil heat diffusion, and (vi) the vortical components of mean flows due to Coriolis effects.
    \end{flushleft}
    \end{footnotesize}
 \end{table*}

Thus, the essence of a GCMM is to model all these levels of complexity at the same time so that the roles played by the different mechanisms involved in the planet's climate can be clearly separated. As such a model has not been developed yet to our knowledge, the present work is a first attempt to design a GCMM dedicated to the study of tide locked rocky planets. For simplicity, we confine ourselves to the slow rotation regime and ignore Coriolis effects in the dynamics. This allows us to avoid the mathematical complications related to the three-dimensional geometry and to speed up calculations. Our GCMM is thus designed to generate models ranging from 0-D configurations to 2-D configurations. Besides, we opt for a finite-volume method to solve the hydrostatical primitive equations (HPEs), following the approaches detailed by \cite{YS1987} and implemented in standard finite-volume GCMs such as the \texttt{LMDZ} \citep[][]{Hourdin2006} or \texttt{THOR} \citep[][]{Mendonca2016} GCMs. As the finite-volume method divides the atmosphere into cells, it is well appropriate to describe the radiative energy balance models on which the analytic theory is built. For instance, the one-cell configuration ($1 \times 1$ grid) corresponds to the single-layer isothermal atmosphere of Wordsworth's model \citep[][]{Wordsworth2015}. Similarly, increasing numbers of horizontal and vertical grid intervals generate the 1-D ($1 \times 50$ grid), 1.5-D ($2  \times 50$ grid), and 2-D ($32 \times 50$ grid) model setups, respectively.

In order to minimise the size of the parameter space, we confine ourselves to the dry case in this study. The effects of moisture (clouds, latent heat transport, sedimentation, surface condensation or evaporation) are ignored. Radiative transfer is described in the double-gray approximation, meaning that the fluxes are divided into two bands -- shortwave and longwave -- characterised by effective absorption parameters \citep[e.g.][Sect.~4.1]{Heng2017}. We also make the two-stream approximation, and consider that radiative fluxes only travel upwards and downwards \citep[e.g.][Sect.~3.1]{Heng2017}. In addition with radiative transfer, the vertical turbulent diffusivity induced by the interactions between mean flows and the planet surface in the PBL is taken into account by making use of a model based upon the mixing length theory \citep[][]{HB1993}. Finally, the thermal diffusion in the soil is included in the diffusion scheme of the GCMM with a 1-D finite-difference model following the method described by \cite{WCD2016}. As shown by earlier studies \citep[][]{Wordsworth2015,KA2016,ADH2020}, the three above physical ingredients (circulation, radiative transfer, turbulent diffusion) predominantly determine the nightside surface temperature, and thereby the atmospheric stability against collapse. Table~\ref{tab:models} summarises the physics described by the studied instances of the meta-model and \texttt{THOR} 3-D GCM, which is used to benchmark the former. Physical mechanisms are gradually \RV{captured} by the grid formats characterising the models\RV{, as the number of degrees of freedom increases}.

In \sect{sec:preliminary_scalings}, we introduce some control parameters and analytical scaling laws characterising the climate and circulation regime of tide locked planets. In \sect{sec:gcmm}, we detail the main features of the GCMM and the physical setup of the studied Earth-like and pure $\carbondiox$ atmospheres. Section~\ref{sec:climate_regime} introduces the four instances of the meta-model used in this work: 0-D, 1-D, 1.5-D, and 2-D. In \sect{sec:stability}, we run grid simulations for these instances to characterise the atmospheric stability of Earth-sized synchronous planets against collapse as a function of the stellar flux and surface pressure. Particularly, this vertical inter-comparison highlights the influence of the planetary boundary layer on climate, day-night advection, and surface conditions. In \sect{sec:slow_fast_rotation}, we investigate the limitations of the zero-spin rate approximation assumed in this approach by running simulations with \texttt{THOR} 3-D GCM. We show that the approximation holds from the moment that the dimensionless equatorial Rossby deformation length is greater than 2. Finally, in \sect{sec:conclusions}, we summarise the conclusions of the study.

\section{Preliminary scalings}
\label{sec:preliminary_scalings}

The circulation regime and thermal state of equilibrium of tide locked planets is controlled by a few parameters and scaling laws derived either from the shallow water approximation \citep[e.g.][]{Vallis2006,PH2019} or from the weak temperature gradient -- or WTG -- approximation \citep[e.g.][]{Pierrehumbert1995,Sobel2001}. One ought to recall these scalings before introducing the physical setup of the numerical approach.  
\subsection{Circulation regimes of synchronous planets}

If we assume that the planet surface is isotropic, the whole physics and dynamics governing the atmospheric circulation are symmetric about the star-planet axis except Coriolis terms. As a consequence, the circulation regime is essentially characterised by one control parameter depending upon the planet's spin angular velocity, which determines whether mean flows are bi-dimensional and symmetric about the star-planet axis, or if they are sufficiently deviated by the planet's rotation to become three-dimensional. 

Such a parameter appears naturally in analyses making use of the Buckingham-Pi theorem \citep[][]{Buckingham1914} in the primitive equations of fluid dynamics and thermodynamics \citep[e.g.][]{KA2015}. In the present study, following \cite{Leconte2013} and \cite{ADH2020}, we define the dimensionless equatorial Rossby deformation length $\Rossbyradnorm$ from the equatorial Rossby deformation radius $\Rossbyrad$ as \citep[e.g.][]{Menou2003}
\begin{equation}
\Rossbyradnorm \define \frac{\Rossbyrad}{\Rpla} = \sqrt{\frac{\cwave}{2 \spinrate \Rpla}},
\label{Ronorm}
\end{equation}
where $\Rpla$ designates the planet radius, and $\cwave $ the speed of horizontally propagating gravity waves. The dimensionless equatorial Rossby deformation length is essentially the square root of the distance -- in radius unit -- that fast gravity waves can travel before they feel the Coriolis effects and geostrophically adjust \citep[][]{Vallis2006}. 

If $\Rossbyradnorm > 1 $, the Coriolis effects are small and the two-way force balance between advection and pressure-gradient accelerations leads to a day-night overturning circulation symmetric about the star-planet axis \citep{Leconte2013,PH2019,HL2021}. In this regime, the dynamics of mean flows is the same in all planes containing the star-planet axis, with high-altitude winds blowing from the dayside to nightside and near-surface winds blowing from the nightside to dayside. This essentially corresponds to the steady state expected in the Weak Temperature Gradient theory, where small Coriolis forces, friction, and nonlinearities make the heat advection unable to annihilate completely the day-night temperature gradient \citep[][]{Sobel2001,HP2017,PH2019}. 

Conversely, for $\Rossbyradnorm \lesssim 1$, \cite{SP2011} demonstrated that the formation of standing planetary-scale equatorial Rossby and Kelvin waves \citep[i.e. waves restored by the Coriolis acceleration; see e.g.][]{LS1997} favours the emergence of super-rotation by pumping angular momentum from the mid-latitudes towards the equator. In this regime, the equatorial Rossby deformation radius ($\Rossbyrad$) essentially corresponds to the latitudinal width of the produced eastward equatorial jet, and mean flows take the form of the Matsuno-Gill standing wave pattern \citep[][]{Matsuno1966b,Gill1980,SP2011,Tsai2014}.

The dimensionless equatorial deformation length introduced in \eq{Ronorm} can be related to the atmospheric parameters by considering the properties of gravity waves. Gravity waves are restored by the Archimedean force associated with the fluid buoyancy and their typical speed is given by $\cwave = \Hpress \BVfreq$, where $\Hpress$ designates the characteristic vertical scale length of the atmosphere and $\BVfreq$ the Brunt-Vaisala frequency, which scales the strength of the atmospheric stratification against convection \citep{GZ2008}. In a dry stably stratified atmosphere, this frequency is expressed as \citep[][]{GZ2008}
\begin{equation} 
\BVfreq^2 = \frac{\ggravi}{\temperature} \left( \frac{\ggravi}{\Cp} + \dd{\temperature}{\zz} \right),
\label{BVfreq}
\end{equation}
where we have introduced the gravity $\ggravi$, the heat capacity per unit mass of the gas $\Cp$, the temperature $\temperature$, and the partial derivative operator over the altitude $\dd{}{\zz}$. In the idealised case of the vertically isothermal atmosphere (constant temperature), $\BVfreq =\ggravi / \sqrt{ \Cp \temp }$, and the vertical scale is the pressure height \citep[][Sect.~1.4]{Vallis2006},
\begin{equation}
\Hpress \define \frac{\Rspec \temperature}{\ggravi},
\label{Hpress}
\end{equation}
where $\Rspec \define \Rgp / \Mmolatm$ designates the specific gas constant for dry air, $\Rgp$ being the ideal gas constant and $\Mmolatm$ the mean molecular weight of the atmosphere. Thus, in this configuration \citep[e.g.][]{Leconte2013}, 
\begin{equation}
\Rossbyradnorm = \sqrt{ \frac{\Rspec \temperature^{1/2}}{2 \spinrate \Rpla \Cp^{1/2}}},
\label{Roiso}
\end{equation}
which highlights the fact that the circulation regime depends on the the planet's spin rotation, thermal state, and atmospheric composition. 

The dimensionless equatorial Rossby deformation length conveys exactly the same information as the WTG parameter~$\WTGpar$ introduced in the Weak Temperature Gradient theory \citep[see e.g.][]{PH2019}, which is defined as the ratio of the Rossby radius of deformation\RV{ -- distinct from the equatorial deformation radius --} normalised by the planet radius \citep[][Sect.~3.8.2]{Vallis2006}. The two parameters are linked together by the relationship $\Rossbyradnorm = \sqrt{\WTGpar/2}$ \citep[][]{PH2019}, meaning that either the former or the latter can be chosen to characterise the circulation regime. In the present study, we confine ourselves to the configuration of the WTG theory ($\Rossbyradnorm > 1$) and consider thereby that mean flows are symmetric about the star-planet axis. 

In addition with the slow and fast rotators regimes, there exists a third dynamical state that is proper to intermediate stellar cases in the range of 3000-3300~K and that is described as the Rhines rotation regime \citep[][]{HaqqMisra2018,Sergeev2020}. This regime is related to the Rhines length, which determines the maximum extent of zonally elongated turbulent structures \citep[][]{Rhines1975}. It occurs when the non-dimensional Rossby deformation radius is greater than one but the non-dimensional Rhines length less than one \citep[][]{HaqqMisra2018}. The slow rotation and fast rotation regimes occur when both the non-dimensional Rhines length and Rossby deformation radius are greater or less than one, respectively. The Rhines rotation regime is not considered here, meaning that we focus on the configuration where both the non-dimensional Rhines length and Rossby deformation radius are greater than one. 

\subsection{Thermal states predicted by radiative box models}

Over the past decade, analytic solutions and scalings characterising the thermal state of equilibrium of tide locked planets have been obtained both for hot Jupiters \citep[][]{KS2016,ZS2017,KK2018}, lava planets \citep[][]{HP2017}, and cooler rocky planets orbiting in the habitable zone of their host star \citep[][]{Showman2013book,Wordsworth2015,KA2016,PD2016,Koll2022,ADH2020}. Most of them were derived in the framework of the WTG theory and involve simplified atmospheric physics and structure. The present study builds on these works, and particularly those based on box model approaches, where the atmosphere and surface are reduced to large scale energy reservoirs exchanging heat with each other \citep[][]{Wordsworth2015,KA2016,ADH2020}. Although they are based on strong simplifications (isothermal atmosphere, large-scale averages, no self-consistent coupling between the dynamics and the thermodynamics), these models provide scalings that capture the behaviour of the thermal state of tide locked rocky planets with a minimum set of physical parameters. Particularly, they lead to closed-form solutions for the nightside surface temperature $\Tnight$, which determines the whole atmospheric stability against collapse. 

By considering a globally isothermal atmosphere interacting with dayside and nightside surface hemispheres, \cite{Wordsworth2015} shows that the pure radiative equilibrium of the surface-atmosphere system corresponds, in the optically thin layer approximation (i.e. transparent in the visible and optically thin in the infrared), to the nightside temperature scaling \citep[][Eq.~(29)]{Wordsworth2015}
\begin{equation}
 \Tnightlow \define \Teq \left[ \frac{1}{2} \left( 1 - \Asurfsw \right) \taugrlw  \right]^{\frac{1}{4}},
 \label{Tnlow}
\end{equation}
which can be generalised to optically thick atmospheres with scattering \citep[][]{ADH2020}. In the above expression, $\Asurfsw$ designates the surface albedo in the shortwave, $\taugrlw$ the longwave optical depth at planet's surface, and $\Teq$ the black body equilibrium temperature, which is defined as 
\begin{equation}
\Teq \define \left( \frac{\Fstar}{4 \sigmaSB} \right)^{\frac{1}{4}},
\label{Teq}
\end{equation}
with $\Fstar$ the incident stellar flux and $\sigmaSB = 5.670367 \times 10^{-8} $\units{W~m^{-2}~K^{-4}} the Stefan-Boltzmann constant \citep[][]{codata2014}. 

Since it ignores all types of energy exchanges except radiative transfer, the estimate given by \eq{Tnlow} can be interpreted as a lower bound for the nightside surface temperature of a rocky tide locked planet. In reality, the strong convection generated by the thermal forcing of the atmosphere in the dayside planetary boundary layer increases surface-atmosphere heat fluxes, which significantly affects the nightside temperature \citep[][]{Sergeev2020}. The friction of the flow against the surface generates sensible heat exchanges in dry thermodynamics. Additionally, in moist atmospheres, surface evaporation generates latent heat exchanges, which results from the energy taken from or released in the fluid during the changes of phases of the component \citep[][]{Pierrehumbert2010}. 

We consider here that the atmosphere is dry and thereby ignore the latter contribution. Sensible heat exchanges can be introduced in the radiative box model by including, in the energy balance equations, the hemisphere-averaged sensible heat flux given by \citep[e.g.][Eq.~(6.11), p. 396]{Pierrehumbert2010}
\begin{equation}
\meanDconv = \Cd \Cp \rhoatm \Vconv \left(\Tday - \Tatm \right),
\label{Fturb}
\end{equation}
where $\rhoatm$ is the atmospheric density at planet's surface, $\Cd$ the bulk drag coefficient characterising the strength of friction in the surface layer, and $\Vconv$ the typical horizontal wind speed of the flow. Among these parameters, $\Vconv$ accounts for the circulation, meaning that it cannot be self-consistently related to the thermal state of the system in this simplified approach. Nevertheless, it can be scaled from a dimensional analysis of the thermodynamic equation \citep[e.g.][]{Wordsworth2015}, or by making use of the heat engine theory \citep[e.g.][]{KA2016,KK2018,ADH2020}. For instance, by modelling the overturning circulation as an ideal heat engine and using Carnot's theorem \citep[][]{Bruhat1968book}, \cite{KA2016} found an upper bound of the dayside average surface wind speed \citep[][Eq.~(12)]{KA2016},
\begin{equation}
\Vconv = \left\{ \left[ \Tday - \left( 1 - \Asurfsw \right)^{\frac{1}{4}} \Teq \right] \left( 1 - \expo{- \taugrlw } \right) \left( 1 - \Asurfsw \right) \frac{ \Rspec \Fstar}{2 \Cd \psurf}    \right\}^{\frac{1}{3}},
\label{Vconv}
\end{equation}
which agrees well with the values obtained numerically from 3-D GCM simulations \citep[][]{KA2016,KK2018}.

For an isentropic cycle (i.e. an idealised Carnot's heat engine), the weight of dayside sensible heating\footnote{Sensible heating designates surface-atmosphere heat exchanges due to the vertical turbulent diffusion generated by friction of mean flows against the planet's surface within the surface layer in dry thermodynamics. This mechanism is distinct from latent heat exchanges, which designates the energy exchanges resulting from the change of phase of a condensable substance \citep[e.g.][Sect.~6.3]{Pierrehumbert2010}.} relatively to radiative heating is controlled by the dimensionless parameter \citep[][Eq.~(63)]{ADH2020}
\begin{equation}
\Lconv \define \frac{2 \Cp \Cd \psurf }{\taugrlw \Rspec \Fstar} \left[ \frac{\Qconv \Rspec }{\Cd \psurf} \left( \frac{\Fstar}{2 \sigmaSB} \right)^{\frac{1}{4}} \right]^{\frac{1}{3}},
\label{Lconv}
\end{equation}
which is written here for an atmosphere optically transparent in the shortwave and thin in the longwave. The notation $\Qconv \scale \Vconv^3$ \citep[e.g.][]{KA2016} designates the amount of power per unit area available to drive atmospheric motion. Looking at the zero-convection limit ($\Lconv = 0$) we recover the purely radiative regime, while the opposite asymptotic regime ($\Lconv =+ \infty$) implies that $\Tday = \Tatm$ and provides an upper bound for the nightside surface temperature of a tide-locked rocky planet \citep[e.g.][Eq.~(85)]{ADH2020},
\begin{equation}
\Tnightup \define \Teq \left[ 2 \left( 1 - \Asurfsw \right) \taugrlw \right]^{\frac{1}{4}} = \sqrt{2} \Tnightlow,
\label{Tnup}
\end{equation}
which is valid in the optically thin layer approximation as well as \eq{Tnlow}. Therefore, for a globally isothermal and optically thin atmosphere, the nightside surface temperature falls into the interval 
\begin{equation}
\Tnightlow \leq \Tnight \leq \sqrt{2} \Tnightlow.
\label{Tnight_bounds}
\end{equation}
However, we shall remark that $\Tday = \Tatm$ actually corresponds to an extreme regime that is never reached in standard configurations, and we shall thus consider $\Tnightup$ as a theoretical upper limit.   

Similarly as the dayside convective planetary layer, the nightside atmospheric structure alters the nightside equilibrium temperature. Its effect can be quantified by relaxing the isothermal approximation and by dividing the atmosphere into dayside and nightside air columns, which is the essence of two-column models \citep[][]{YA2014,KA2016,ADH2020}. As shown by \cite{KA2016}, the nightside subsidence induced by the day-night overturning circulation generates a temperature inversion in the lowest layers of the atmosphere if the subsidence timescale is slightly greater than the radiative cooling timescale. The resulting atmospheric structure leads to large day-night differences. 

\section{A General Circulation Meta-model (GCMM)}
\label{sec:gcmm}

We introduce in this section the main features of the meta-model and the used physical setup.

\subsection{Primitive equations}

At a given time $\time$, the dynamical core of the GCMM solves the hydrostatical primitive equations over the Cartesian rectangular domain defined by the colatitude $\col \in \left[ 0^\degree , 180^\degree \right] $ of the tidally locked coordinates \citep[or TLC; see][Appendix~B]{KA2015} and the mass-based vertical coordinate defined, in the absence of the topography, as \citep[e.g.][]{YS1987,Carone2014}

\begin{equation}
\sig \define \frac{\press - \ptop}{\psurf - \ptop} \in \left[ 0 , 1 \right] ,
\label{sig}
\end{equation}
where we have introduced the pressure $\press$, the surface pressure $\psurf$, and the pressure at the top of the atmosphere $\ptop$. In these coordinates, $\col = 0^\degree$ and $\col = 180^\degree$ correspond to the substellar and anti-stellar points, respectively, while $\sig = 0 $ and $\sig = 1$ correspond to the top and the bottom of the atmosphere, respectively. While $\psurf$ and $\press$ vary over time and spatial coordinates, $\ptop$ is set to $\ptop = 0$ in the model, which corresponds to the usual sigma-coordinate $\sig = \press / \psurf$. The vertical coordinate given by \eq{sig} is well suited to the study of the tide locked planets since it follows the distortion of the atmosphere induced by the differential day-night thermal forcing: the pressure of an altitude level may differ by orders of magnitude between the dayside and nightside, which would possibly generate numerical issues with the altitude coordinate. 

The relationship between the altitude ($\zz$) and the generalised vertical coordinate ($\sig$) is contained in the so-called pseudo-density \citep[e.g.][]{Kasahara1974},
\begin{equation}
\pisurf \define - \ggravi \density \dd{\zz}{\sig},
\label{pisurf}
\end{equation}
where $\density$ denotes the density. The pseudo-density is proportional to the mass contained in a generalised volume where the vertical dimension is not a length but an interval of the generalised coordinate~$\sig$. With the chosen mass-based coordinate (\eq{sig}) and the assumed hydrostatic balance, it is simply expressed as \citep[e.g.][]{YS1987}
\begin{equation}
\pisurf = \psurf - \ptop.
\end{equation}
We note that $\pisurf$ would be the density to a constant factor if the vertical coordinate were the altitude. Using the pseudo-density instead of the density allows for writing the HPEs given further in the same form for any chosen vertical coordinate. 

The HPEs are the mass continuity equation \citep[e.g.][]{Kasahara1974},
\begin{equation}
\dd{\pisurf}{\time} + \divx{\sig} \left(  \pisurf \Vvecth \right) + \dd{}{\sig} \left( \pisurf \Vsig  \right) = 0,
\label{hpe1}
\end{equation}
the horizontal momentum equation,
\begin{align}
& \dd{}{\time} \left(  \pisurf \Vtheta \sin \col \right) + \frac{1}{\Rpla} \dd{}{\col} \left( \pisurf \Vtheta^2 \sin \col \right)  & \\
 &+ \dd{}{\sig} \left( \pisurf \Vtheta \Vsig \sin \col \right) + \frac{1}{\Rpla} \pisurf \sin \col \left[ \dd{\geopot}{\col} +  \teta \dd{\Exner}{\col} \right] = \pisurf \sin \col  \forceh. \nonumber
\end{align}
the potential temperature equation,
\begin{equation}
\dd{\pisurf \teta}{\time}   + \divx{\sig} \left( \pisurf \teta \Vvecth \right) + \dd{}{\sig} \left( \pisurf \teta \Vsig \right) = \pisurf  \frac{\Qheat}{ \Exner},
\end{equation}
and the hydrostatic equation combined with the ideal gas law, 
\begin{equation}
\dd{\geopot}{\sig} +  \teta \dd{\Exner}{\sig} = 0,
\label{hpe4}
\end{equation}
where we have introduced the horizontal velocity vector $\Vvecth \define \Vtheta \, \etheta$, the sigma-velocity $\Vsig \define \DD{\sig}{\time} $ (with $\DD{}{\time} $ the material time derivative), the geopotential $\geopot \define \ggravi \zz$, the potential temperature~$\teta$, the Exner function $\Exner$ \citep[e.g.][Sect.~3.9]{Vallis2006}, and the horizontal divergence operator at constant $\sig$,
\begin{equation}
\div \define \frac{1}{\Rpla} \etheta \scalprod \dd{}{\col}. 
\end{equation}
The potential temperature and Exner function are defined as 
\begin{align}
\label{teta_exner}
& \teta \define \temp \left( \frac{\press}{\prefteta} \right)^{-\rcp}, & \Exner \define \Cp  \frac{\temp}{\teta} =  \Cp \left( \frac{\press}{\prefteta} \right)^{\rcp},
\end{align}
where $\temp$ is the temperature, $\prefteta$ a constant reference pressure set to $\prefteta = 1$~bar, and $\rcp \define \Rspec / \Cp$. We note that the Exner function is a proxy for pressure. It is used for convenience, as it facilitates the integration of the hydrostatic equation. In right-hand members of \eqsto{hpe1}{hpe4}, the source-sink terms are the force per unit mass $\forceh$, and the heat power per unit mass $\Qheat$. We note that the primitive equations are written in their conservative forms, which involve mass flows and mass-integrated quantities rather than the original quantities themselves. Besides, these equations are given here for any vertical coordinate varying monotonically with altitude for the sake of generality. 

The nondimensional HPEs derived from \eqsto{hpe1}{hpe4} (see \append{app:nondimensional_hpes}) are solved for $\left\{ \pisurf, \Vtheta, \teta, \geopot \right\} $ on a staggered Arakawa~C grid \citep[][]{AL1977} with uniformly spaced horizontal intervals, and $\sig$-dependent vertical intervals refined near the model bottom and top (see \append{app:dynamical_core}). Following the method implemented in the \texttt{LMDZ} GCM \citep[][]{Hourdin2006}, the integration is done using a leapfrog scheme with a periodic predictor/corrector timestep. The source-sink terms associated with the physics, $\left\{ \forceh, \Qheat \right\}$, as well as other physical variables, are updated periodically using implicit schemes except radiative transfer equations, which are solved directly from the current thermodynamical state. In the 2-D GCM-like configurations, integrating the HPEs on a discrete domain generates numerical instabilities that develop at grid scale and may strongly disrupt calculations. In GCMs, this concern is usually handled by introducing horizontal hyper-diffusion, which is ideally designed to damp efficiently the numerical instabilities at grid scale while leaving the mean flows unchanged \citep[e.g.][]{TC2011}. Therefore we include in the model a fourth-order hyper-diffusion (or bi-harmonic diffusion) using an anisotropic super-diffusivity that vanishes at the substellar and anti-stellar points (see \append{app:hyper_diff}) in order to avoid the stability concerns associated with isotropic diffusion near the poles \citep[][Sect.~13.3]{Lauritzen2011}. \RV{The corresponding hyperdiffusion terms for the momentum and temperature equations are given by}
\begin{align}
& \Fdiffv = - \Kdiffbi \sin^{2 \canis} \left( \col \right) \nabnh{4} \Vtheta, \\
& \Fdifftemp = - \Kdiffbi \sin^{2 \canis} \left( \col \right) \nabnh{4} \temp,
\end{align}
\RV{where $\nabnh{4} \define \laph  \laph $ designates the second order horizontal hyper-Laplacian operator, $\canis$ the anisotropy exponent ($\canis=1$ in the model), and $\Kdiffbi$ the super-diffusivity defined by \eq{Kdiffq}. Validation test simulations were run to verify that the mean flow and temperature distribution are insensitive to the hyper-diffusion scheme (see \fig{fig:validation_hydiff}).}

In very hot cases, exponentially growing gravity waves propagating upwards and reflected by the upper boundary of the model can lead to extreme fluctuations of the dynamical quantities in the upper atmosphere. To address these instabilities, it can be necessary to use a sponge layer in addition with horizontal hyper-diffusion. In the present work, we introduce a linear Rayleigh friction sponge \citep[e.g.][Sect.~13.4.5]{Lauritzen2011} following the formulation proposed by \cite{PK2002} for the vertical profile of the Rayleigh coefficient (see \append{app:sponge_layer}). The sponge layer is thus modelled by a Rayleigh damping increasing with the altitude between a critical level $ \sig = \sigspon $ and the top of the atmosphere, $\sig = 0$, which tends to annihilate horizontal winds in the vicinity of the upper boundary. Finally, the strong convection generated by the thermal forcing on the dayside can induce superadiabatic vertical temperature gradients\RV{ (i.e. $\partial \teta / \partial \zz<0$)}, which is another source of numerical errors. This behaviour can be prevented by introducing a convective adjustment scheme in the model. The convective adjustment scheme regularises the atmospheric structure by dynamically correcting the tendencies to adjust superadiabatic regions to adiabatic profiles. We implemented a standard scheme \citep[e.g.][]{Hourdin1993,MB2020} that can be activated when necessary (see \append{app:convadj}). However, we did not have to use the convective adjustment scheme nor the sponge layer in the simulations performed for this work.  

\subsection{Radiative transfer}

In the model, radiative transfer is described through the double-gray approximation, which consists in (i) decoupling the stellar radiation (shortwave flux) and the planet radiation (longwave flux) -- each band being characterised by an effective optical depth --, and in (ii) assuming that radiative fluxes only propagate upwards and downwards, which is the essence of the two-stream approximation \citep[e.g.][Sects.~3.1 and~4.1]{Heng2017}. This allows for calculating fluxes in the shortwave and in the longwave separately (see \append{app:rad_transfer}). We denote upward and downward fluxes by $\Fup$ and $\Fdown$, respectively. The equations governing the propagation of the wavelength-integrated total flux $\Ftot \define \Fup + \Fdown$ and net flux $\Fnet \define \Fup - \Fdown$ have the same formulation in both bands. They are written as \citep[][]{Heng2017}
\begin{align}
\label{eqrt1}
& \DD{\Ftot}{\optdepth} = \frac{1}{\betascat} \Fnet, \\
\label{eqrt2}
& \DD{\Fnet}{\optdepth} = \betascat \left( \Ftot - 2 \Bblackbody \right),
\end{align}
where $\Bblackbody \define \sigmaSB \temp^4$ designates the black body radiation of the gas, which is zero in the shortwave since the atmosphere is assumed to radiate in the infrared only. 

In these equations, $\optdepth$ designates the optical depth of the associated band, and $\betascat$ the scattering parameter ($\betascat = 1$ for pure absorption; $0 < \betascat <1 $ in the presence of scattering). The fluxes equations given by \eqs{eqrt1}{eqrt2} are solved numerically in the code by computing the transmission functions of each atmospheric layer as a first step, and by solving the boundary condition problem as a second step. We note that the solution obtained numerically for the single-layer atmosphere this way exactly corresponds to that derived in radiative box models based on the isothermal atmosphere approximation \citep[see e.g.][]{Wordsworth2015,ADH2020}. The optical depths in the shortwave $\optdepthsw$ and longwave $\optdepthlw$ are both assumed to scale linearly with pressure, and are defined as
\begin{align}
\label{optical_depths}
& \optdepthsw \define \frac{\ksw \press}{\ggravi}, & \optdepthlw \define \frac{\klw \press}{\ggravi},
\end{align}
where we have introduced the effective absorption coefficients of the gas in the short- and longwave, $\ksw$ and $\klw$, respectively. 

\RV{We remark that the optical depths defined by \eq{optical_depths} depict horizontally averaged vertical profiles rather than local profiles varying as functions of the propagation angles of radiative fluxes, $\pangsw$ and $\panglw$. Therefore, the effective absorption coefficients $\ksw$ and $\klw$ introduced in \eq{optical_depths} are related both to the absorption properties of the gas and to the mean cosine of the propagation angles in the visible and in the infrared, $\mean{\cos \pangsw}$ and $\mean{\cos \panglw}$. Typically, these coefficients are related to Wordsworth's absorption coefficients \citep[][Eq.~(12)]{Wordsworth2015} -- denoted by $ \kswww$ and $\klwww$ -- by the equations\footnote{\RV{In \cite{Wordsworth2015}, the atmosphere is assumed to be transparent in the visible and the mean cosine of the propagation angle in the infrared is set to $\mean{\cos \panglw} = 0.5$. As a consequence, the effective absorption coefficient $\klw$ of \eq{optical_depths} is exactly twice larger than Wordsworth's value, the optical depths being equal.}} $\ksw = \kswww / \, \mean{\cos \pangsw}$ and $\klw = \klwww /\, \mean{\cos \panglw}$. Therefore, changing the value of the mean cosine of the propagation angle in this definition amounts to changing the value of the absorption coefficient in \eq{optical_depths}. One shall also bear in mind that the absorption coefficients defined in the double-gray approach are not fundamental parameters of the gas but parameters that mimic the averaged effect of highly frequency-dependent atmospheric opacities, as shown by \cite{Wordsworth2010a} for $\carbondiox$-dominated atmospheres. }

\RV{Although it does not fully account for the complex physics of radiative transfer, the adopted double-gray approximation with average optical depths captures the dependence of optical depths upon pressure, and it allows for fast numerical computation of radiative fluxes. The radiative transfer scheme may be refined in future works by using more sophisticated approaches such as the correlated-k distribution method \citep[e.g.][]{LO1991}, but this goes beyond the scope of the present study.}

\subsection{Planetary boundary layer}
\label{ssec:pbl}
In the planetary boundary layer, the shear instability generates turbulence, which acts to mix the flow. This turbulent mixing induces a vertical diffusion of momentum, heat, and potentially moisture in moist cases, near the planet's surface. The associated eddy diffusivities are controlled by the gradient Richardson number $\Riz$ defined as \citep[][]{Vallis2006}
\begin{equation}
\Riz \define \frac{\ggravi \dd{\teta}{\zz} }{\teta \left( \dd{\Vtheta}{\zz} \right)^2 },
\label{Ri_gradient}
\end{equation}
which characterises fluid stratification. The Richardson number is essentially the ratio of the production of turbulent energy due to the shear instability over the restoring force induced by buoyancy. For a given quantity $\xvar = \Vtheta, \Cp \teta$ (or $\qmoist$ in moist cases, $\qmoist$ being the specific humidity), the vertical diffusion equation is written, in the gradient-flux theory \citep[e.g.][]{Garratt1994}, as
\begin{equation}
\DD{\xvar}{\time} = \frac{1}{\density} \dd{}{\zz} \left[ \density \Kdiffx \dd{\xvar}{\zz} \right] ,
\label{diffusive_term}
\end{equation}
the upward diffusive flux being given by
\begin{equation}
\Fdiffx = - \density \Kdiffx \dd{\xvar}{\zz}. 
\end{equation}
In these equations, $\Kdiffx$ is the eddy diffusivity associated with turbulent mixing for $\xvar$. This parameter is a function of the mean fields. The lower boundary condition is a continuity condition determined by the exchanges with the planet's surface, while the upper condition is a zero-flux condition. In a dry atmosphere, the tendencies for the momentum and potential temperature equations in the dry case are expressed from the diffusive terms given by \eq{diffusive_term} as 
\begin{equation}
\begin{array}{ll}
\displaystyle \forceh = \DD{\Vtheta}{\time} , & \displaystyle \Qheat = \left( \frac{\press}{\prefteta} \right)^{\rcp} \DD{}{\time} \left( \Cp \teta \right). 
\end{array}
\end{equation}
To calculate the eddy diffusivities, we make use of the formulation given by \cite{HB1993}. In that model, the eddy diffusivities of \eq{diffusive_term} are expressed as functions of a mixing length scale $\mixingL$, the local shear $\abs{\dd{\Vtheta}{\zz}}$, and the gradient Richardson number $\Riz$ defined by \eq{Ri_gradient}. They read \citep[e.g.][Eq.~(3.2)]{HB1993}
\begin{equation}
\Kdiffx = \mixingL^2  \abs{\dd{\Vtheta}{\zz}} \fdiffx \left( \Riz \right),
\label{Kdiffx}
\end{equation}
where $\fdiffx \left( \Riz \right) $ describes the functional dependence of $\Kdiffx$ on the gradient Richardson number. The form of $\fdiffx$ is determined by the turbulent regime, which can be either stable ($\Riz \geq 0 $) or unstable ($\Riz < 0$). The mixing length is expressed as \citep[][]{Blackadar1962}
\begin{equation}
\label{mixing_length}
\mixingL = \frac{ \mixingLmax \karman \zz  }{\karman \zz + \mixingLmax }, 
\end{equation}
the parameter $\karman \approx 0.4$ being the von Karman constant \citep[e.g.][]{Garratt1994}, and $\mixingLmax$ the asymptotic length scale ($\mixingL \approx \mixingLmax$ for $\karman \zz \gg \mixingLmax$), which varies as a function of $\zz$ (see \append{app:stability_functions}). 

The turbulent friction of mean flows against the soil in the surface layer leads to sensible momentum and heat exchanges that are described in the form of parametrised surface fluxes. Denoting by M and H the subscripts for the momentum and heat components, respectively, we write the upward momentum and heat surface fluxes as
\begin{align}
\label{Fsmom}
& \Fsmom = -  \Cmom \rhofirst  \abs{\Vvecthfirst} \Vthetafirst, \\
\label{Fsheat}
& \Fsheat = -  \Cheat \rhofirst \Cp  \abs{\Vvecthfirst} \left( \tetafirst - \tetasurf \right),
\end{align}
where the subscripts $\isurf$ and $\ilayone$ refer to values at the surface and at the top of the surface layer\footnote{The top of the surface layer is taken at the middle of the lowest model layer. Therefore, it corresponds to the lowest model level.}, respectively. The surface-layer exchange coefficients $\Cmom$ and $\Cheat$ are defined as \citep[][]{HB1993} 
\begin{align}
\label{flux_mom}
& \Cmom \define \Cneutral \fmom \left( \Ribulk \right) , \\
\label{flux_heat}
& \Cheat \define \Cneutral \fheat \left( \Ribulk \right).
\end{align}
Here, $\Cneutral$ designates the neutral exchange coefficient \citep[e.g.][]{HB1993},
\begin{equation}
\Cneutral \define \left[ \frac{\karman}{\ln \left( 1 + \zzfirst / \zrough  \right) } \right]^2,
\end{equation}
where $\zrough$ denotes the roughness height, while $\fmom$ and $\fheat$ are two functions of the bulk Richardson number,
\begin{equation}
\Ribulk \define \frac{\ggravi \zzfirst \left( \tetafirst - \tetasurf \right)}{\tetafirst \abs{\Vvecthfirst}^2 },
\label{Ribulk}
\end{equation}
which controls the stability of the surface layer. We note that the bulk Richardson number $\Ribulk$ corresponds to the local gradient Richardson number $\Riz$ (\eq{Ri_gradient}) characterising the surface layer. The functions $\fmom$ and $\fheat$, as well as the functions $\fdiffx$ introduced in \eq{Kdiffx}, are detailed in \append{app:stability_functions}. The physical tendencies resulting from turbulent diffusion are evaluated every physical time step using an implicit scheme (see \append{app:discretisation_diff}).

\RV{As it accounts for the dependence of eddy diffusivities on the gradient Richardson number, the above turbulent diffusion scheme describes both the regime of strong convection occurring on dayside, and the regime of stable stratification associated with the nightside temperature inversion (see Figs.~\ref{fig:snapshots_HB1993_diff} and \ref{fig:potential_temp} in the following). It thus captures the evolution of turbulent diffusivities in the PBL between the dayside -- where they are high --, and the nightside -- where they are low. However, we note that the used standard formulation of turbulent diffusion is not sophisticated enough to account properly for the heat and momentum exchanges due to turbulent flows in the case of strong stratification ($\Riz \gg 1$). In this regime, the vertical momentum mixing continues even at relatively high $\Riz$ due to the momentum transport of vertically propagating internal gravity waves, which may increase the surface-atmosphere sensible heat exchanges and warm up the nightside surface \citep[e.g.][]{Sukoriansky2005,Joshi2020}. Considering this effect, the used turbulent scheme might tend to underestimate the atmospheric stability against collapse. Nevertheless, we adopt it as a convenient compromise between simplicity and realism, which can be refined in the future by using more advanced turbulent diffusion models \citep[e.g.][]{Sukoriansky2005}. }

\begin{table*}[h]
\centering
 \textsf{\caption{\label{tab:param_reference_case} Values of parameters used in the two reference cases of the present work. The acronyms SW and LW are used in place of 'shortwave' and 'longwave', respectively. }}
\begin{small}
    \begin{tabular}{l l c c c}
      \hline
      \hline
      \textsc{Symbol} & \textsc{Description}  & \textsc{Units} & \textsc{Earth-like case} & \textsc{Pure $\carbondiox$ case}  \\ 
      \hline
      \\[-0.2cm]
     \multicolumn{5}{c}{\textit{Planet characteristics}} \\
      $\Rpla$ & Planet radius$^{\rm a}$  &  $\Rearth$ & $1.0 $ & $1.0 $  \\
      $\ggravi$ & Surface gravity$^{\rm a}$ &  ${\rm m \ s^{-2}}$ &  $9.8 $ & $9.8 $   \\[0.2cm]
      \multicolumn{5}{c}{\textit{Atmospheric properties}} \\
      $\Rspec$ & Gas constant for dry air$^{\rm a}$&  ${\rm J~kg^{-1}~K^{-1}}$ & $287$ &  $188.9$ \\
      $\Cp$ & Heat capacity per unit mass$^{a}$ & ${\rm J~kg^{-1}~K^{-1}}$ & $1005$ & $909.3$  \\
      $\ksw$ &SW absorption coefficient$^{\rm b}$ &  ${\rm m^2~kg^{-1}}$ & $10^{-6}$ & $10^{-6}$  \\
      $\klw$ & LW absorption coefficient$^{\rm b}$ &  ${\rm m^2~kg^{-1}}$ & $10^{-4} $ &  $2.5 \times 10^{-4} $  \\
      $\betasw$ & SW scattering parameter & --  & $1.0$ &  $1.0$ \\
      $\betalw$ & LW scattering parameter & -- & $1.0$ & $1.0$ \\[0.2cm]
      \multicolumn{5}{c}{\textit{Surface properties}} \\
      $\emissurf$ & Surface emissivity & -- & $1.0$ & $1.0$  \\ 
      $\Asurfsw$ & Surface albedo$^{\rm c}$ & -- & $0.2$ & $0.2$ \\
      $\inertiagr$ & Thermal inertia$^{\rm c}$ &  ${\rm J \ m^{-2} \ s^{-1/2} \ K^{-1}}$ & $2000 $ & $2000 $  \\
      $\zrough$ & Roughness height$^{\rm d}$ & m & $3.21 \times 10^{-5}$ & $3.21 \times 10^{-5}$ \\ 
       \hline
    \end{tabular}
    \end{small}
    \begin{footnotesize}
    \begin{flushleft}
    $^{\rm a}$ Values for the Earth's atmosphere given by \cite{Deitrick2020}, Table~2, in the Earth-like case. In the pure $\carbondiox$ case, $\Rspec$ is calculated from \cite{Meija2013} and $\Cp$ is evaluated for $\temp = 350$~K from \cite{Yaws1996book}, Appendix~E.   \\ 
    $^{\rm b}$ Defined so that the optical depths in the longwave and in the shortwave at $\press = 1$~bar are $\optdepthlw = 1$ and $\optdepthsw = 10^{-2} \optdepthlw$, respectively, in the Earth-like case. In the pure $\carbondiox$ case, the optical depth in the longwave is set to an effective value for which the collapse pressure computed using the 2-D instance of the meta-model in \sect{sec:stability} corresponds to that obtained by \cite{Wordsworth2015} from numerical simulations performed with a 3-D GCM with correlated-$k$ radiative transfer \citep[][]{LO1991}.    \\
    $^{\rm c}$ Typical values for Venus-like soils \citep[e.g.][]{Lebonnois2010}. \\
    $^{\rm d}$ Defined so that $\Cneutral = 10^{-3}$ for a 10~m-thick surface layer, following \cite{Frierson2006}.
    \end{flushleft}
 \end{footnotesize}
 \end{table*}

\subsection{Soil heat transfer}

The soil heat transfer is solved by using a classical 1-D soil heat conduction approach \citep[e.g.][]{Hourdin1993,WCD2016}. The evolution of the ground temperature $\temp$ due to vertical diffusion is governed by the heat equation 
\begin{equation}
\Cvgr \dd{\temp}{\time} =  - \dd{\Fcond}{\zd} ,
\end{equation}
where $\zd = - \zz$ is the depth from surface ($\zd \geq 0$), $\Cvgr$ the heat capacity of the ground per unit volume, and $\Fcond$ the conductive flux propagating downwards. The conductive flux is expressed as 
\begin{equation}
\Fcond = - \condgr \dd{\temp}{\zd},
\label{Fcond_soil}
\end{equation}
the parameter $\condgr$ being the thermal conductivity of the material. Both $\Cvgr$ and $\condgr$ are assumed to be constants in the model. 

We introduce the normalised pseudo-depth
\begin{equation}
\zdn \define \zd \sqrt{\frac{\Cvgr}{\condgr}},
\end{equation}
which has dimensions of ${\rm s^{1/2}}$. Expressed in terms of $\zdn$, the vertical diffusion is controlled by one parameter solely, the thermal inertia
\begin{equation}
\label{inertiagr}
\inertiagr \define \sqrt{\condgr \Cvgr}.
\end{equation}
The downward conductive flux given by \eq{Fcond_soil} then becomes
\begin{equation}
 \Fcond  = - \inertiagr \dd{\temp}{\zdn}
\end{equation}
and the vertical diffusion equation simply reads
\begin{equation}
 \dd{\temp}{\time}  = \ddn{\temp}{\zdn}{2}.
 \label{heat_conduction}
\end{equation}
The above equation is solved by means of the finite difference method between the surface ($\zdn = 0$) and an inner boundary ($\zdn = \zdn_{\rm bot} $) using an implicit time scheme. At the surface, a continuity condition is applied: the incoming heat fluxes equalise the outcoming fluxes. At the inner boundary, a zero-flux condition is applied, which enforces the assumption that the planet interior is in thermal equilibrium. These two conditions are respectively formulated as 
\begin{equation}
\begin{array}{rl}
\displaystyle - \Fcond + \Fdown - \Fupsw - \Fsheat - \emissurf \sigmaSB \Tsurf^4 = 0 & \displaystyle \ \mbox{at} \ \zdn = 0, \\
\displaystyle \Fcond = 0 & \displaystyle \  \mbox{at} \ \zdn = \zdn_{\rm bot},
\end{array}
\end{equation}
where $\Fdown$ designates the downward radiative flux (i.e. the sum of the shortwave and longwave contributions), $\Fupsw$ the reflected shortwave flux, $\Tsurf$ the surface temperature, and $\emissurf$ the surface emissivity in the longwave. The scheme used to solve the soil heat transfer is detailed in \append{app:soil_heat_transfer}. Similarly as the radiative transfer and turbulent diffusion equations, the soil heat transfer equation is discretised and integrated as a boundary condition problem by means of the tridiagonal matrix algorithm (see \append{app:thomas_algorithm}). 

\RV{Several simulations were run in order to assess the sensitivity of mean fields to the heat transfer scheme. The obtained results are discussed in \append{app:soil_heat_transfer}. They show that mean flows and temperatures do not depend much on the ground thermal inertia that parametrises the soil heat transfer scheme (\eq{inertiagr}). Particularly, the nightside surface temperature is almost insensitive to the value chosen for $\inertiagr$. Nevertheless, we recall that horizontal diffusion is ignored in the scheme since we focus of dry rocky planets, while it might play an important role for an ocean planet due to heat advection by oceanic flows, as discussed by \cite{Wordsworth2015}. }

\def\wpanel{0.28\textwidth}
\def\wlegend{0.32\textwidth}
\def\hraisebox{0.20\textwidth}
\begin{figure*}[t]
   \centering
  \hspace{2cm} \textsc{Pressure (bar)} \hspace{3.0cm} \textsc{Temperature (K)} \hspace{2.5cm} \textsc{Vertical wind speed (${\rm m \ s^{-1}}$)} \\[0.3cm]
     \raisebox{\hraisebox}[1cm][0pt]{%
   \begin{minipage}{1cm}%
   \textsc{2-D}
\end{minipage}}
   \includegraphics[width=\wpanel,trim = 2.5cm 0.0cm 0.8cm 0.9cm,clip]{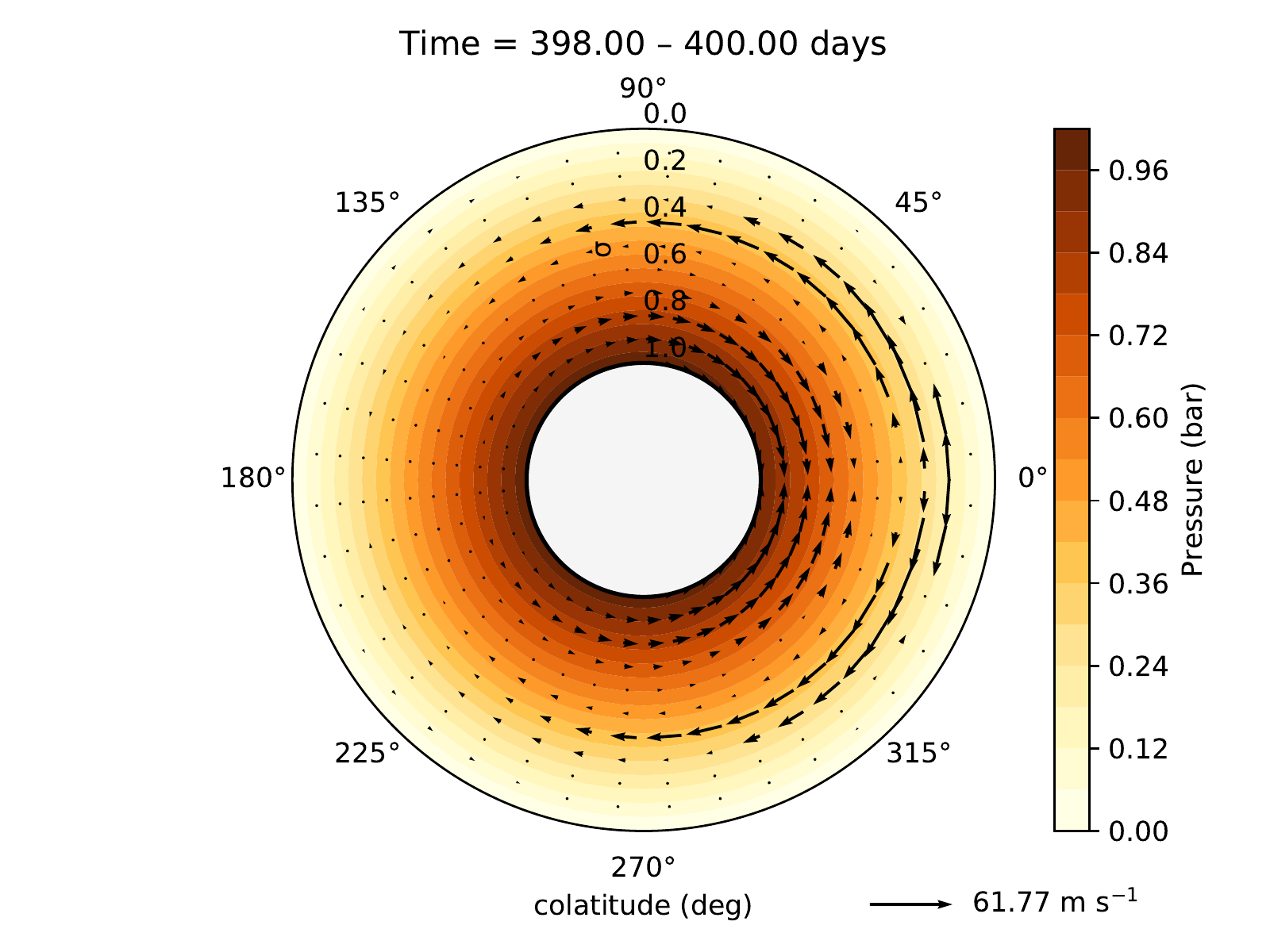}
   \includegraphics[width=\wpanel,trim = 2.5cm 0.0cm 0.8cm 0.9cm,clip]{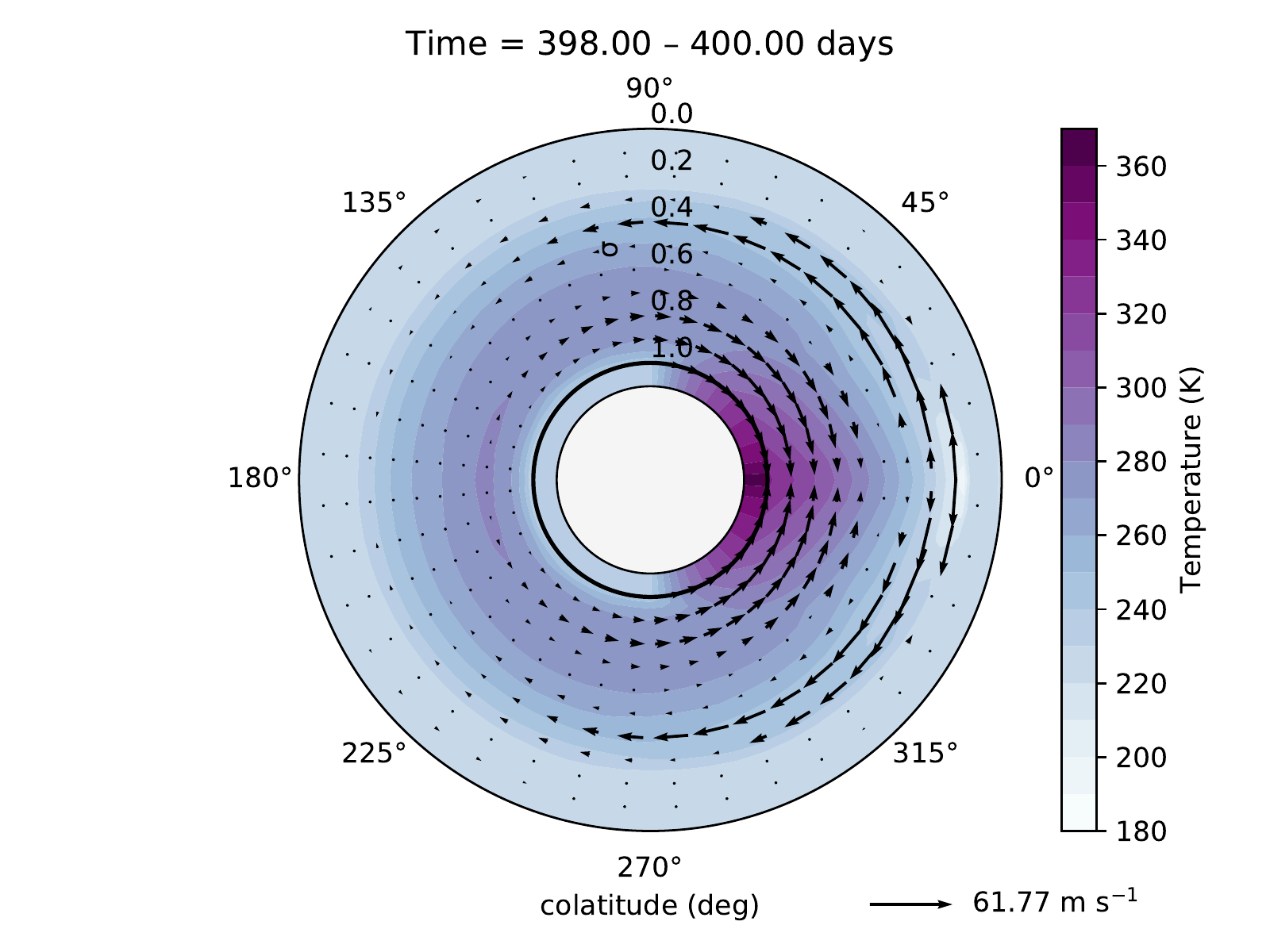} 
   \includegraphics[width=\wpanel,trim = 2.5cm 0.0cm 0.8cm 0.9cm,clip]{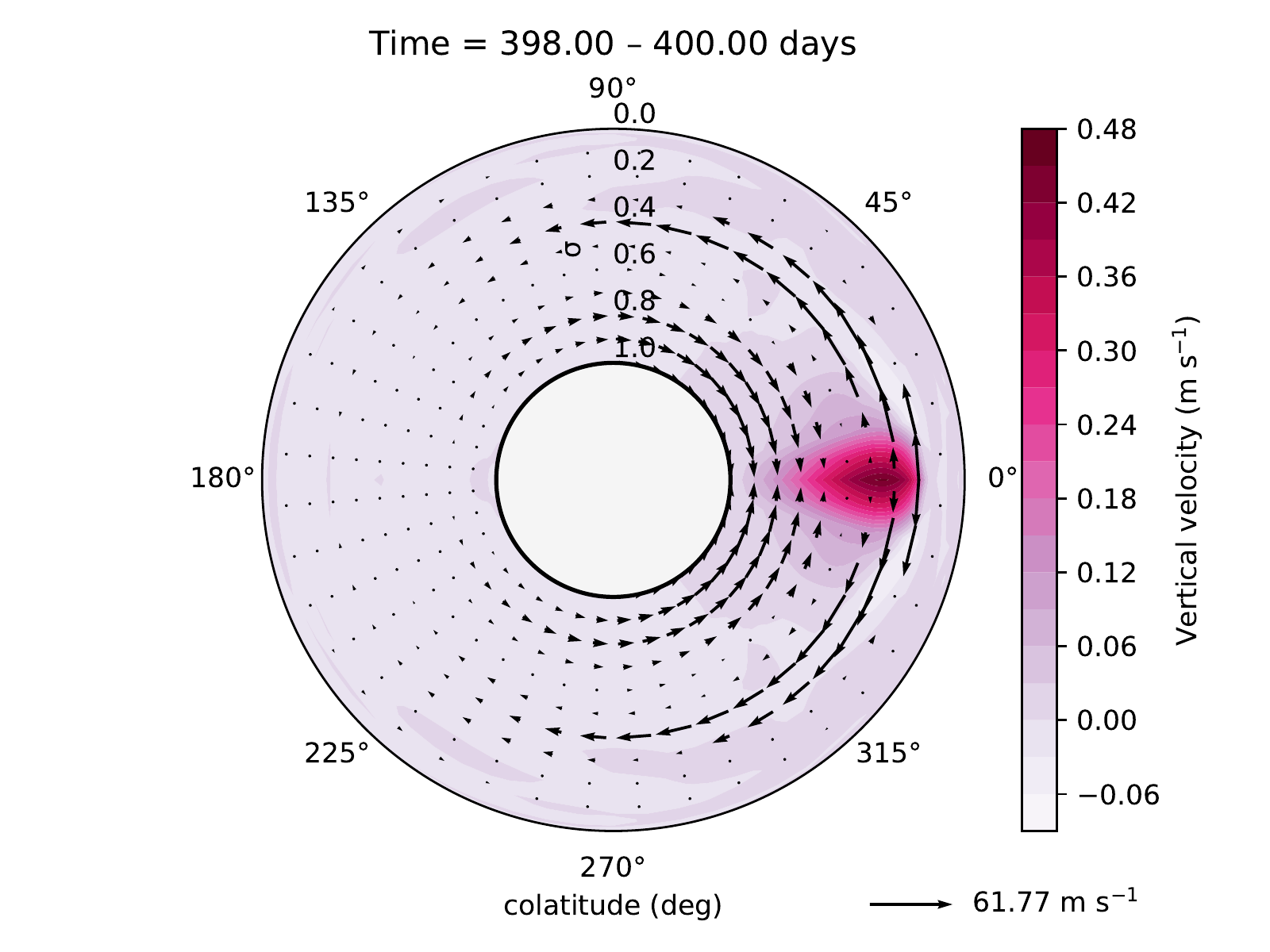} \\
     \raisebox{\hraisebox}[1cm][0pt]{%
   \begin{minipage}{1cm}%
   \textsc{1.5-D}
\end{minipage}}
   \includegraphics[width=\wpanel,trim = 2.5cm 0.0cm 0.8cm 0.9cm,clip]{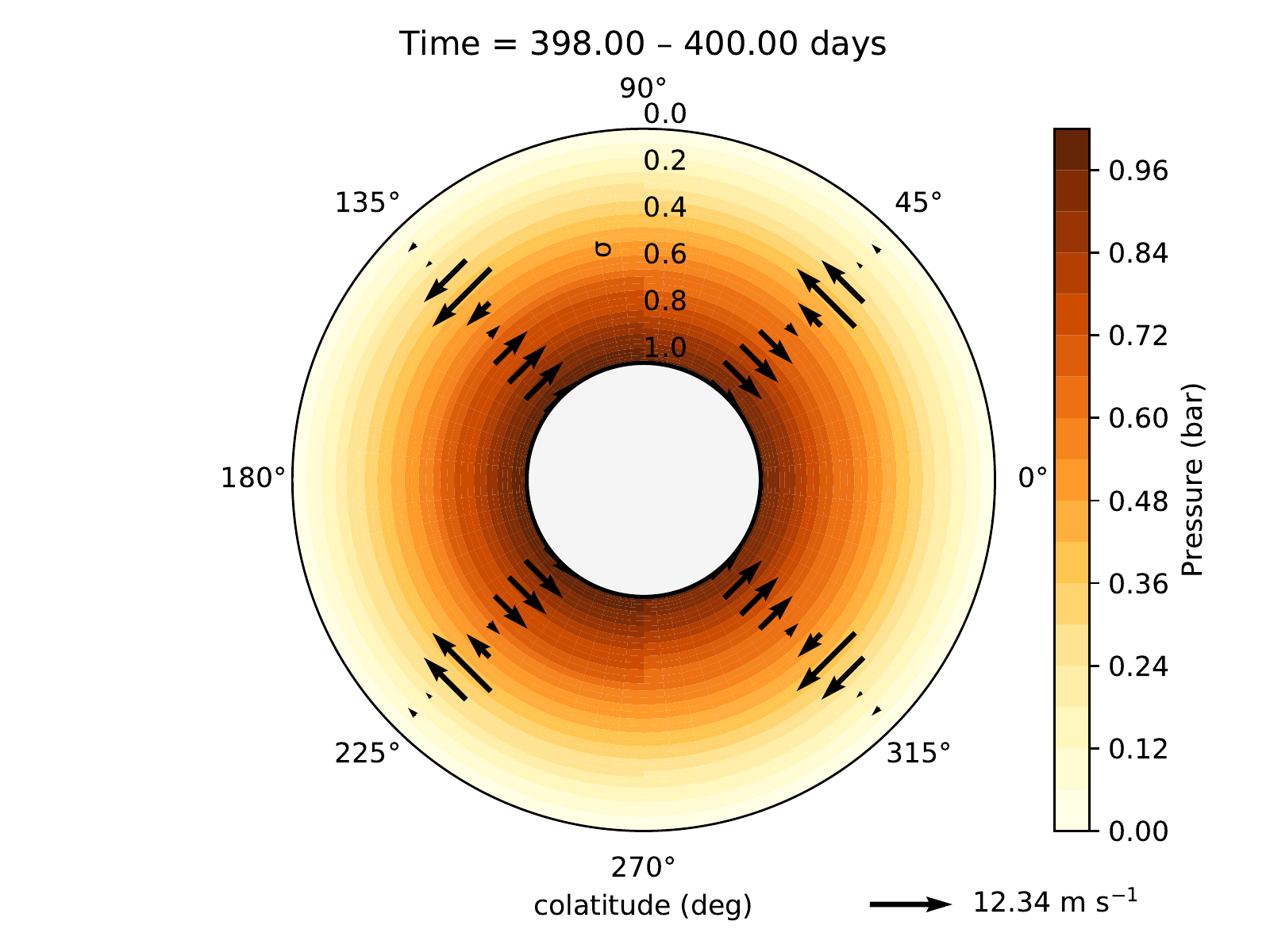}
   \includegraphics[width=\wpanel,trim = 2.5cm 0.0cm 0.8cm 0.9cm,clip]{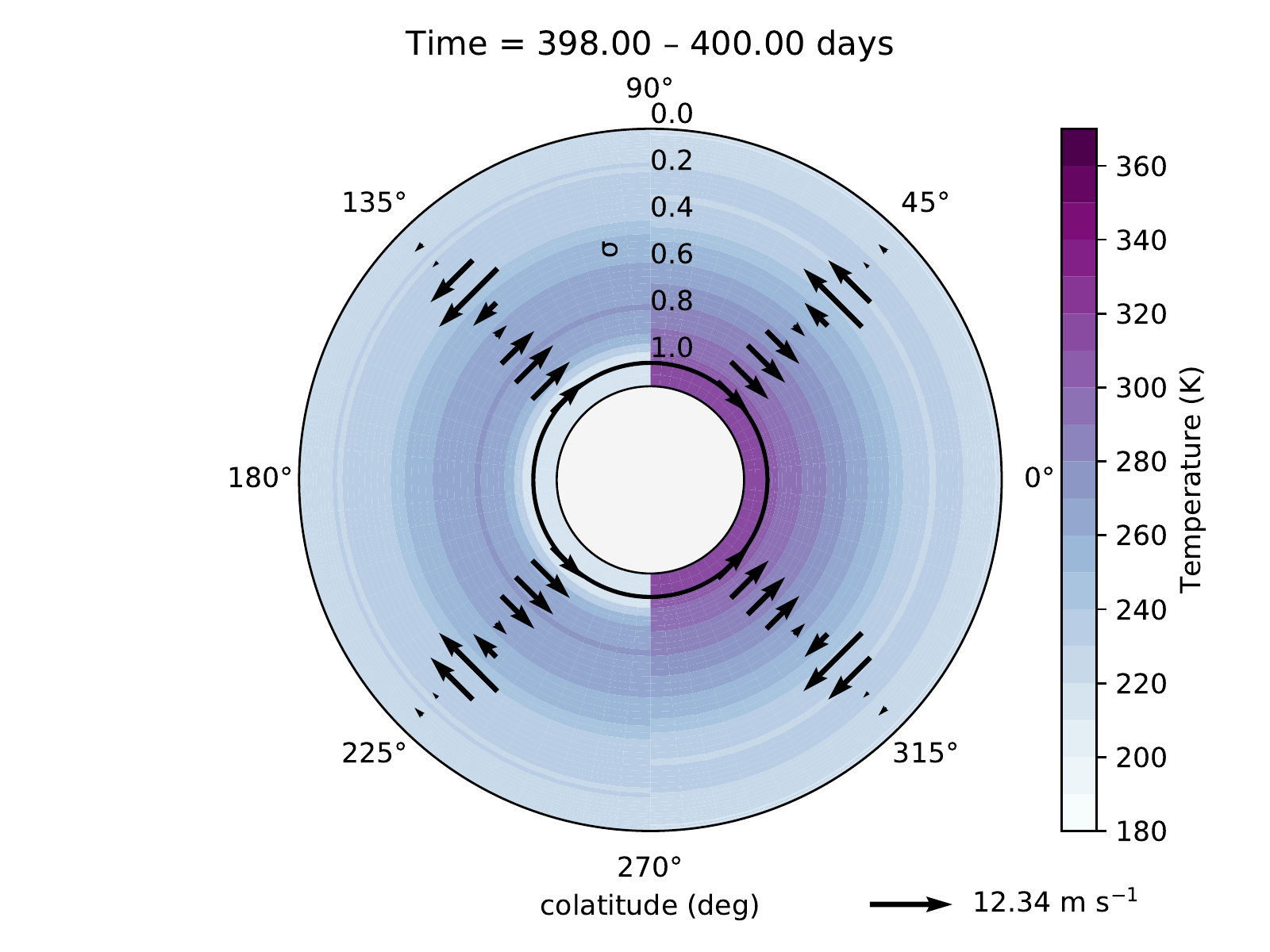}
   \includegraphics[width=\wpanel,trim = 2.5cm 0.0cm 0.8cm 0.9cm,clip]{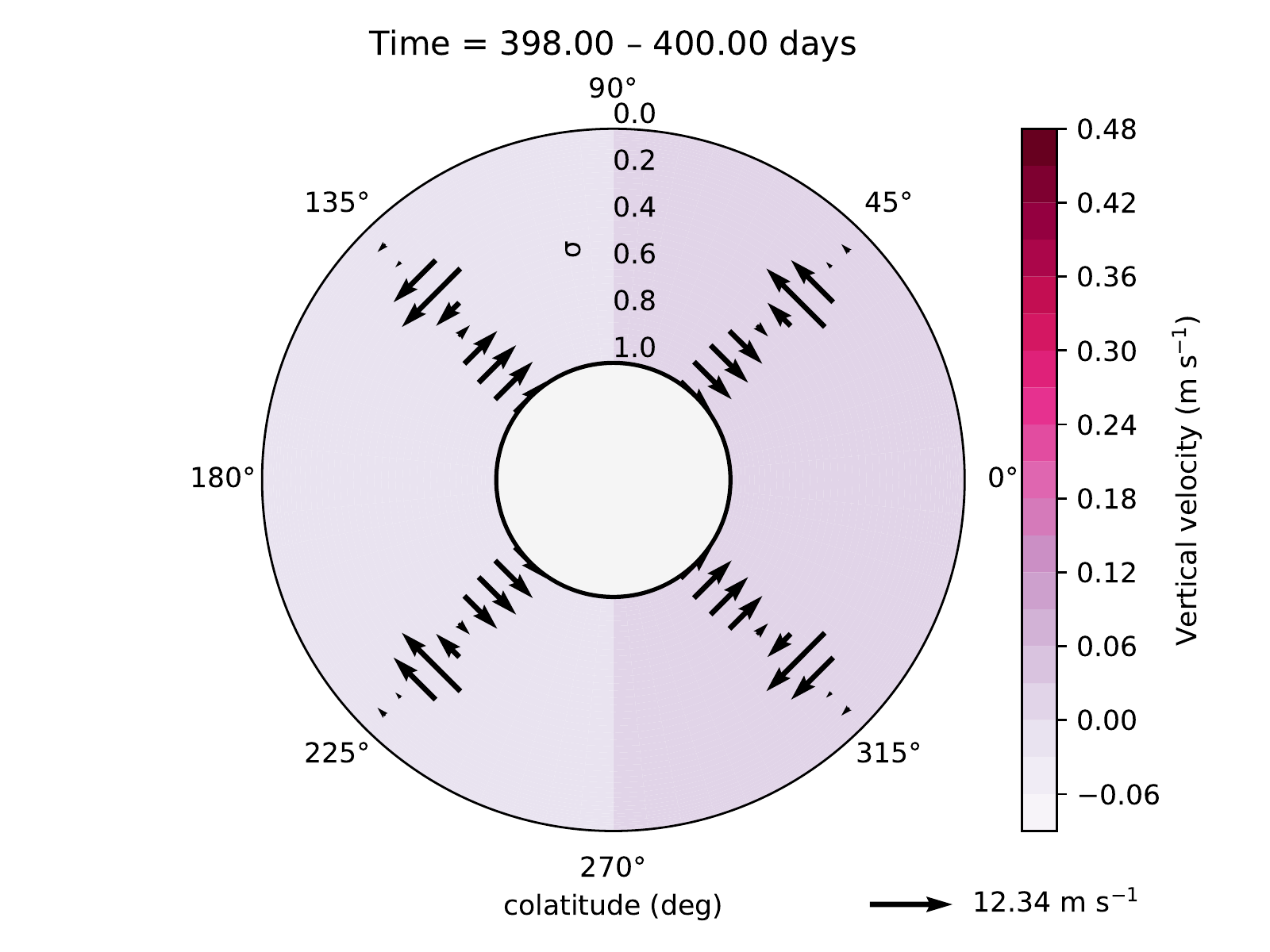} \\
      \raisebox{\hraisebox}[1cm][0pt]{%
   \begin{minipage}{1cm}%
   \textsc{1-D}
\end{minipage}}
    \includegraphics[width=\wpanel,trim = 2.5cm 0.0cm 0.8cm 0.9cm,clip]{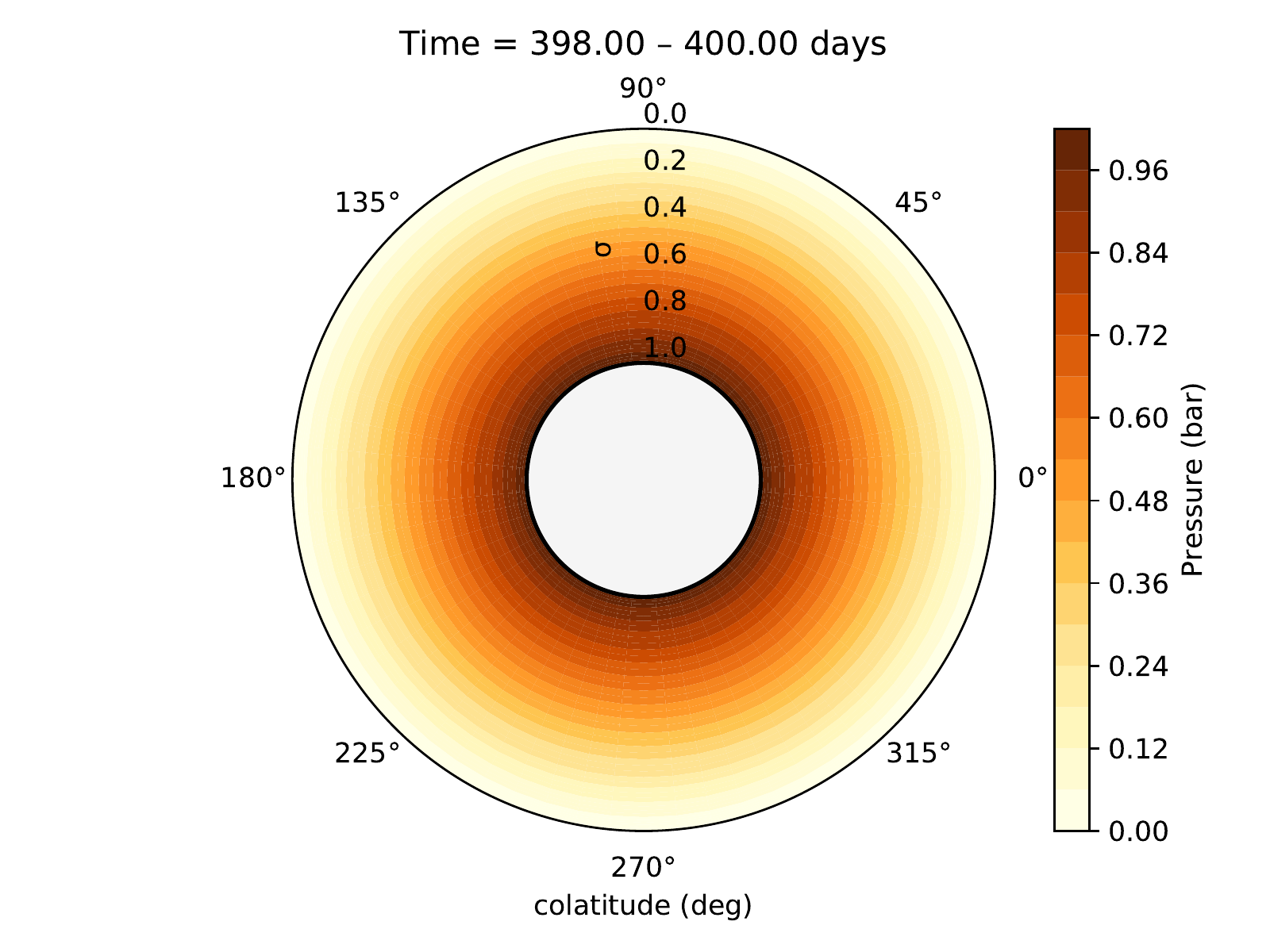}
   \includegraphics[width=\wpanel,trim = 2.5cm 0.0cm 0.8cm 0.9cm,clip]{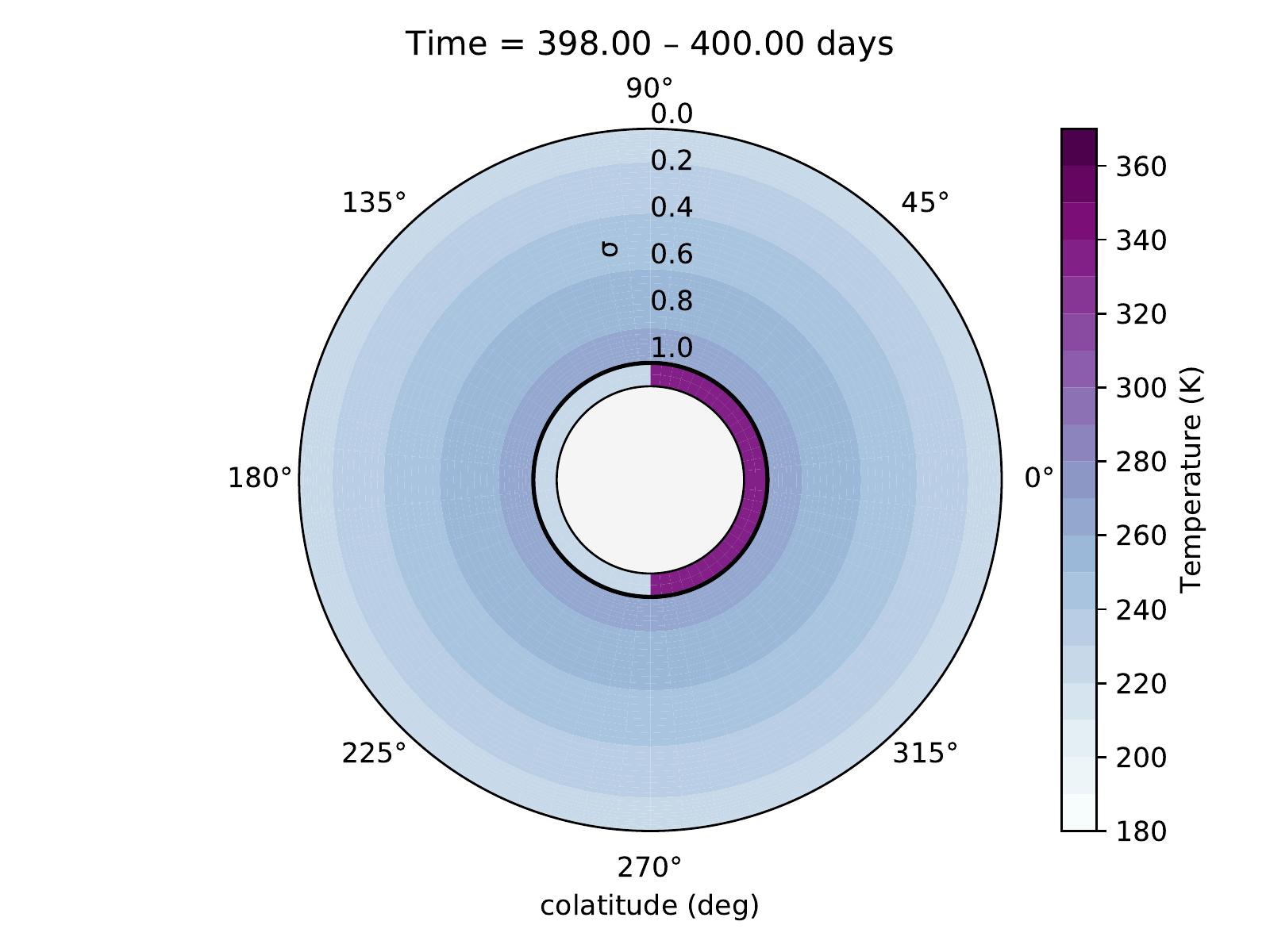} 
   \includegraphics[width=\wpanel,trim = 2.5cm 0.0cm 0.8cm 0.9cm,clip]{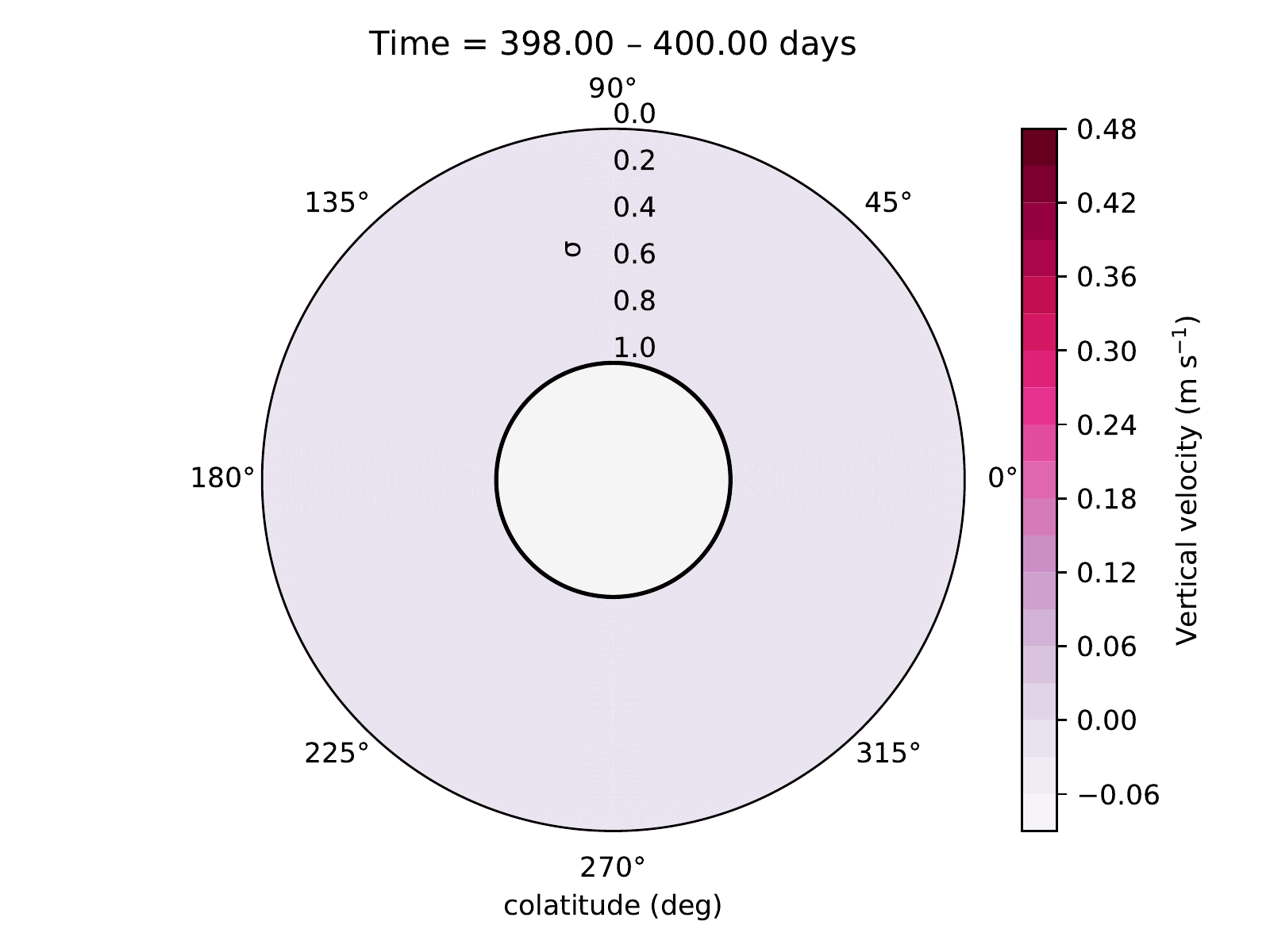} \\
   \raisebox{\hraisebox}[1cm][0pt]{%
   \begin{minipage}{1cm}%
   \textsc{0-D}
\end{minipage}}
   \includegraphics[width=\wpanel,trim = 2.5cm 0.0cm 0.8cm 0.9cm,clip]{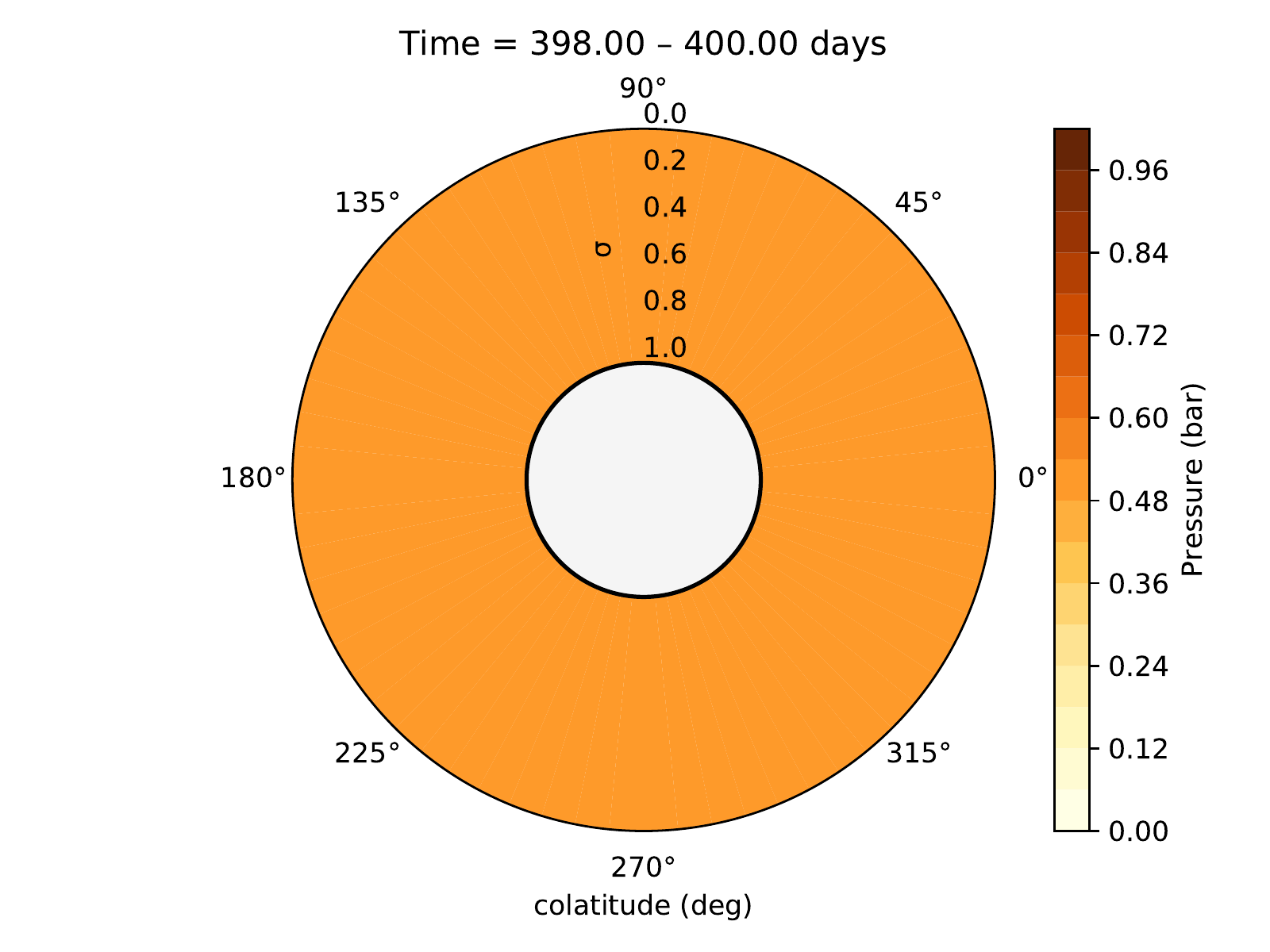}
   \includegraphics[width=\wpanel,trim = 2.5cm 0.0cm 0.8cm 0.9cm,clip]{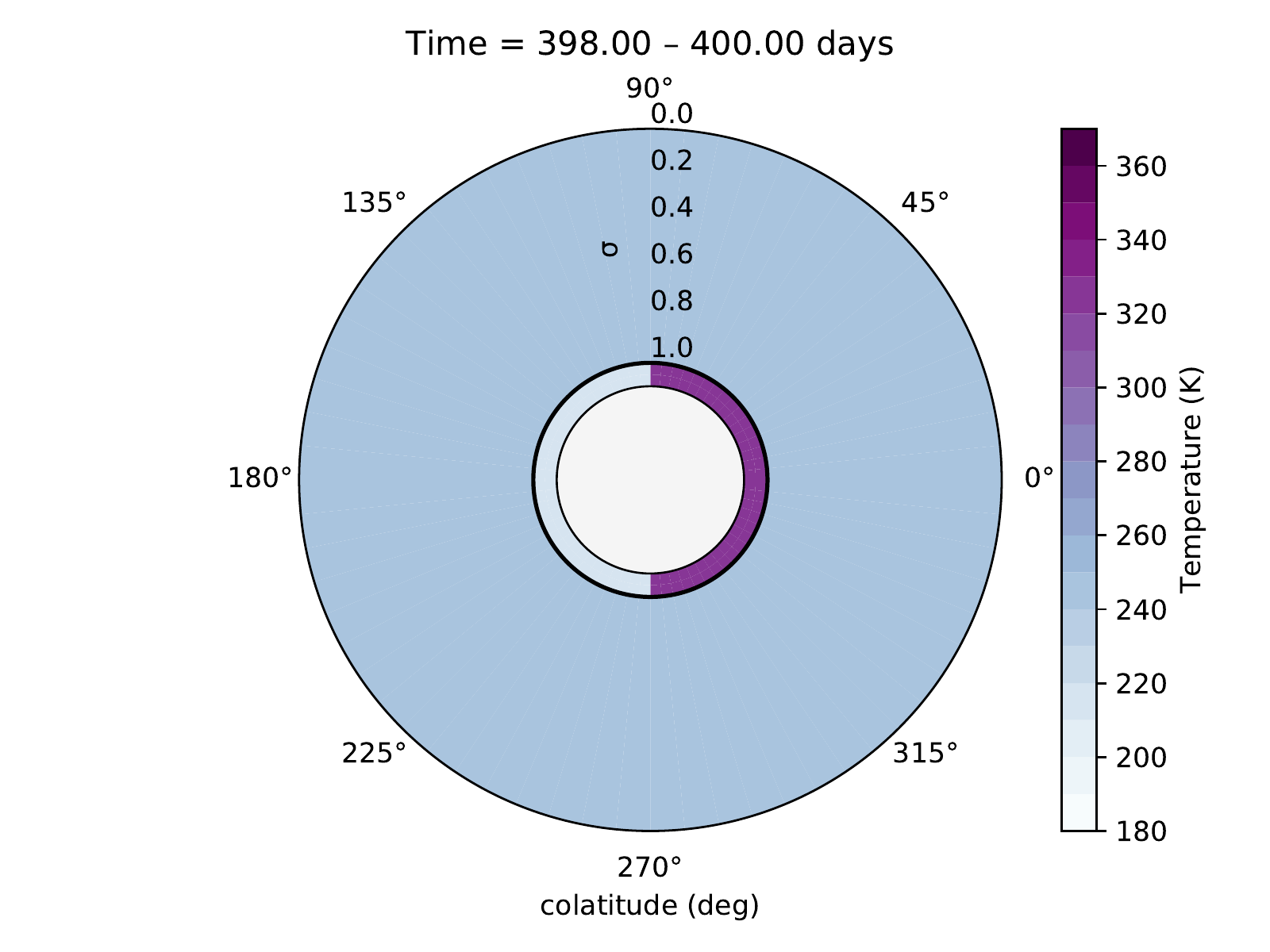} 
   \includegraphics[width=\wpanel,trim = 2.5cm 0.0cm 0.8cm 0.9cm,clip]{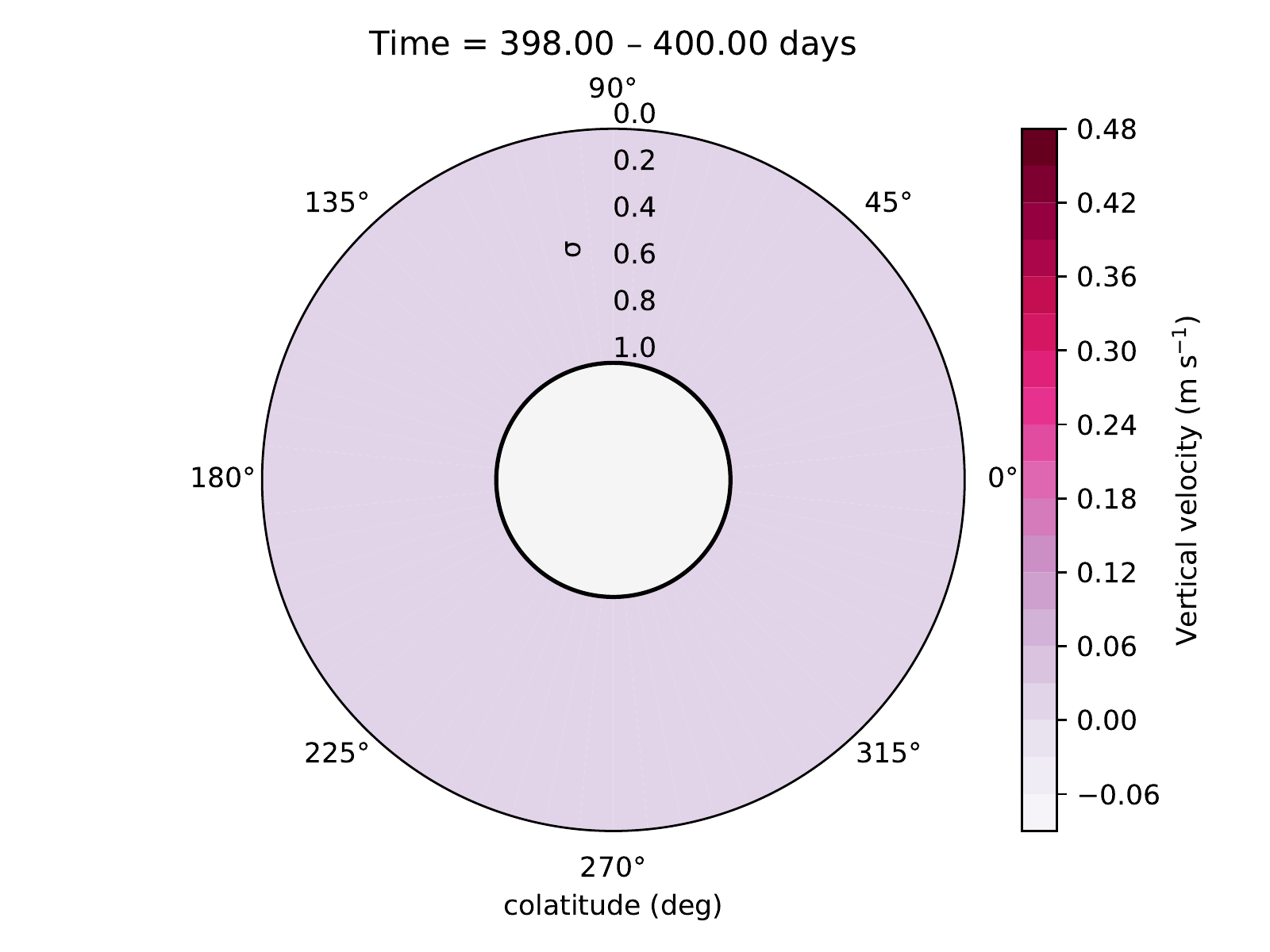} 
      \caption{Two-day averaged snapshots of pressure (left), temperature (middle), and vertical wind speed (right) distributions for the 1~bar-atmosphere of an Earth-sized tidally locked planet (Earth-like case of Table~\ref{tab:param_reference_case}) with a stellar irradiation of 1366~${\rm W \ m^{-2}}$ after convergence ($\time = 400$~days). {\it From bottom to top:} 0-D ($1 \times 1$~grid), 1-D ($1 \times 50$~grid), 1.5-D ($2 \times 50$~grid), and 2-D ($32 \times 50$~grid) instances of the meta-model. The substellar point corresponds to $\col = 0^\degree$ and the anti-stellar point to $\col = 180^\degree$.}
       \label{fig:snapshots_HB1993_diff}%
\end{figure*}

\subsection{Physical setup}

In the following, we perform simulations for the two cases defined in Table~\ref{tab:param_reference_case}: (i) and Earth-like atmosphere, and (ii) a pure $\carbondiox$ atmosphere. For the Earth-like case, we use the values given by \cite{Deitrick2020} in the synchronous Earth case \citep[see][Table~2]{Deitrick2020}. These values correspond to a tide-locked Earth-sized planet with an atmosphere having the thermodynamical properties of the Earth's atmosphere. Following \cite{Wordsworth2015}, we assume that $\optdepthlw = 1$ at $\press = 1$~bar. Similarly, we fix $\optdepthsw = 0.01$ at $\press = 1$~bar to enforce the assumption that the atmosphere is optically thin in the visible. The effective absorption coefficients introduced in \eq{optical_depths} are set accordingly. Besides, we consider the case of pure absorption (no scattering). The scattering parameter is therefore set to $\betascat = 1$ both for the shortwave and longwave. Assuming that the planet's surface is made of bare rocks, we use the typical values commonly assumed for Venus-like soils \citep[e.g.][]{Lebonnois2010} to set the planet's surface properties. Following \cite{Frierson2006}, the roughness height is set to $\zrough = 3.21 \times 10^{-5}$~m so that $\Cneutral = 10^{-3}$ for $\zzfirst = 10$~m. 

The pure $\carbondiox$ case is similar to the Earth-like case except for the specific gas constant, $\Rspec = 188.9 \ {\rm J \ kg^{-1} \ K^{-1}}$ \citep[calculated from][]{Meija2016}, the heat capacity per unit mass, $\Cp = 909.3 \ {\rm J \ kg^{-1} \ K^{-1}}$ \citep[evaluated for $\temp = 350$~K from][Appendix~E]{Yaws1996book}, and the absorption coefficient in the longwave, $\klw = 2.5 \times 10^{-4} \ {\rm m^2 \ kg^{-1}} $. The latter is an effective value of $\klw$ for which the 2-D instance of the GCMM presented in \sect{sec:climate_regime} approximately reproduces the stability diagram obtained by \cite{Wordsworth2015} from 3-D GCM simulations with correlated-$k$ radiative transfer \citep[][Fig.~12]{Wordsworth2015}, as shown in \sect{sec:stability}. We note that this value is less than that obtained by \cite{Wordsworth2015} by adjusting the analytic solution given by \eq{Tnlow} to GCM simulations ($\klw = 3.2 \times 10^{-4} \ {\rm m^2 \ kg^{-1}}$ by taking into account the factor~2 difference between the definition of $\klw$ given by Eq.~(12) of the article and ours, given by \eq{optical_depths}). As highlighted by \sect{sec:stability}, this discrepancy is due to the fact that the planetary boundary layer acts to warm up the nightside surface in the general case, which consequently requires a weaker greenhouse effect to reach the same thermal state. 

\section{Climate regime: from 0-D to 2-D models} 
\label{sec:climate_regime}
The GCMM described in the preceding section can be used at the same time for various grid configurations, each of them being a possible instance of the meta-model. This allows for comparing the results obtained from a bench of models of various complexities albeit sharing the same intrinsic theoretical background and physical setup. Four instances of the meta-model are examined in the present study (Table~\ref{tab:models}). There are introduced below in ascending order of complexity, while the resulting pressure, temperature, and vertical wind speed snapshots are plotted in \fig{fig:snapshots_HB1993_diff}. 

\RV{We note that, for legibility, the arrows representing wind speeds are uniformly spaced both horizontally and vertically in the figure, rather than being centred at the points where winds speeds are really evaluated. For instance, in the 1.5-D model, the arrows are determined from a linear interpolation and located at $\col = 45^\degree, 135^\degree$, while horizontal wind speeds are evaluated at $\col = 90^\degree$. In the range $180^\degree \leq \col \leq 360^\degree$, the distributions are plotted by applying a symmetric transformation to the distributions calculated in the range $0^\degree \leq \col \leq 180^\degree$. The additional inner ring in temperature snapshots (middle column of \fig{fig:snapshots_HB1993_diff}) corresponds to the soil temperature.}

The 0-D model ($1 \times 1$~grid) is the simplest configuration. Here the atmosphere is vertically and horizontally isothermal, and it exchanges heat with dayside and nightside isothermal surface hemispheres through radiative transfer only. This configuration corresponds exactly to that of the two-layer grey radiative model described in Sect.~3 of \cite{ADH2020}, which provides closed-form solutions. In the optically thin limit, it reproduces the analytic model introduced by \cite{Wordsworth2015}.

The next level of complexity is the 1-D model ($1 \times 50$~grid). In this configuration, the vertical structure of the atmosphere is allowed to adjust with radiative transfer although it is still horizontally isothermal. Similarly as in the 0-D~configuration, the planet's surface is divided into the dayside and nightside hemispheres, and the surface-atmosphere heat exchanges are purely radiative. The 1-D~model thus provides a more realistic vertical temperature profile than that assumed in the 0-D~model. This profile can be interpreted as an approximation of the globally averaged atmospheric structure. This instance of the GCMM appears as a simplified version of more sophisticated one-dimensional models \citep[e.g.][]{RC2012}. 

Then we introduce the 1.5-D model ($2 \times 50$~grid). This level corresponds to the two-column approach, where the atmosphere is modelled by dayside and nightside hemispherical air columns, similarly as the planet's surface \citep[e.g.][]{YA2014,KA2016}. Therefore the vertical structure ceases to be horizontally uniform and differences appear between the dayside and the nightside. However these differences are mitigated by the fact that the dayside and nightside planetary boundary layers are both controlled by the horizontal wind speed at the planet's terminator ($\col = 90^\degree$) here, whereas horizontal wind speeds strongly differ between the dayside and nightside in reality. Besides, this configuration allows for taking into account the coupling between the thermodynamics and the day-night overturning circulation, and particularly the contribution of heat advection to the planet's thermal state of equilibrium.

\begin{figure}[t]
   \centering
   \includegraphics[width=0.48\textwidth,trim = 0.3cm 0.3cm 0.3cm 0.8cm,clip]{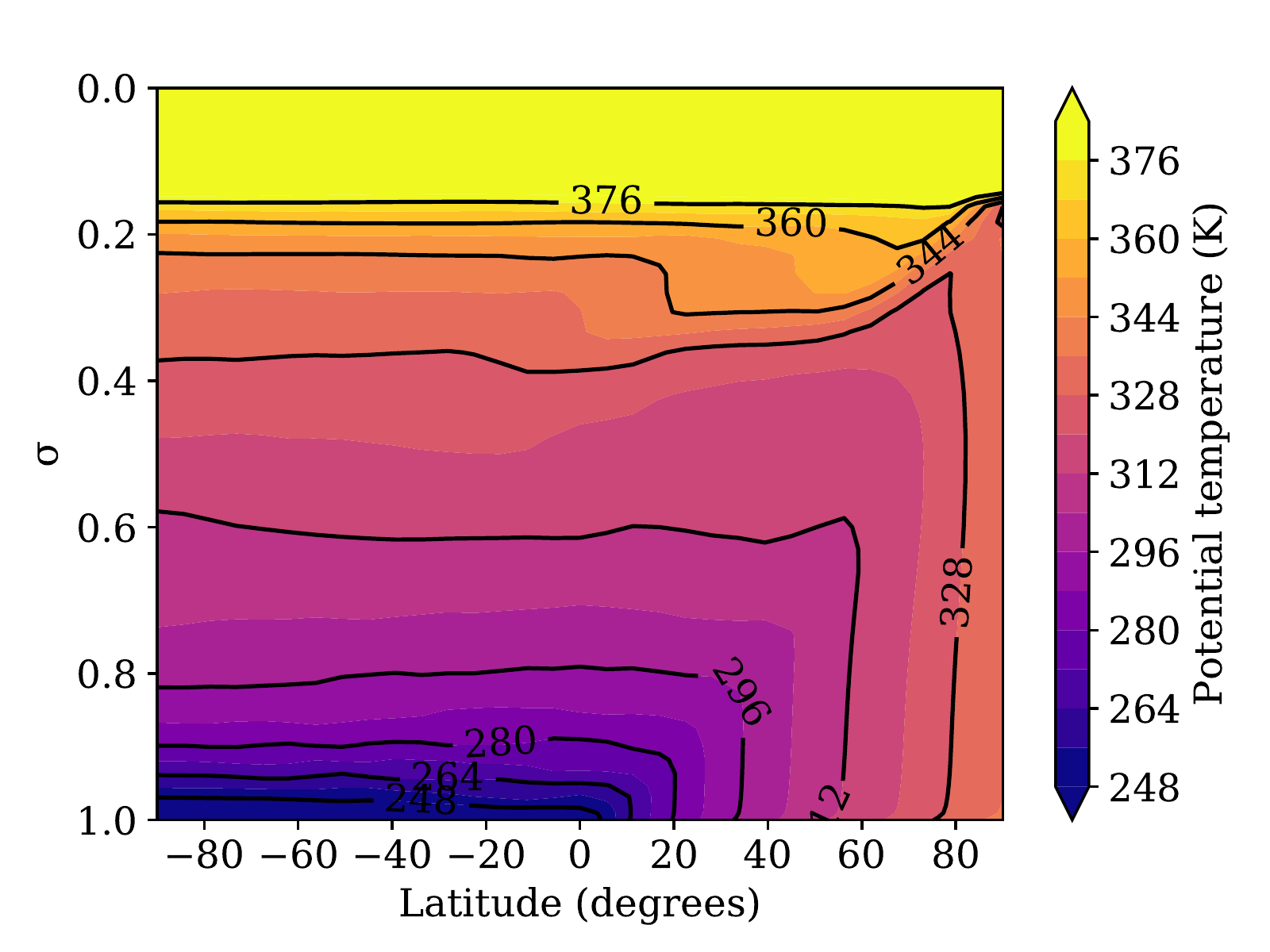}
      \caption{\RV{Two-day averaged potential temperature field in Earth-like case of Table~\ref{tab:param_reference_case} with $\psurf = 1$~bar and $\Fstar = 1366 \ {\rm W \ m^{-2}}$. Plotted values belong to the interval $\left[ 248 \ {\rm K}, 376 \ {\rm K} \right] $. Outside of this interval, the color is set to that of the closest bound. The 2-D instance of the meta-model ($32 \times 50$~grid) is used. The latitudes $90^\degree$ and $-90^\degree$ correspond to the sub- and anti-stellar points, respectively.  }}
       \label{fig:potential_temp}%
\end{figure}

Finally, the 2-D model ($32 \times 50$~grid) is the more complex instance of the meta-model. In this configuration, the latter behaves similarly as a 2-D GCM \citep[e.g.][]{Song2021}, or as a 3-D GCM in the slow rotation regime \citep[e.g.][]{Leconte2013,HaqqMisra2018,PH2019,Turbet2021}. Both the atmospheric structure and mean flows are fully resolved, which describes the global heat engine circulation. The planetary boundary layer strongly differs between the dayside, where it is unstable ($\Riz <0$) due to convection \citep[see e.g.][]{KA2016}, and the nightside, where it is stable ($\Riz \geq 0$). Also, the circulation is highly asymmetric with respect to the terminator: as shown by \fig{fig:snapshots_HB1993_diff} (top panels), it exhibits strong upwelling flows in a small region around the substellar point, and weak subsidence over a large area that spreads from $\col \approx 60^\degree$ to $\col = 180^\degree$ (anti-stellar point). 

\RV{We recover the day-night evolution of the atmospheric structure in the potential temperature distribution shown by \fig{fig:potential_temp}, which was obtained using the 2-D model. In this figure, the latitude is measured with respect to the terminator, located at $0^\degree$. The sub- and anti-stellar point thus correspond to the latitudes $90^\degree$ and $- 90^\degree$, respectively. The nightside stably stratified region is characterised by a positive vertical potential temperature gradient. On dayside, the convective mixing induced by the thermal forcing makes the temperature gradient converge towards the adiabat. As a consequence, convective regions are indicated by vertically uniform profiles of potential temperature. We note that the thickness of the planetary boundary layer grows as the latitude increases, and that it is maximal at the sub-stellar point. }

Owing to spatial bi-dimensionality, the solver is relatively fast even for the 2-D instance, where one day of simulation with a two-minute dynamical timestep is equivalent to 0.61 seconds of CPU time on one CPU. This allows for running grid simulations, which is the purpose of the next section. 

\def\wpanel{0.24\textwidth}
\def\wlegend{0.32\textwidth}
\def\hraisebox{0.20\textwidth}
\def\hraiseboxps{0.12\textwidth}
\def\wcbar{1.05cm}
\begin{figure*}[t]
   \centering
  \hspace{1cm} \textsc{Earth-like atmosphere} \hspace{1.5cm} \textsc{Pure $\carbondiox$ atmosphere} \\[0.3cm]
     \raisebox{\hraisebox}[1cm][0pt]{%
   \begin{minipage}{1cm}%
   \textsc{2-D}
\end{minipage}}
\raisebox{\hraiseboxps}[1cm][0pt]{\rotatebox[origin=c]{90}{\tiny{$\log_{10} \left( \psurf \right)$~(bar)}}}
   \includegraphics[height=\wpanel,trim = 1.0cm 1.0cm 3.0cm 0.75cm,clip]{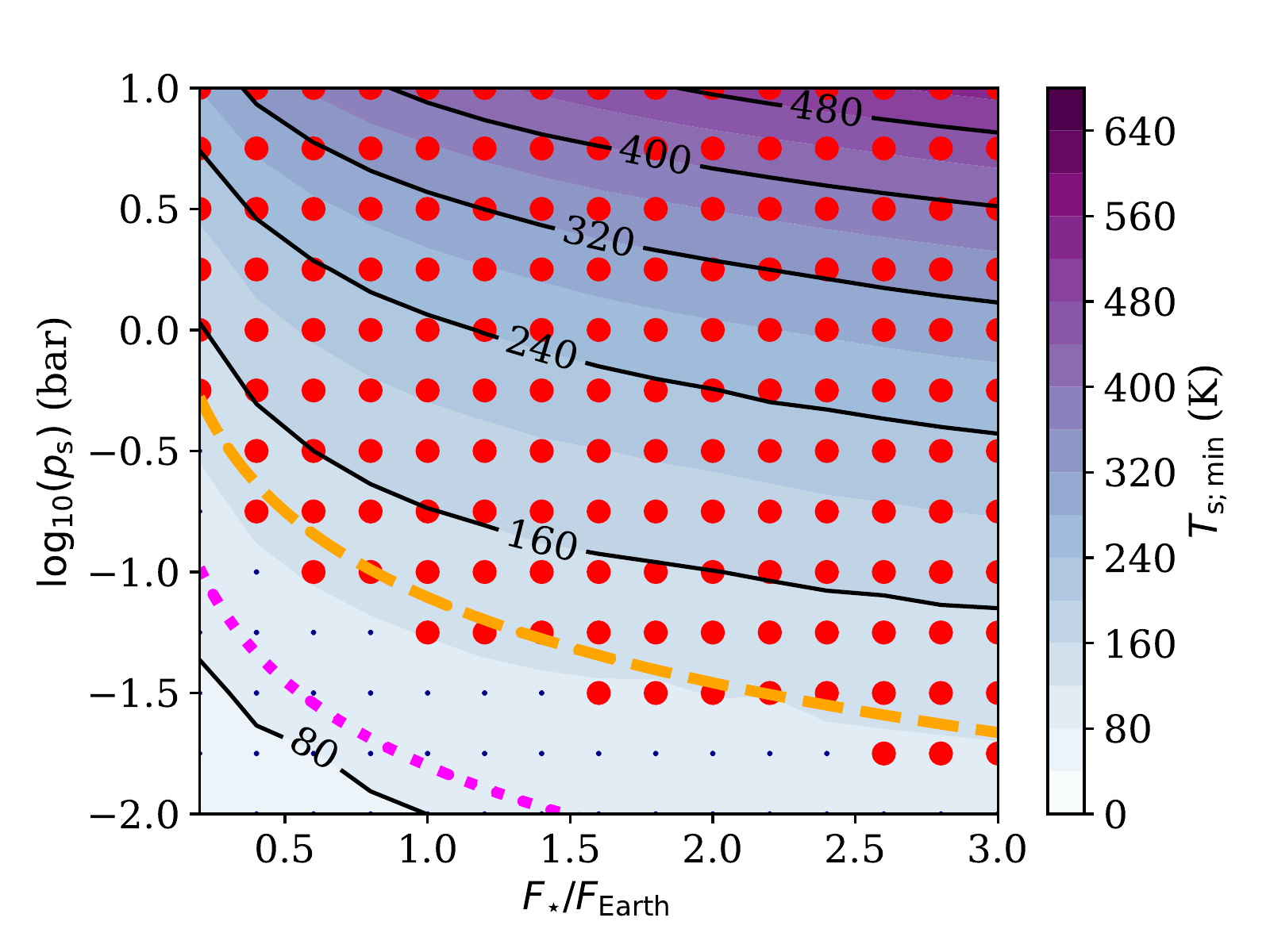}
   \includegraphics[height=\wpanel,trim = 1.0cm 1.0cm 3.0cm 0.75cm,clip]{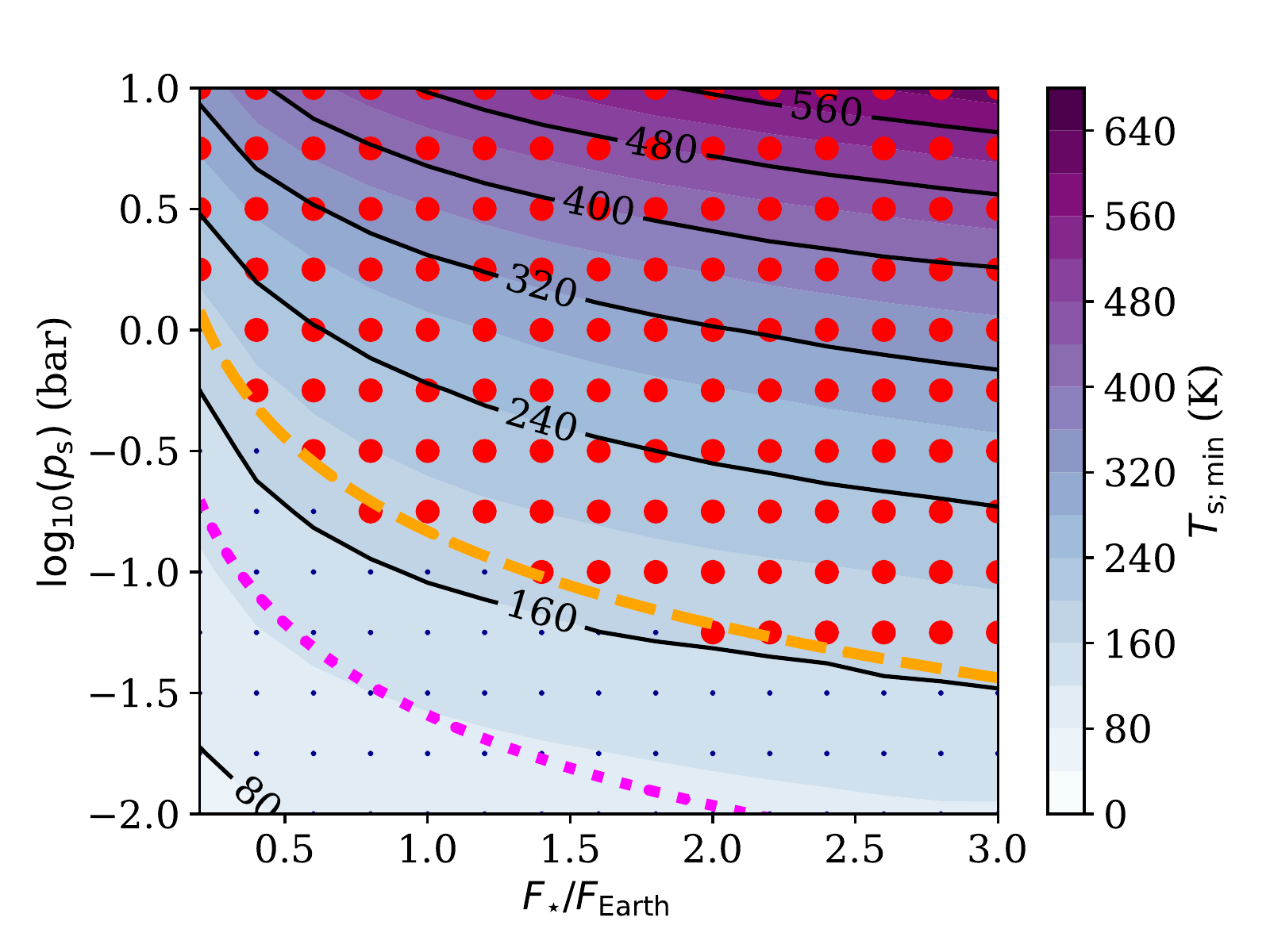} 
   \includegraphics[height=\wpanel,trim = 13.3cm 1.0cm 1.2cm 0.75cm,clip]{auclair-desrotour_fig3b.pdf}
   \raisebox{\hraiseboxps}[1cm][0pt]{\rotatebox[origin=c]{90}{\tiny{$\Tnight$~(K)}}} \\
     \raisebox{\hraisebox}[1cm][0pt]{%
   \begin{minipage}{1cm}%
   \textsc{1.5-D}
\end{minipage}}
\raisebox{\hraiseboxps}[1cm][0pt]{\rotatebox[origin=c]{90}{\tiny{$\log_{10} \left( \psurf \right)$~(bar)}}}
   \includegraphics[height=\wpanel,trim = 1.0cm 1.0cm 3.0cm 0.75cm,clip]{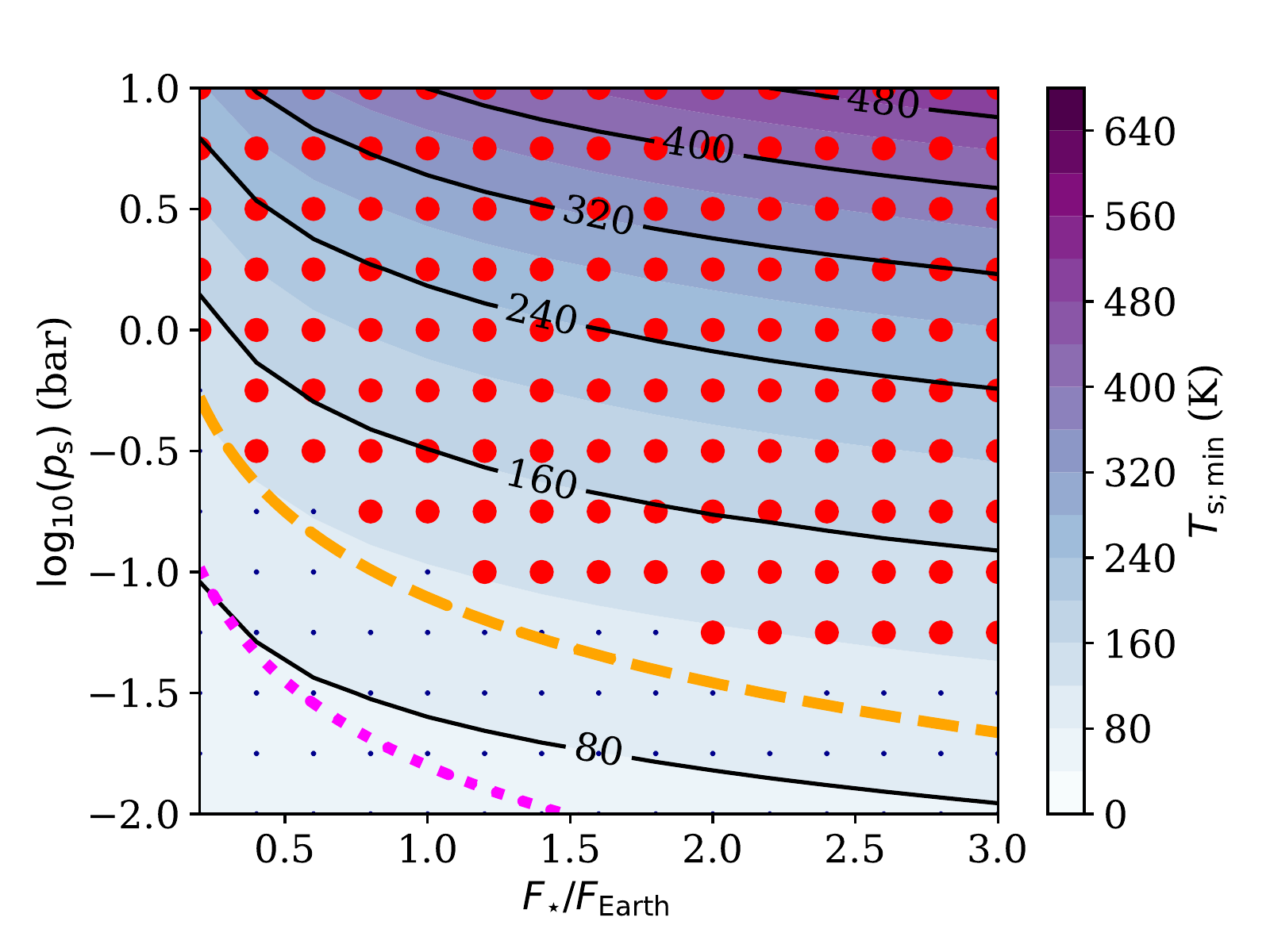}
   \includegraphics[height=\wpanel,trim = 1.0cm 1.0cm 3.0cm 0.75cm,clip]{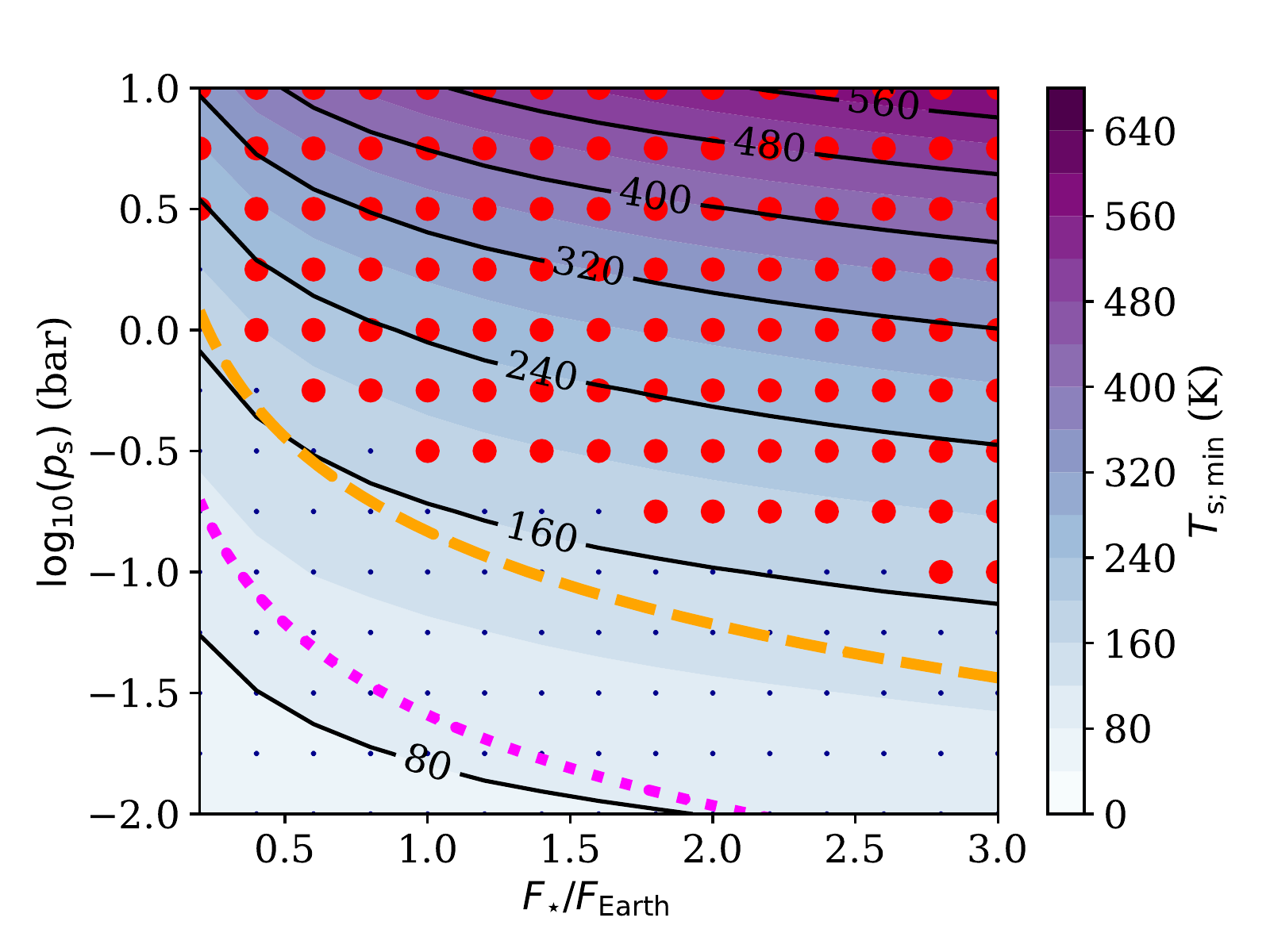} 
\hspace{\wcbar}~ \\
        \raisebox{\hraisebox}[1cm][0pt]{%
   \begin{minipage}{1cm}%
   \textsc{1-D}
\end{minipage}}
\raisebox{\hraiseboxps}[1cm][0pt]{\rotatebox[origin=c]{90}{\tiny{$\log_{10} \left( \psurf \right)$~(bar)}}}
   \includegraphics[height=\wpanel,trim = 1.0cm 1.0cm 3.0cm 0.75cm,clip]{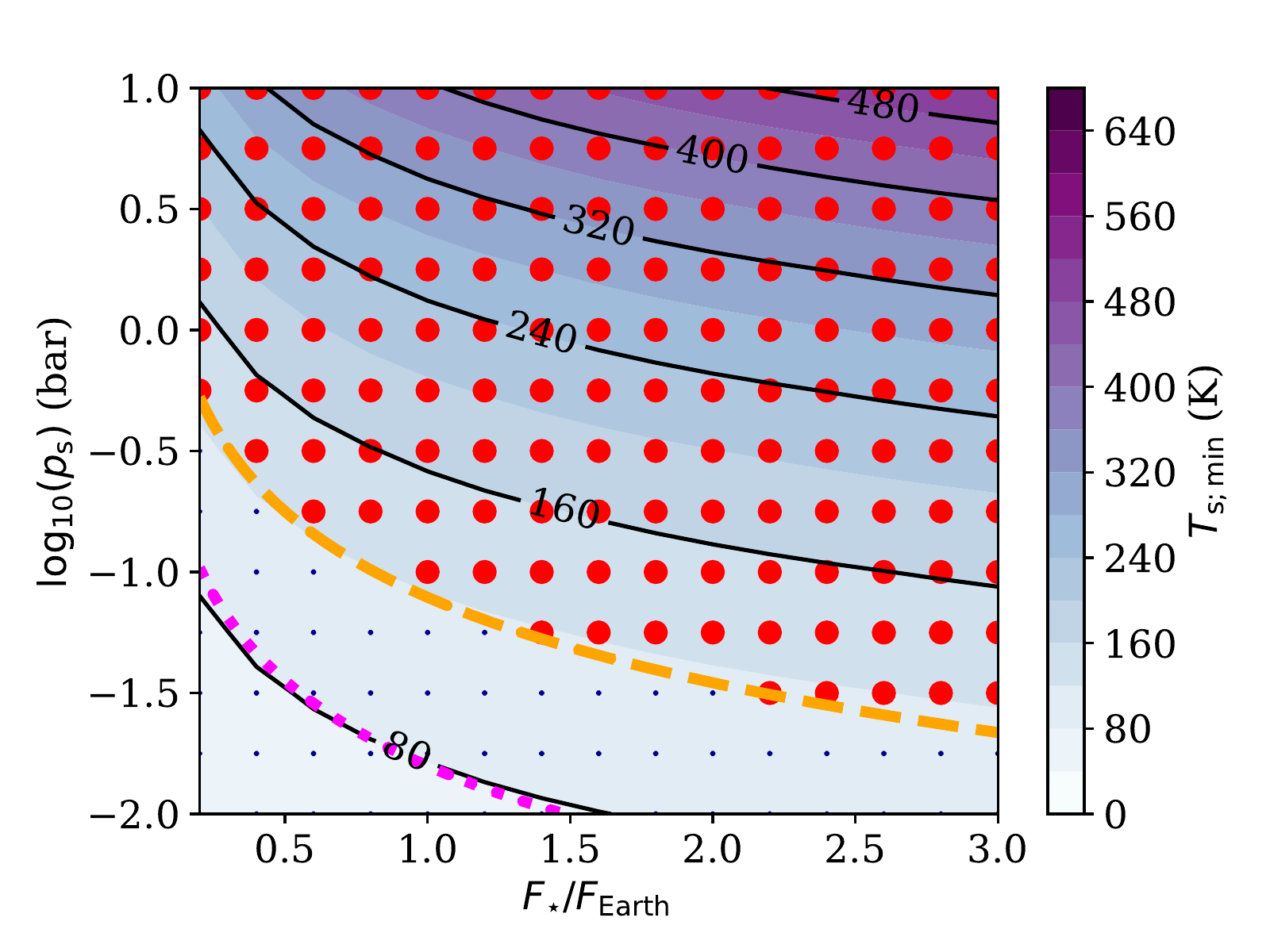}
   \includegraphics[height=\wpanel,trim = 1.0cm 1.0cm 3.0cm 0.75cm,clip]{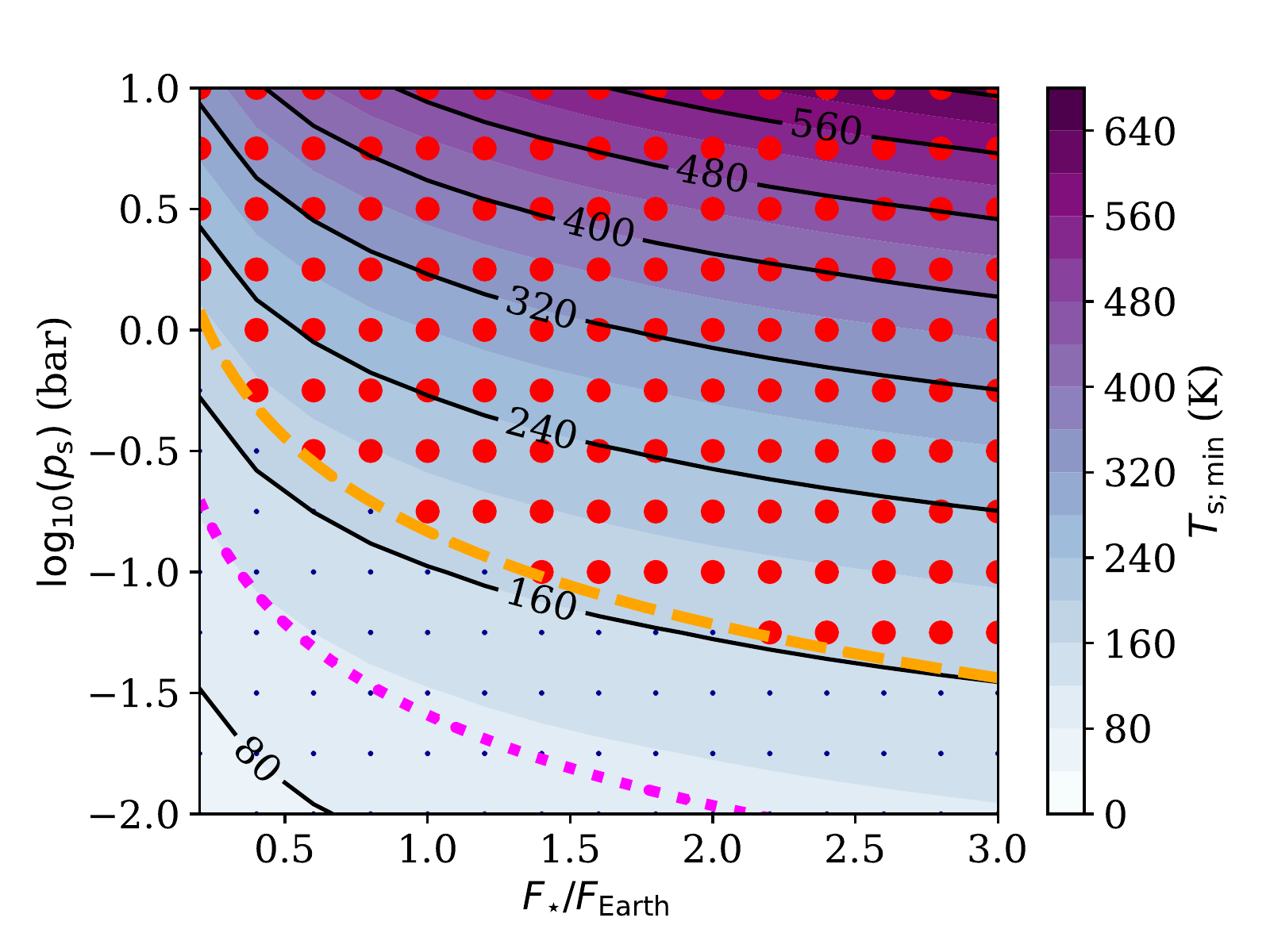} 
 \hspace{\wcbar}~ \\
   \raisebox{\hraisebox}[1cm][0pt]{%
   \begin{minipage}{1cm}%
   \textsc{0-D}
\end{minipage}}
\raisebox{\hraiseboxps}[1cm][0pt]{\rotatebox[origin=c]{90}{\tiny{$\log_{10} \left( \psurf \right)$~(bar)}}}
   \includegraphics[height=\wpanel,trim = 1.0cm 1.0cm 3.0cm 0.75cm,clip]{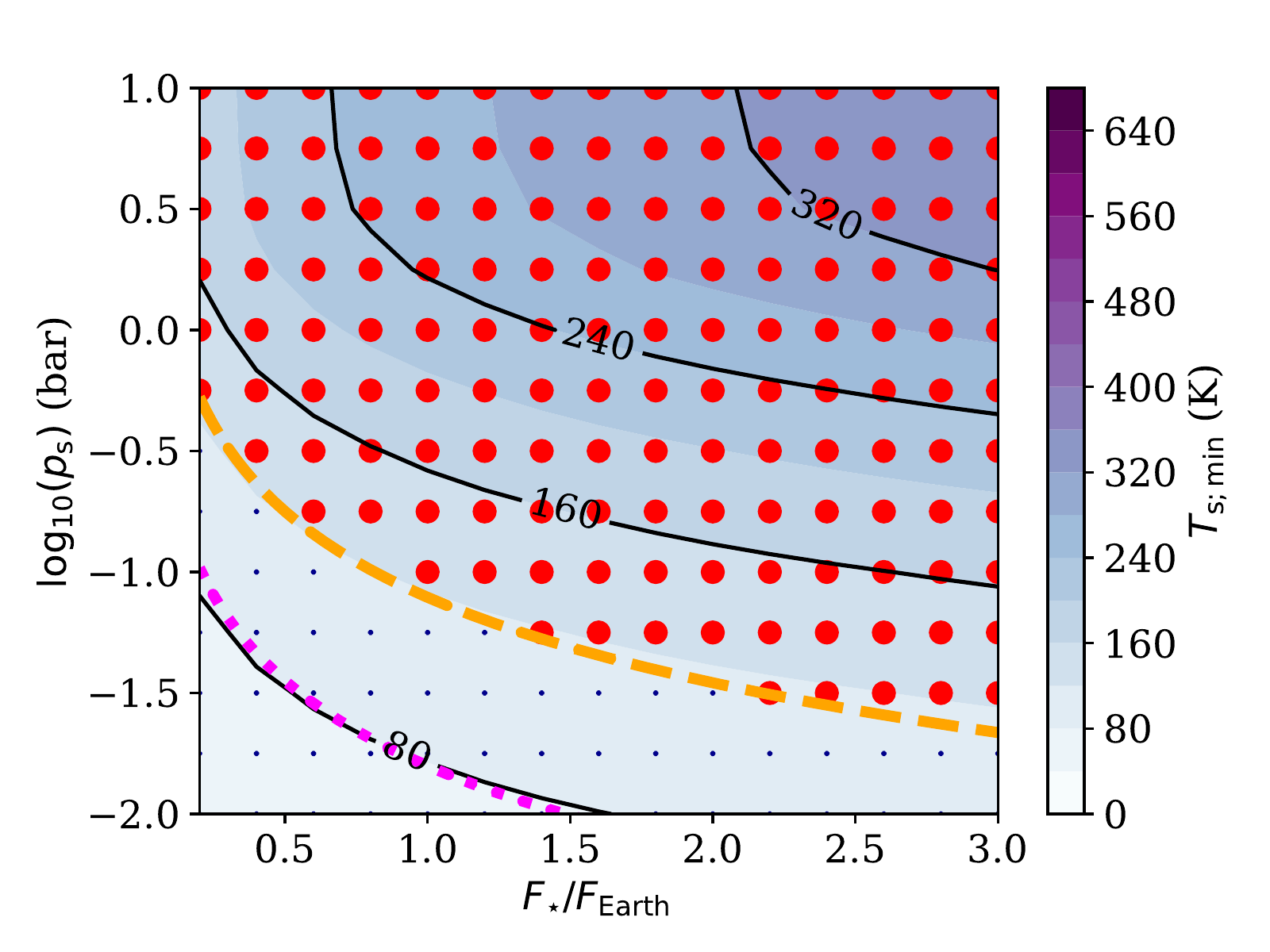}
   \includegraphics[height=\wpanel,trim = 1.0cm 1.0cm 3.0cm 0.75cm,clip]{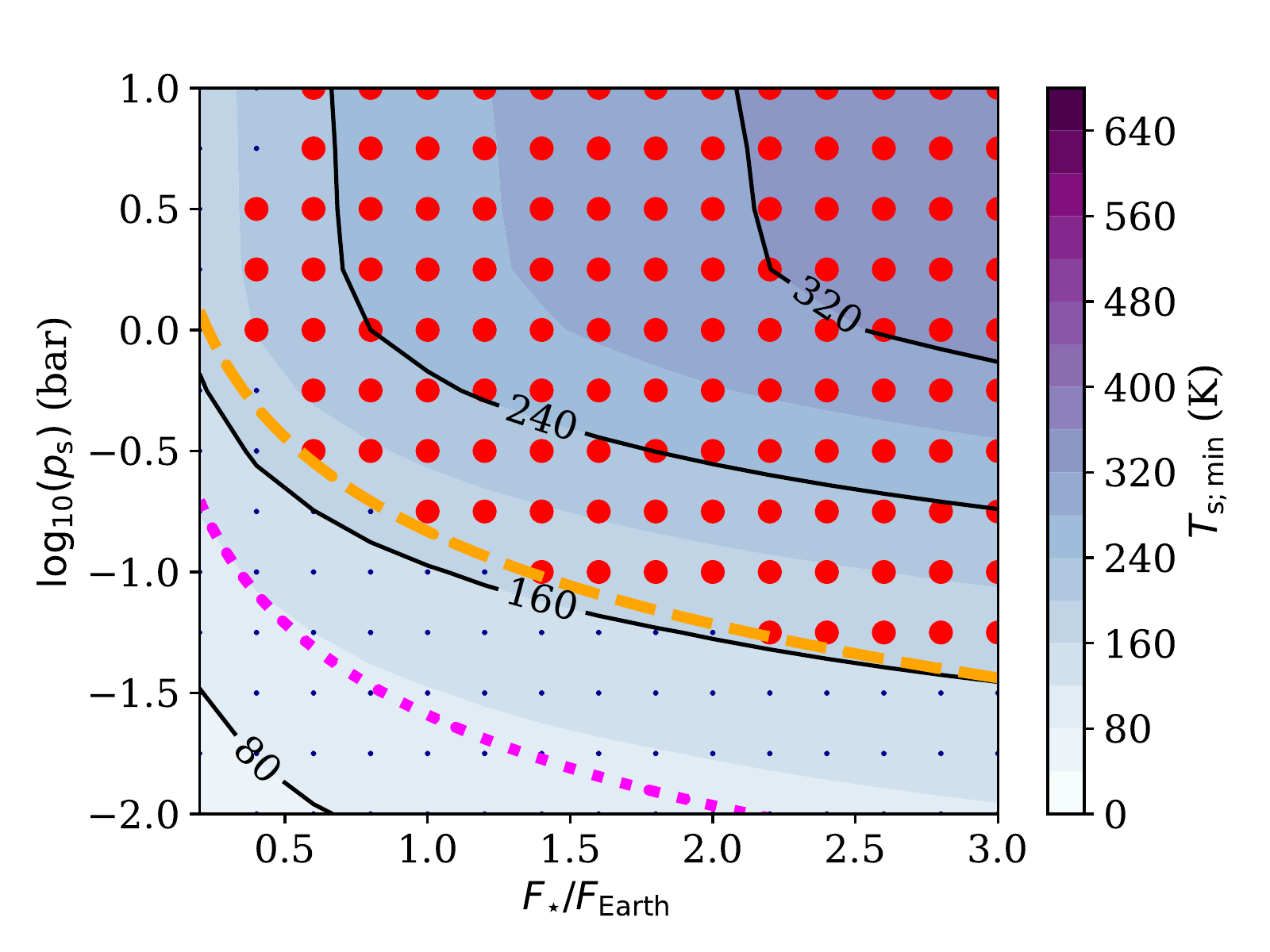} 
\hspace{\wcbar}~ \\
   \hspace{1cm} \tiny{$\Fstar / \Fearth $} \hspace{4cm}  \tiny{$\Fstar / \Fearth $} 
      \caption{Stability diagrams of Earth-sized tidally locked planets with Earth-like (left panels) and pure $\carbondiox$ (right panels) atmospheres, and associated minimum surface temperature. Quantities are plotted as functions of the stellar flux normalised by the Earth's stellar flux, $\Fstar / \Fearth $ (horizontal axis), and the surface pressure $\psurf$ in logarithmic scale (vertical axis). {\it From bottom to top:} 0-D ($1 \times 1$~grid), 1-D ($1 \times 50$~grid), 1.5-D ($2 \times 50$~grid), and 2-D ($32 \times 50$~grid) instances of the meta-model. Large red dots indicate simulations where the atmosphere remained stable, while small blue dots indicate atmospheric collapse. The orange dashed (or pink dotted) line indicates the collapse pressure $\pclow$ (or $\pcup$) corresponding to the lower (or upper) bound of the nightside surface temperature predicted by Wordsworth's analytic model, and given by \eq{Tnight_bounds}. }
       \label{fig:stability_diagrams}%
\end{figure*}

\section{Stability diagrams}
\label{sec:stability}

In this section, we examine both the role played by several physical features in the atmospheric stability, and the sensitivity of model predictions to the simplifications made in the different approaches. To do so, we proceed to a vertical inter-comparison between the four models introduced in \sect{sec:climate_regime} for the two Earth-sized synchronous planets defined in Table~\ref{tab:param_reference_case}. For each instance of the meta-model, simulations were performed on a $15 \times 13$ grid in the space of stellar flux and initial surface pressure, with $0.2 \Fearth \leq \Fstar \leq 3 \Fearth$ and $0.01 \ {\rm bar} \leq \psurf \leq 10 \ {\rm bar} $, $\Fearth = 1366 \ {\rm W \ m^{-2}}$ being the Earth's incident stellar flux. \RV{Starting from isothermal and zero-velocity initial conditions, simulations were run for a period $\trun = \min \left\{ \max \left[ \tradia, \tmin  \right] , \tmax \right\}$ ranging between $\tmin = 300$ and $\tmax = 30\,000$~Earth days, $\tradia$ being an empirical estimate of the timescale necessary to reach radiative equilibrium that includes the dependence of the radiative cooling timescale on surface pressure and stellar flux \citep[e.g.][]{SG2002},}
\begin{equation}
    \tradia =  900 \ {\rm days} \times \left( \frac{\psurf}{1 \ {\rm bar}} \right) \left( \frac{\Fstar}{1366 \ {\rm W \ m^{-2}}} \right)^{-3/4}.
\end{equation}
\RV{At the end of simulations,} two-day averaged distributions of mean fields were computed, as well as the resulting minimum (or nightside) surface temperature ($\Tnight$). 

\RV{As highlighted by \cite{WW2020} in the case of sub-Neptunes, the convergence time of 3-D numerical simulations can be extremely long ($\trun \sim 250 \ 000$~Earth days, typically) for massive atmospheres ($\psurf \gtrsim 80$~bar) owing to the long radiative timescale of deep atmospheric layers. In the present study, the maximum surface pressure ($10$~bar) is smaller than that usually assumed for sub-Neptunes, the circulation described by our 2-D model is simpler than that described by 3-D GCMs -- where complex structures and super-rotating zonal jets can emerge --, and the major part of the incident stellar flux reaches the planet's surface, which tends to facilitate vertical energy transport. Therefore the resulting convergence times are expected to be smaller than those reached in the case of sub-Neptunes by one order of magnitude approximately. However, we verified a posteriori that both the energy balance between the outgoing longwave radiation (OLR) and the absorbed stellar radiation (ASR) on the one hand, and the circulation on the other hand, had reached a steady state at the end of several test simulations. Besides, by running grid simulations with larger values of $\trun$, we noticed that increasing this parameter did not affect the considered mean fields.}  

Following earlier studies \citep[][]{Wordsworth2015,KA2016,ADH2020}, we assume that the greenhouse effect is mainly due to the presence of $\carbondiox$ in the atmosphere. The condensation temperature of $\carbondiox$ is given, in~K, by \citep[][]{Fanale1982,Wordsworth2010,Wordsworth2015}
\begin{equation}
\Tcondcarbdiox \left( \press \right) \! = \!
\left\{
\begin{array}{ll}
\! \! \dfrac{3167.8}{23.23 - \ln \left( 0.01 \press \right) }  & \! \mbox{if} \ \press < \ptriple, \\[0.3cm] 
\! \! 684.2 - 92.3 \ln \left( \press \right) + 4.32 \ln^2 \left( \press \right) & \! \mbox{if} \ \press \geq \ptriple,
\end{array}
\right.
\end{equation}
where the partial pressure of the gas $\press$ is given in Pa, and $\ptriple = 5.18 \times 10^5 $~Pa designates the triple point pressure. The stability diagrams are therefore obtained by comparing the minimum surface temperature calculated from simulations with the condensation temperature of $\carbondiox$ at the planet's surface, $\Tcondcarbdiox \left( \mixratio \psurf \right)$, where $\mixratio$ designates the\RV{ volume} mixing ratio of $\carbondiox$. In the Earth-like case, the\RV{ volume} mixing ratio of $\carbondiox$ is set to the value of Earth at the beginning of the twenty-first century, namely \RV{$\mixratio = 370 $~ppm} \citep[e.g.][]{Etheridge1996}, while $\mixratio = 1.0$ in the pure $\carbondiox$ case. The atmosphere is considered to be stable if $\Tnight \left( \Fstar , \psurf \right) > \Tcondcarbdiox \left( \mixratio \psurf \right) $ and unstable else. We remark that the collapse itself, when it occurs, is not described by the model since the changes of phases of $\carbondiox$ are not taken into account. 

Figure~\ref{fig:stability_diagrams} shows the simulation results. The minimum surface temperature is plotted in both cases as a function of the normalised stellar flux $\Fstar/ \Fearth$ and the surface pressure $\psurf$ in logarithmic scale. The stability diagrams are plotted too, with large red dots indicating stability and small blue dots collapse. The collapse pressures associated with the lower and upper bounds of the nightside temperature given by \eqs{Tnlow}{Tnup}, are obtained by solving for $\psurf$ the equations
\begin{align}
& \Tnightlow \left( \Fstar , \psurf \right) = \Tcondcarbdiox \left( \mixratio \psurf \right) , \\ 
& \Tnightup \left( \Fstar , \psurf \right) = \Tcondcarbdiox \left( \mixratio \psurf \right), 
\end{align}
and are denoted by $\pclow$ (orange dashed line) and $\pcup$ (pink dotted line), respectively.

With the 0-D model, we recover the behaviour of the nightside temperature predicted by the closed-form solutions of \cite{ADH2020} in the purely radiative regime. This is due to the fact that the two models are actually the same in this configuration, the temperatures being computed numerically here instead of analytically. Compared with the other models, the 0-D model tends to underestimate the nightside temperature in the high surface pressure regime. As the surface pressure increases, $\Tnight$ reaches a plateau corresponding to the planet's equilibrium temperature $\Teq \left( \Fstar \right)$ given by \eq{Teq}, and does not evolve with $\psurf$ any more. This unrealistic behaviour is a consequence of the isothermal approximation, which does not account for the strong vertical temperature gradient characterising thick atmospheres, especially their convective regions. In the low stellar flux regime, the 0-D model captures the stability decrease observed for pure $\carbondiox$ atmospheres in the 3-D GCM simulations performed by \cite{Wordsworth2015}. However, this feature is due to the isothermal temperature profile too since it vanishes from the moment that the atmospheric structure is allowed to adjust with radiative transfer, in models of higher dimensions. This effect is a caveat of the limitations of \RV{idealised} models in explaining predictions of much more sophisticated 3-D GCMs. 


The 1-D model exhibits the same behaviour as the 0-D model for surface pressures less than 1~bar, which corresponds to the regime where the vertically isothermal approximation holds. Beyond $\psurf \approx 1$~bar, the nightside surface temperature increases as a function of both the stellar flux and surface pressure, which makes the collapse pressure associated with $\Tnightlow$ capture the threshold of the stability region for the whole stellar flux interval. The two-column configuration (1.5-D model) relaxes the horizontally isothermal atmosphere approximation. As a consequence, the nightside temperature becomes dependent upon the efficiency of the interhemispheric heat redistribution, and can thereby be less than the lower bound obtained in the horizontally isothermal atmosphere approximation, $\Tnightlow$. The coarse spatial resolution of the 1.5-D model for the horizontal direction does not account for the strong convection generated in the substellar region. The wind speed is therefore underestimated, and so the strength of the overturning circulation and the heat advected from dayside to nightside are. This leads to the observed stability decrease: the collapse pressure is approximately increased by ${\sim} 25\%$ with respect to $\pclow$ in the Earth-like case, and by ${\sim} 80\% $ in the pure $\carbondiox$ case. 

Conversely, the 2-D model predicts a wider stability region. The collapse pressure is lowered by ${\sim} 10{-}40 \%$ with respect to $\pclow$ for $\Fstar \gtrsim 0.7 \Fearth$, which is significant albeit less than the $75\%$ maximum decrease predicted by the analytic theory \citep[$\pcup / \pclow = 1/4$, see][Eq.~(86)]{ADH2020} in the case of intense sensible heating ($\Lconv \rightarrow + \infty$, see \eq{Lconv}). This stability increase results from the effect of the planetary boundary layer. The vertical turbulent diffusion generated by the friction of mean flows against the planet's surface in the planetary layer acts both (i) to increase the thermal forcing of the atmosphere by intensifying sensible exchanges, and (ii) to enhance the day-night heat advection by strengthening the overturning circulation. As a consequence, the nightside surface is warmer by ${\sim} 4{-}14$~K in the vicinity of the threshold between the stability and collapse regions. 

\def\wpanel{0.23\textwidth}
\def\wpanelm{0.24\textwidth}
\def\wlegend{0.32\textwidth}
\def\hraisebox{0.20\textwidth}
\def\hraiseboxps{0.12\textwidth}
\def\wbox{2.5cm}
\def\wepsi{0.0cm}
\begin{figure*}[t]
   \centering
  \hspace{2cm} \textsc{Earth-like atmosphere} \hspace{1.5cm} \textsc{Pure $\carbondiox$ atmosphere} \\[0.3cm]
     \raisebox{\hraisebox}[1cm][0pt]{%
   \begin{minipage}{\wbox}%
   \textsc{$\Tntd - \Tnntd$~(K)}
\end{minipage}}
\raisebox{\hraiseboxps}[1cm][0pt]{\rotatebox[origin=c]{90}{\tiny{$\log_{10} \left( \psurf \right)$~(bar)}}}
   \includegraphics[height=\wpanel,trim = 1.0cm 1.0cm 2.8cm 0.75cm,clip]{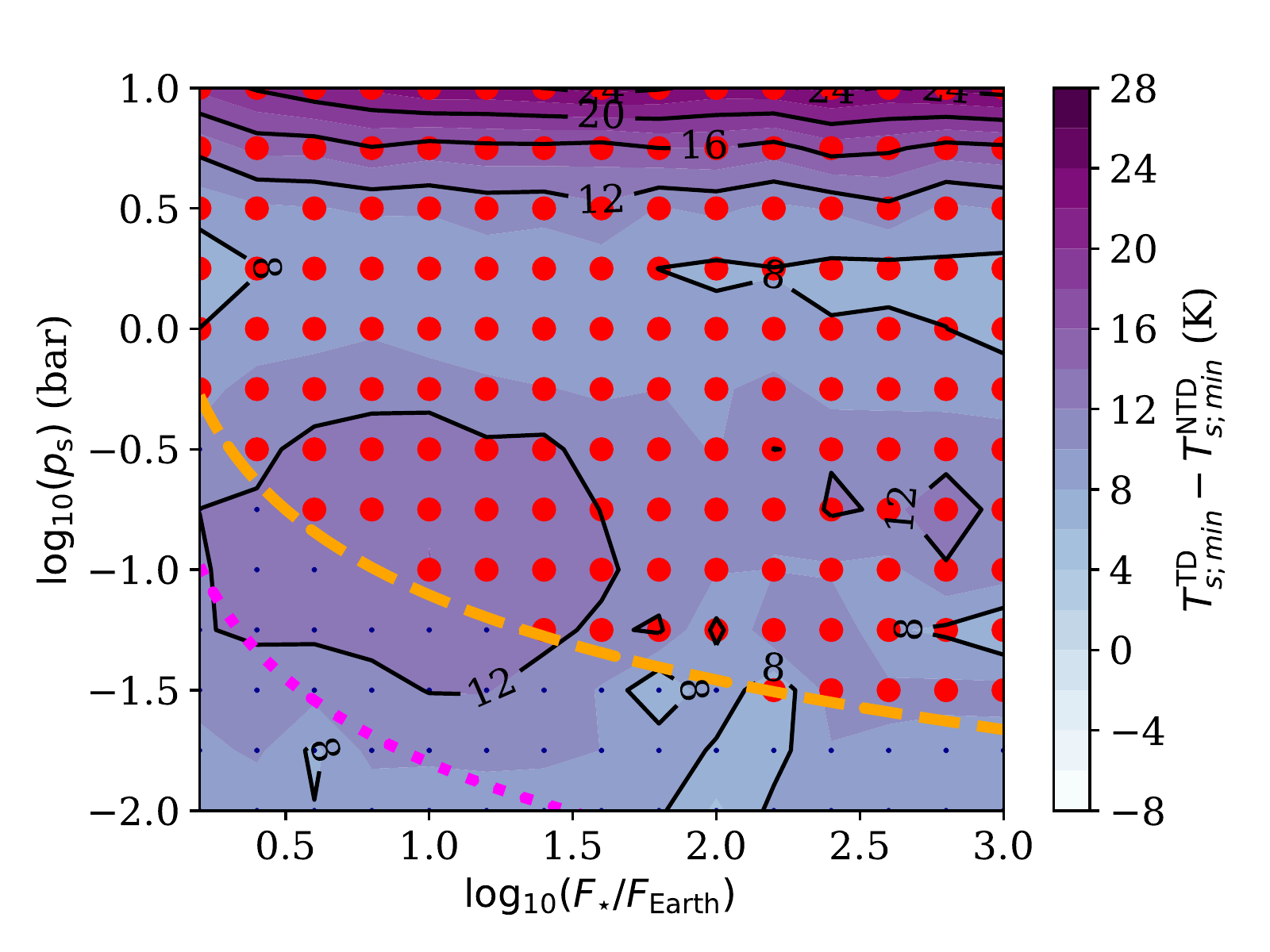}
   \includegraphics[height=\wpanel,trim = 1.0cm 1.0cm 2.8cm 0.75cm,clip]{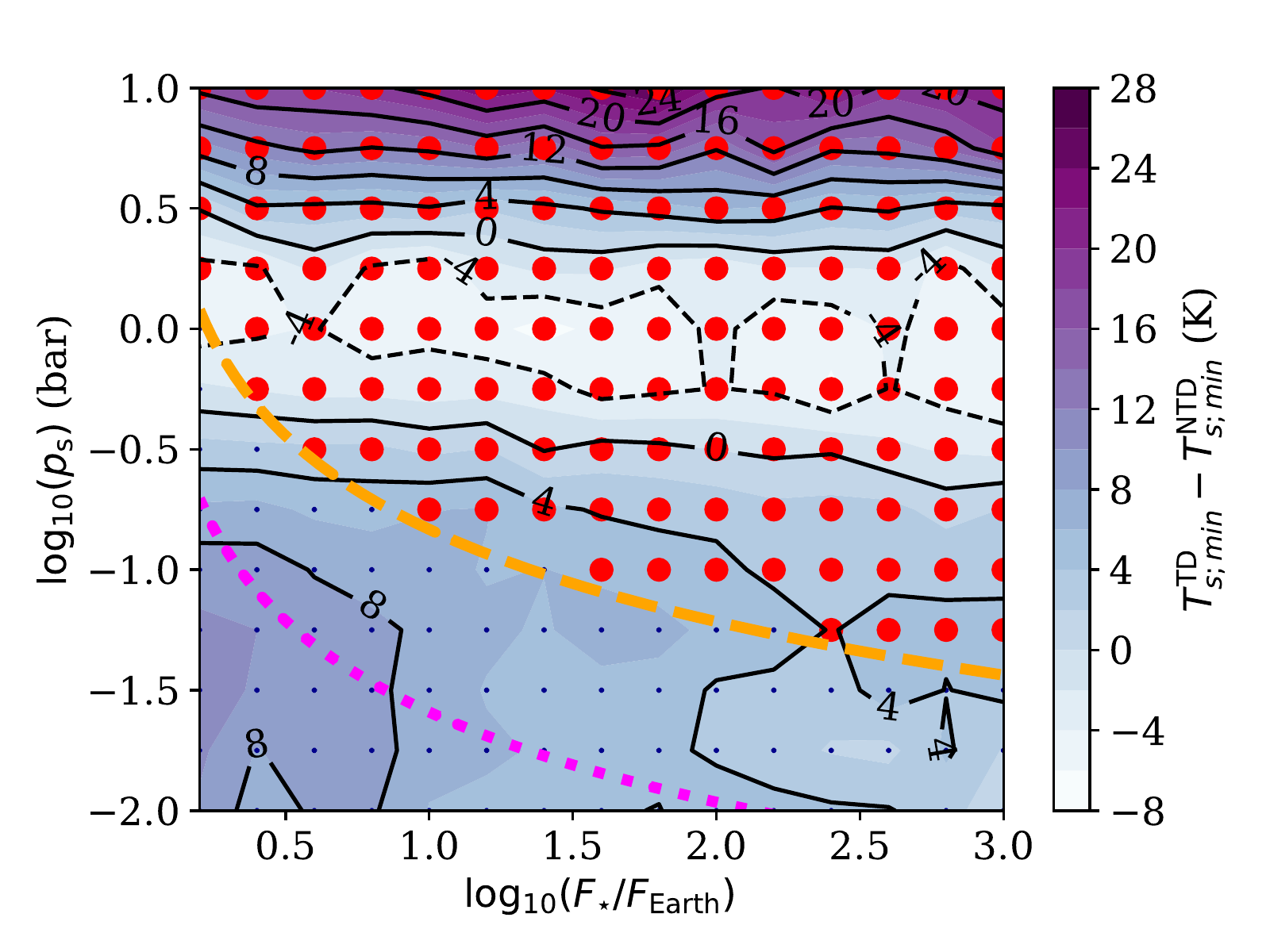} 
   \includegraphics[height=\wpanel,trim = 13.4cm 1.0cm 1.3cm 0.75cm,clip]{auclair-desrotour_fig4b.pdf} 
  \raisebox{\hraiseboxps}[1cm][0pt]{\rotatebox[origin=c]{90}{\tiny{$\Tntd - \Tnntd$~(K)}}} \\
  \hspace{\wepsi} 
     \raisebox{\hraisebox}[1cm][0pt]{%
   \begin{minipage}{\wbox}%
   \textsc{$\log_{10} \left( \tadvtd / \tadvntd \right)$}
\end{minipage}}
\raisebox{\hraiseboxps}[1cm][0pt]{\rotatebox[origin=c]{90}{\tiny{$\log_{10} \left( \psurf \right)$~(bar)}}}
 \hspace{\wepsi}  \includegraphics[height=\wpanelm,trim = 1.0cm 1.0cm 3.5cm 0.75cm,clip]{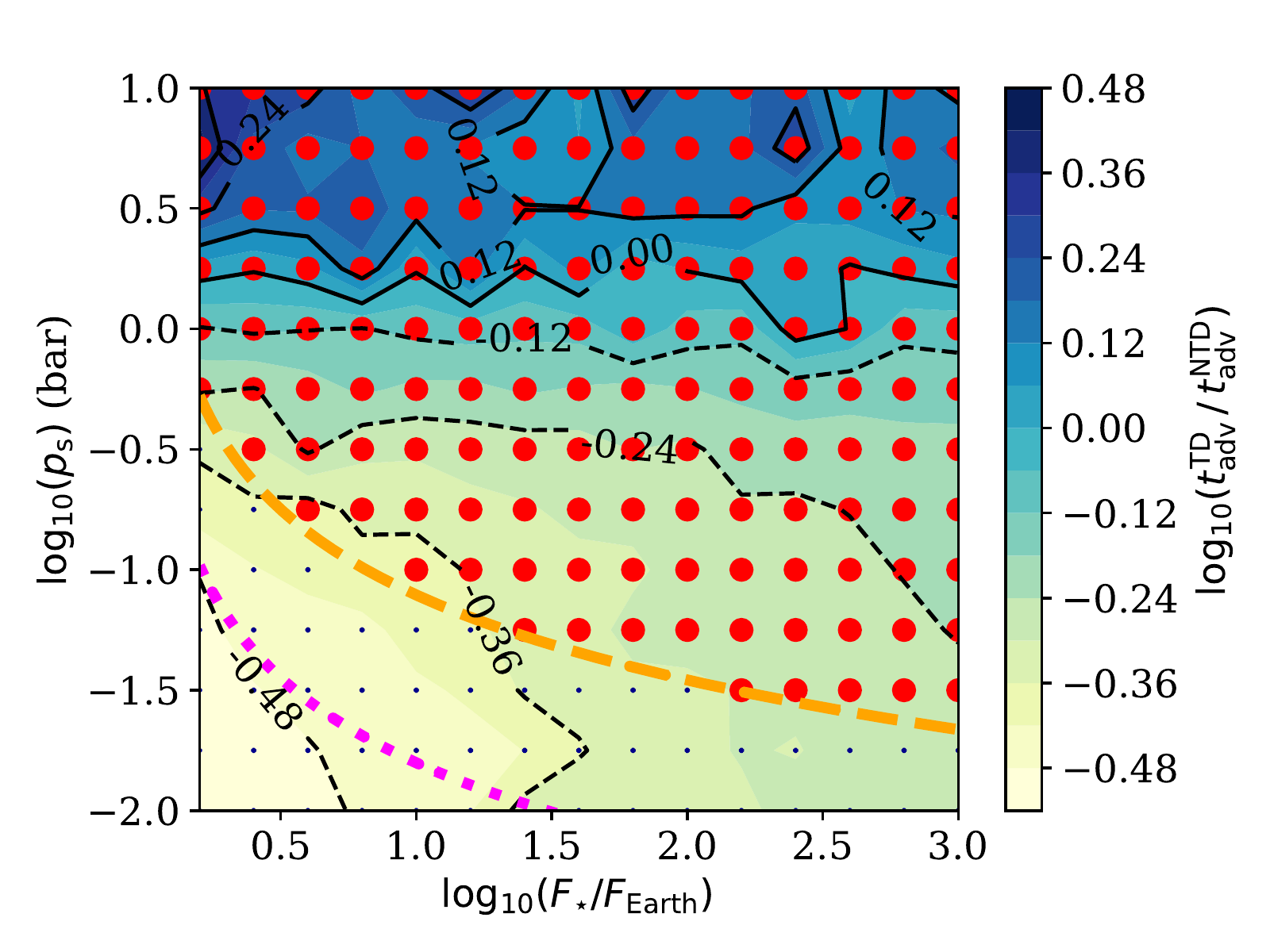}
\hspace{\wepsi}    \includegraphics[height=\wpanelm,trim = 1.0cm 1.0cm 3.5cm 0.75cm,clip]{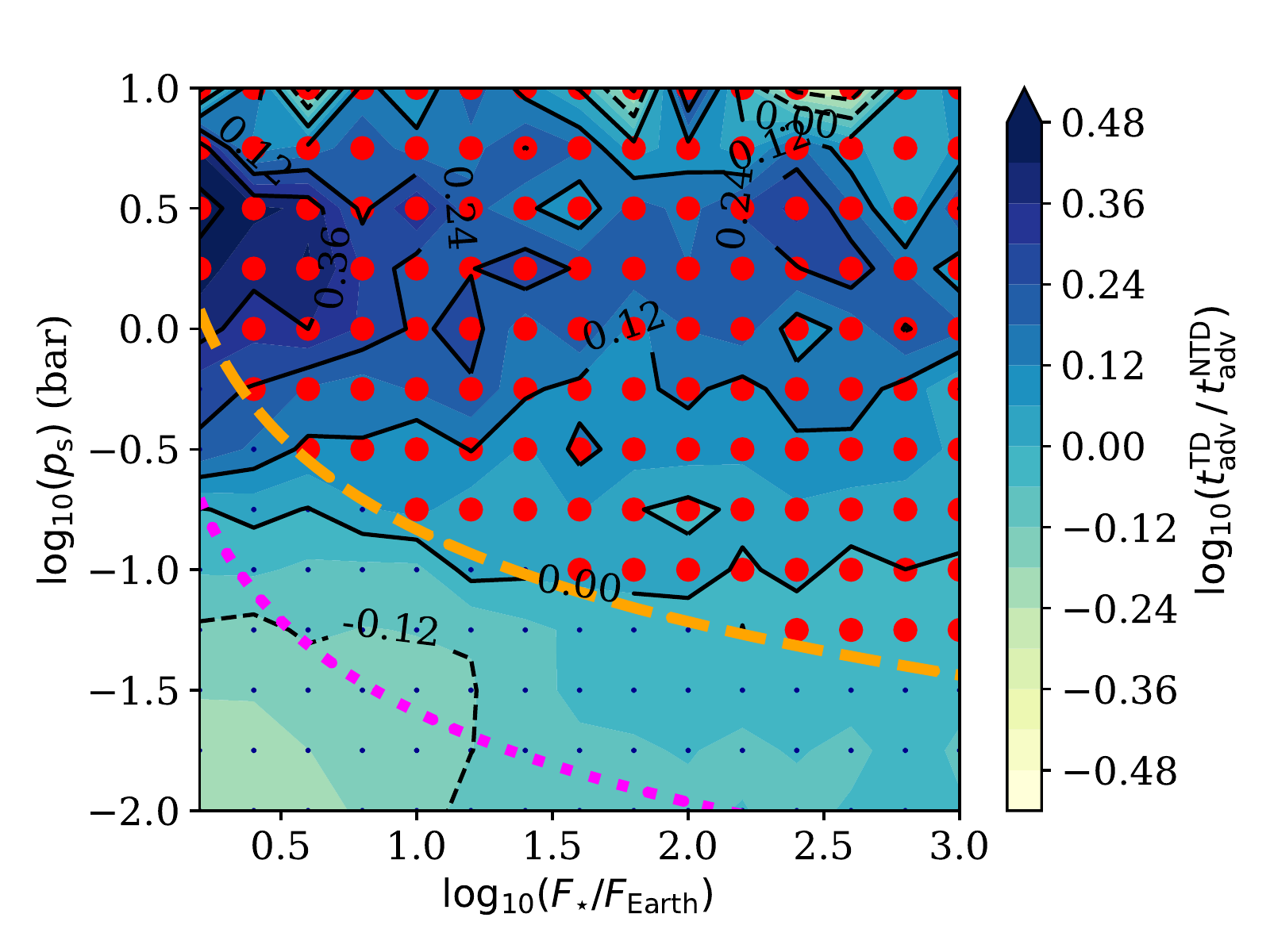} 
\hspace{\wepsi}    \includegraphics[height=\wpanelm,trim = 12.8cm 1.0cm 1.2cm 0.75cm,clip]{auclair-desrotour_fig4d.pdf}
   \raisebox{\hraiseboxps}[1cm][0pt]{\rotatebox[origin=c]{90}{\tiny{$\log_{10} \left( \tadvtd / \tadvntd \right)$}}} \\
   \raisebox{\hraisebox}[1cm][0pt]{%
   \begin{minipage}{\wbox}%
   \textsc{$\log_{10} \left( \tadvtd / \tadvKA \right)$}
\end{minipage}}
\raisebox{\hraiseboxps}[1cm][0pt]{\rotatebox[origin=c]{90}{\tiny{$\log_{10} \left( \psurf \right)$~(bar)}}}
   \includegraphics[height=\wpanel,trim = 1.0cm 1.0cm 2.8cm 0.75cm,clip]{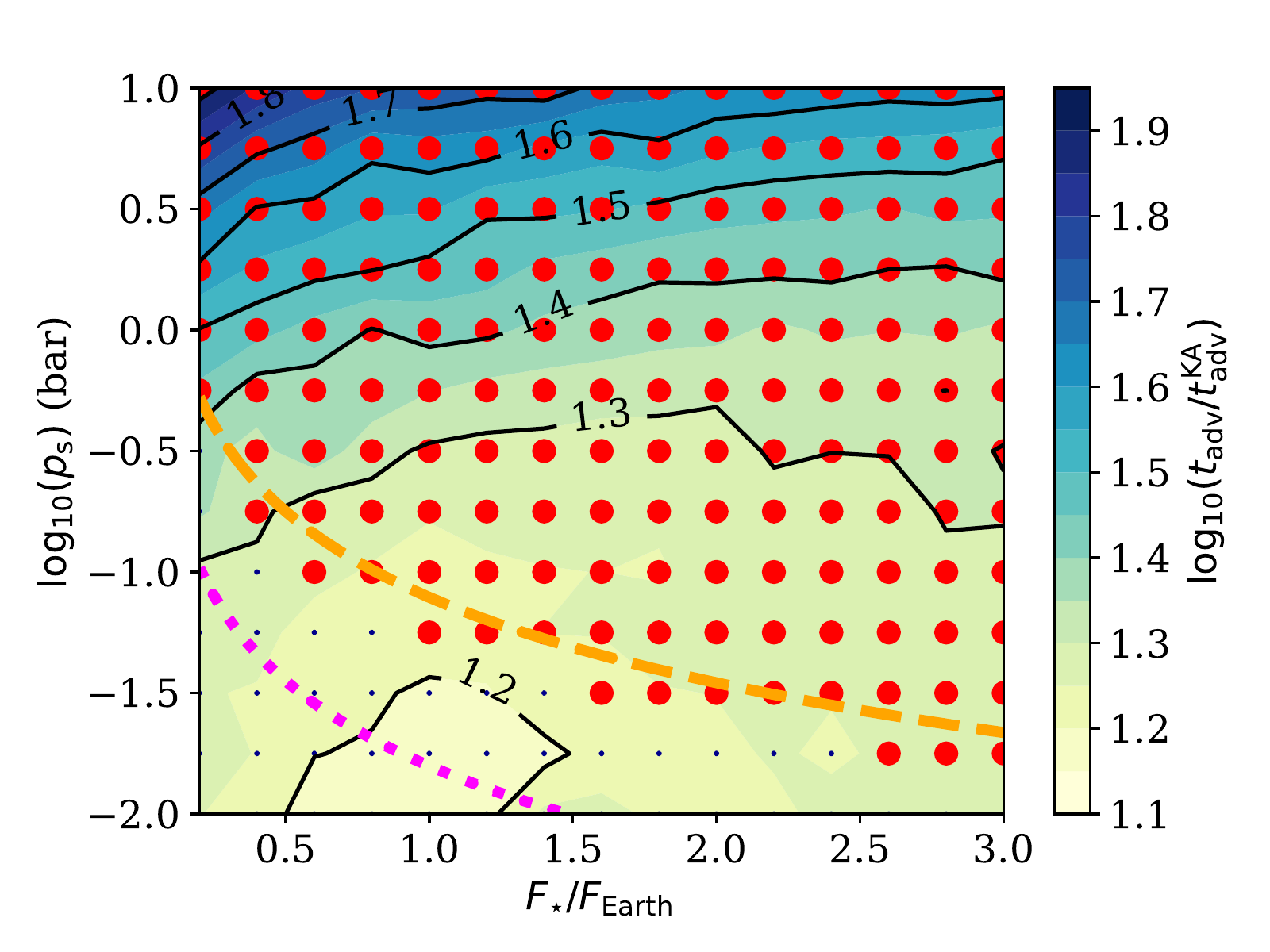}
   \includegraphics[height=\wpanel,trim = 1.0cm 1.0cm 2.8cm 0.75cm,clip]{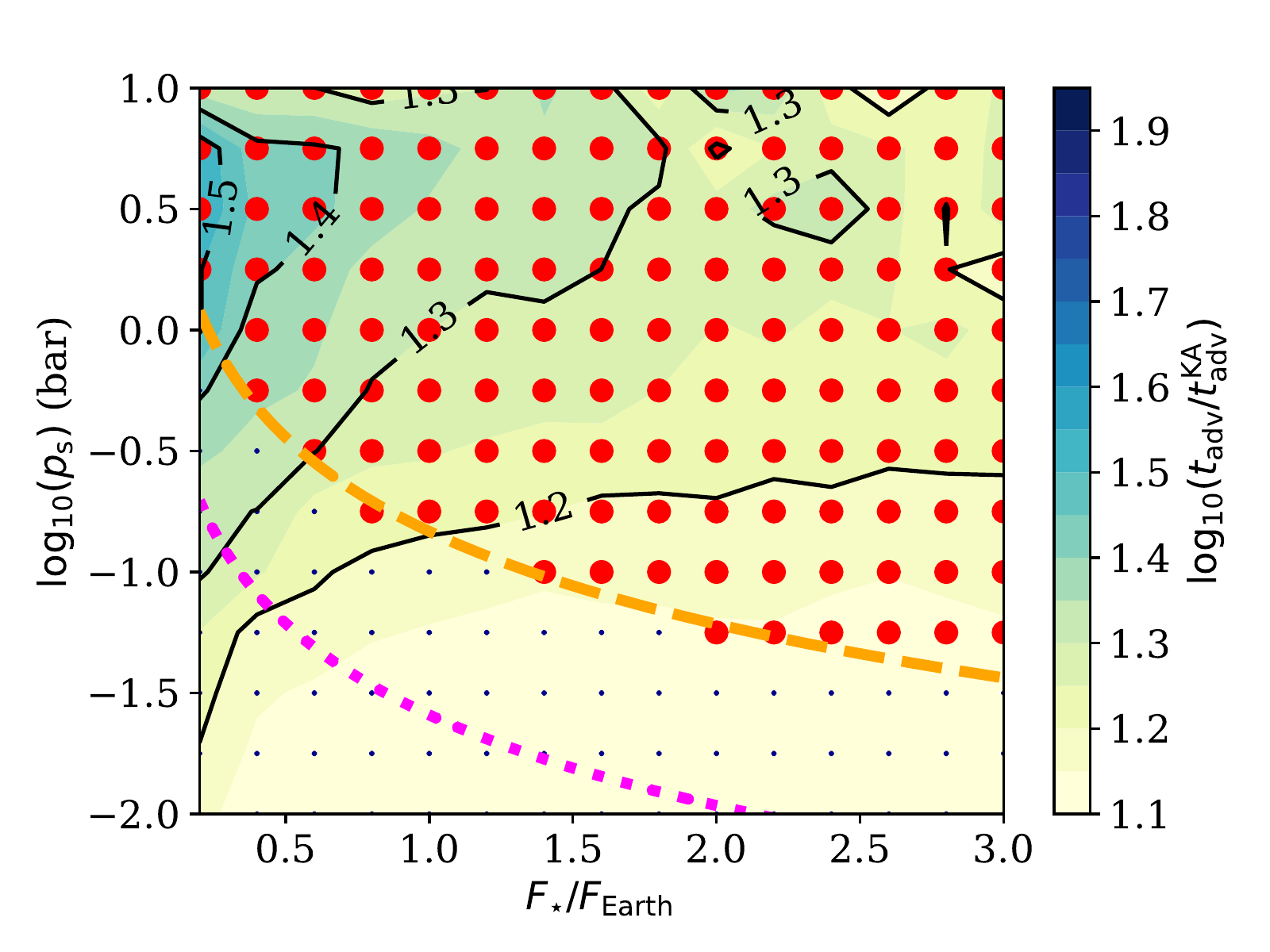} 
   \includegraphics[height=\wpanel,trim = 13.4cm 1.0cm 1.3cm 0.75cm,clip]{auclair-desrotour_fig4f.pdf}
   \raisebox{\hraiseboxps}[1cm][0pt]{\rotatebox[origin=c]{90}{\tiny{$\log_{10} \left( \tadvtd / \tadvKA \right)$}}} \\
    \raisebox{\hraisebox}[1cm][0pt]{%
   \begin{minipage}{\wbox}%
   \textsc{$\log_{10} \left( \tadvtd \right)$~(days)}
\end{minipage}}
\raisebox{\hraiseboxps}[1cm][0pt]{\rotatebox[origin=c]{90}{\tiny{$\log_{10} \left( \psurf \right)$~(bar)}}}
   \includegraphics[height=\wpanel,trim = 1.0cm 1.0cm 2.8cm 0.75cm,clip]{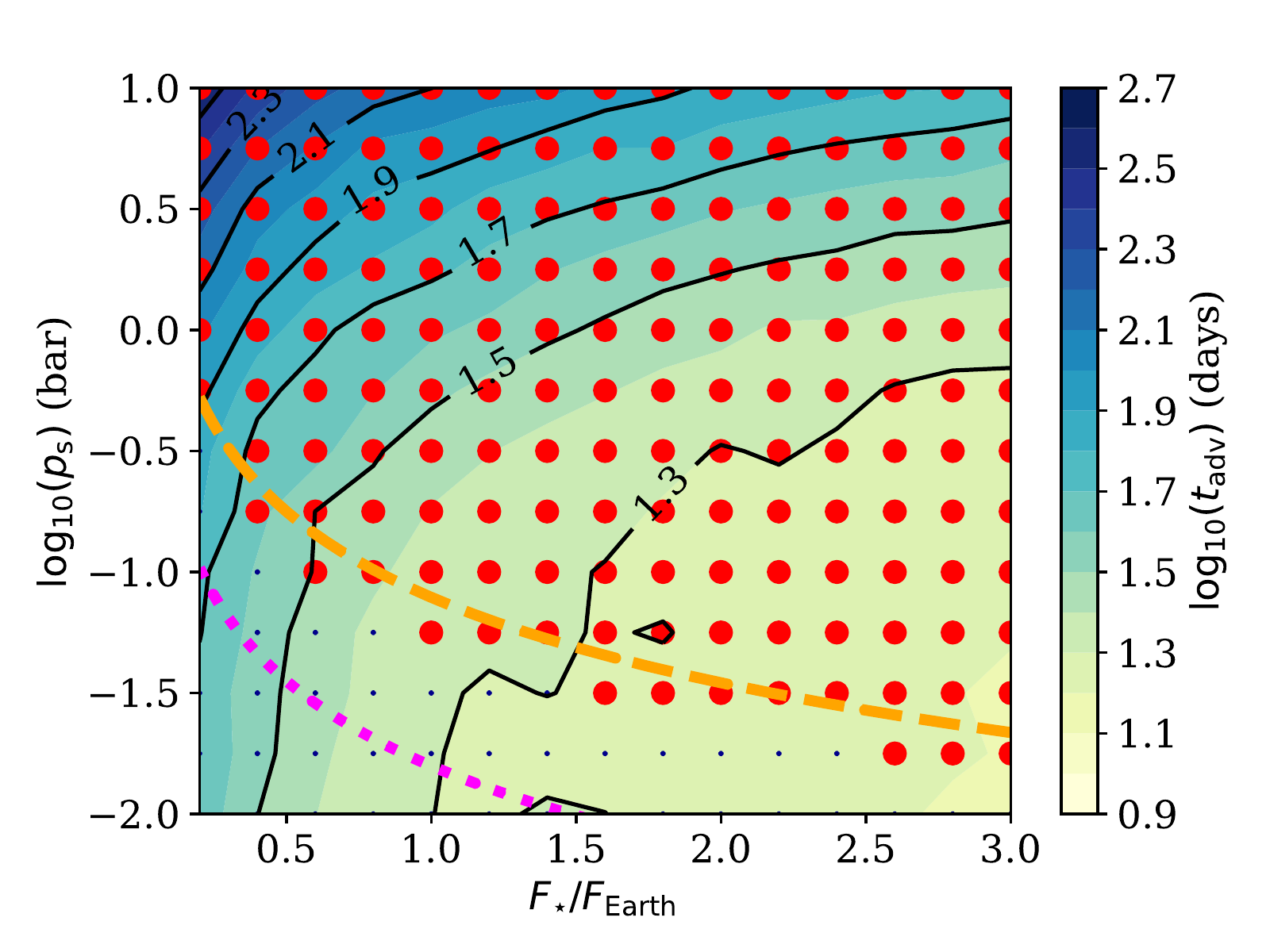}
   \includegraphics[height=\wpanel,trim = 1.0cm 1.0cm 2.8cm 0.75cm,clip]{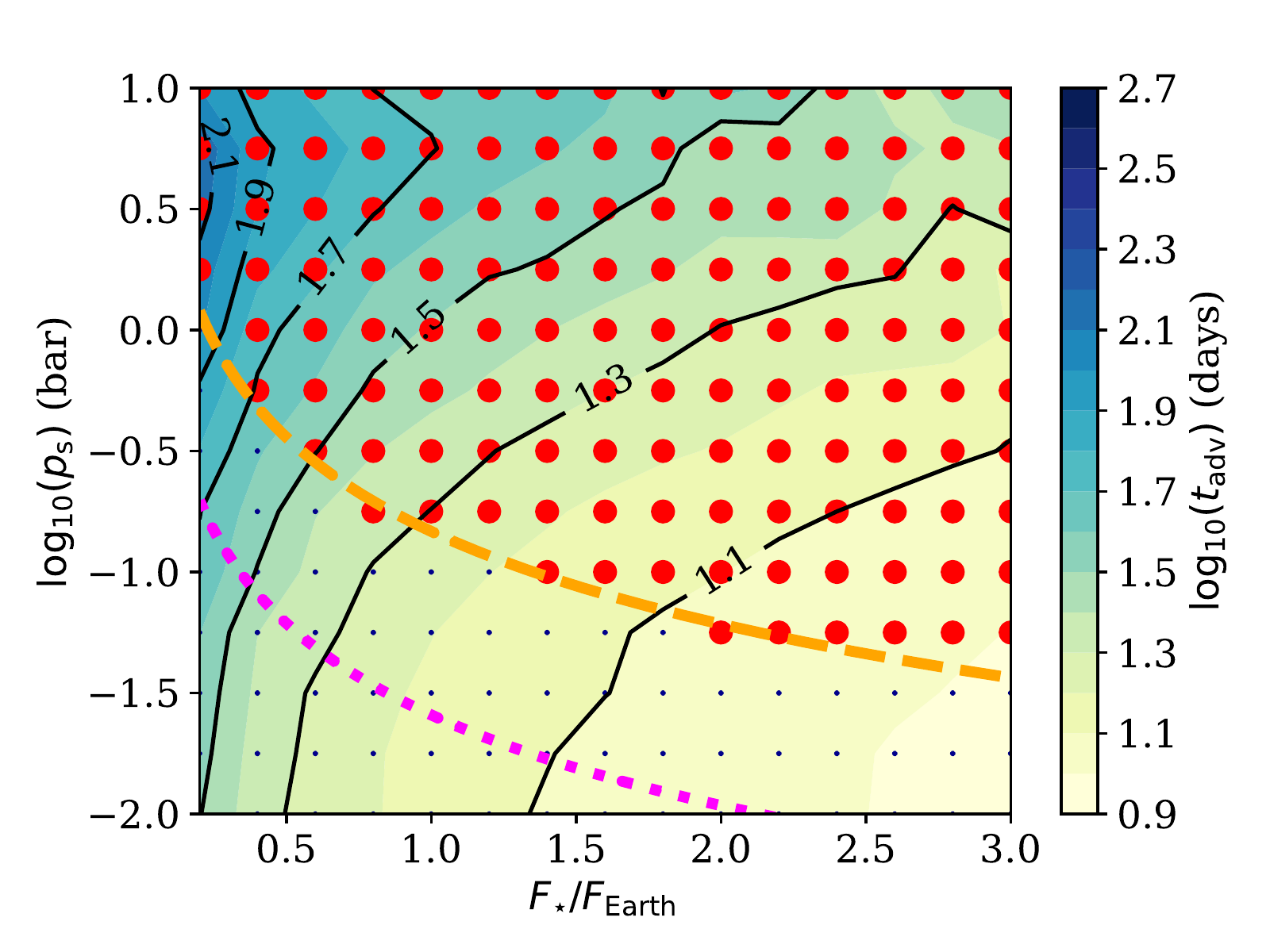} 
   \includegraphics[height=\wpanel,trim = 13.5cm 1.0cm 1.2cm 0.75cm,clip]{auclair-desrotour_fig4h.pdf}
   \raisebox{\hraiseboxps}[1cm][0pt]{\rotatebox[origin=c]{90}{\tiny{$\log_{10} \left( \tadvtd \right)$~(days)}}} \\
   \hspace{\wbox} \tiny{$\Fstar / \Fearth $} \hspace{4cm}  \tiny{$\Fstar / \Fearth $} 
      \caption{Stability diagrams of Earth-sized tidally locked planets hosting Earth-like (left panels) and pure $\carbondiox$ (right panels) atmospheres, with (TD, four bottom panels) and without turbulent diffusion (NTD, four top panels)\RV{, superimposed on the maps of various quantities}. {\it From bottom to top:} Day-night advection timescale with turbulent diffusion, day-night advection timescale with turbulent diffusion over the lower bound estimated analytically by \cite{KA2016} using the heat engine theory, ratio of day-night advection timescales with and without turbulent diffusion, nightside surface temperature difference between the cases with and without turbulent diffusion. \RV{The curves and symbols are the same as in \fig{fig:stability_diagrams}.}}
       \label{fig:pbl_effect}%
\end{figure*}

The nightside temperature increase induced by the planetary boundary layer was quantified by running simulations in the 2-D configuration without turbulent diffusion both for Earth-like and pure $\carbondiox$ atmospheres. In these simulations, the surface-atmosphere heat exchanges are induced by radiative transfer only, and there is no friction of mean flows against the surface. Figure~\ref{fig:pbl_effect} shows the resulting stability diagrams, as well as the corresponding nightside surface temperature difference between the cases with and without turbulent diffusion, denoted by the superscripts TD (Turbulent Diffusion) and NTD (No Turbulent Diffusion), respectively (top panels). In addition with the nightside surface temperature, we consider the day-night advection timescale $\tadv$, which is defined here as the mean period necessary for a fluid parcel to accomplish one full cycle of the day-night overturning circulation (see \append{app:mass_flow_rate}),
\begin{equation}
\label{tadv}
\tadv \define \frac{4 \Rpla}{\left[ \integ{\abs{\Vtheta}}{\sig}{0}{1} \right]_{90^\degree}},
\end{equation}
where the subscript $90^\degree$ indicates that the integral of the mass flow rate is performed over the terminator annulus ($\col = 90^\degree$). The day-night advection timescale also corresponds to the mean renewal time of the air contained in one atmospheric hemisphere (dayside or nightside). This timescale can be compared to the advection timescales introduced in earlier studies, such as the analytic expression of the lower bound obtained by \cite{KA2016} from the heat engine theory \citep[][Eq.~(12)]{KA2016},
\begin{equation}
\tadvKA \define \frac{\Rpla}{\Vconv},
\end{equation}
where the typical speed $\Vconv$, given by \eq{Vconv}, is a function of the stellar flux, the surface pressure, the surface albedo, the dayside surface temperature, the optical depth in the longwave at surface, the specific gas constant, and the bulk drag coefficient of the surface layer. The timescale $\tadvKA$ is estimated by setting the bulk drag coefficient to the typical value $\Cd = 10^{-3}$ for convenience and by taking the maximum surface temperature for the dayside temperature. Thus, in addition with the nightside temperature difference mentioned above, we plot in \fig{fig:pbl_effect} the day-night advection timescale in the case with turbulent diffusion (bottom panels), the ratio of this timescale over $\tadvKA$, and the ratio between advection timescales in the cases with and without turbulent diffusion (middle panels).

We first consider the stability diagrams obtained in the absence of turbulent diffusion (\fig{fig:pbl_effect}, top and middle panels). Similarly as in the 1-D configuration, the threshold of the stability region coincides with the lower bound of $\Tnight$ associated with the purely radiative regime in the radiative box model, namely $\pclow$. This indicates that the bulk atmosphere is horizontally isothermal for $\psurf \gtrsim 0.1$~bar, which corresponds to an efficient interhemispheric heat redistribution. As highlighted by temperature differences (\fig{fig:pbl_effect}, top panels), turbulent diffusion tends to warm up the nightside surface in the general case with, for instance a temperature increase of $6-28$~K in the Earth-like case. However this temperature increase does not vary monotonically with the stellar flux and surface pressure, but instead it exhibits a bi-modal behaviour with maxima and minima depending on surface pressure. 

Particularly, turbulent diffusion in the PBL somehow counterintuitively acts to decrease the nightside surface temperature instead of increasing it in a region centred on $\psurf \sim 1$~bar for pure $\carbondiox$ atmospheres, with a minimum of $ - 8$~K for $\psurf = 1$~bar and $\Fstar = 1.2 \Fearth$. A similar -- although slightly smaller -- negative difference ($-5$~K) was obtained by running gray gas simulations with the fully global 3-D GCM \texttt{THOR} \citep[][]{Mendonca2016,Deitrick2020} in this configuration using the values given by Table~\ref{tab:param_reference_case} and assuming a zero-spin angular velocity, which tends to corroborate the prediction of the 2-D model. This effect of the planetary boundary layer can be analysed through the interplay between the day-night advection timescale given by \eq{tadv} and the dayside radiative timescale, $\trad$, which is the typical timescale needed for a warm fluid parcel located in the upper atmosphere to cool down radiatively. Both parameters are altered by the turbulent diffusion taking place within the planetary boundary layer. 

Notwithstanding the high pressure -- and optically thick -- regime, where strong convection develops, the effect of turbulent diffusion reaches a maximum around $\psurf \sim 0.1$~bar for Earth-like atmospheres and around $\psurf \sim 0.03 $~bar for pure $\carbondiox$ atmospheres (\fig{fig:pbl_effect}, top panels). This maximum is consistent with the fact that the additional thermal forcing due to turbulent diffusion is all the more significant as the radiative absorption is weak, which tends to maximise the impact of turbulent diffusion for small optical thicknesses. However, the radiative timescale of the atmosphere scales as $\trad \scale \psurf / \Tatm^3$ \citep[e.g.][Eq.~(10)]{SG2002} in the optically thin regime, meaning that the heat surplus provided by sensible exchanges is radiated towards space over timescales that become extremely short as the surface pressure tends to zero. Thus, in spite of the decay of $\tadv$ induced by turbulent diffusion, a fluid parcel is radiatively cooled before being advected to nightside by mean flows in the optically thin limit ($\trad \ll \tadv$), which mitigates the impact of the PBL in this regime and makes the temperature difference decay as $\psurf \rightarrow 0$. 

Similarly, as the surface pressure increases, the atmosphere switches from the optically thin regime to the optically thick regime, with a transition occurring at lower pressures in the pure $\carbondiox$ case than in the Earth-like case due to the difference between optical depths in the infrared. In this transition regime, the PBL strongly affects the advection timescale, which is increased up to three times (\fig{fig:pbl_effect}, middle panels) and becomes thereby greater than the radiative timescale. As a consequence, less heat is advected towards nightside than in the absence of PBL although the latter generates a heat surplus on dayside, and the nightside temperature difference falls to the observed minimum valley where it reaches negative values in the pure $\carbondiox$ case. 

\RV{We note that this behaviour could be significantly altered by the presence of gas, dust, and aerosols inducing the so-called anti-greenhouse effect by increasing the shortwave scattering and absorption, as observed on Titan \citep[][]{McKay1991}. The anti-greenhouse effect refers to the cooling of the planet surface resulting from the fact that a substantial part of the incident stellar flux is absorbed and re-radiated towards space in the infrared by the upper layers of the atmosphere, and that only a fraction of it reaches the surface \citep[e.g.][]{Pierrehumbert2010}. This effect is likely to play a major role on rocky planets with thick atmospheres similar to Venus, where only $\sim 0.1 - 1 \%$ of the incident solar flux reaches the surface \citep[][]{Lacis1975}. }

\def\wpanel{0.23\textwidth}
\def\wpanelm{0.235\textwidth}
\def\wlegend{0.32\textwidth}
\def\hraisebox{0.20\textwidth}
\def\hraiseboxps{0.12\textwidth}
\def\wbox{3.2cm}
\def\wepsi{0.0cm}
\begin{figure*}[t]
   \centering
  \hspace{2cm} \textsc{Earth-like atmosphere} \hspace{1.5cm} \textsc{Pure $\carbondiox$ atmosphere} \\[0.3cm]
     \raisebox{\hraisebox}[1cm][0pt]{%
   \begin{minipage}{\wbox}%
   \textsc{$\log_{10} \left( \sfmax \right)$~(${\rm Gkg \ s^{-1}}$)}
\end{minipage}}
\raisebox{\hraiseboxps}[1cm][0pt]{\rotatebox[origin=c]{90}{\tiny{$\log_{10} \left( \psurf \right)$~(bar)}}}
   \includegraphics[height=\wpanel,trim = 1.0cm 1.0cm 2.8cm 0.75cm,clip]{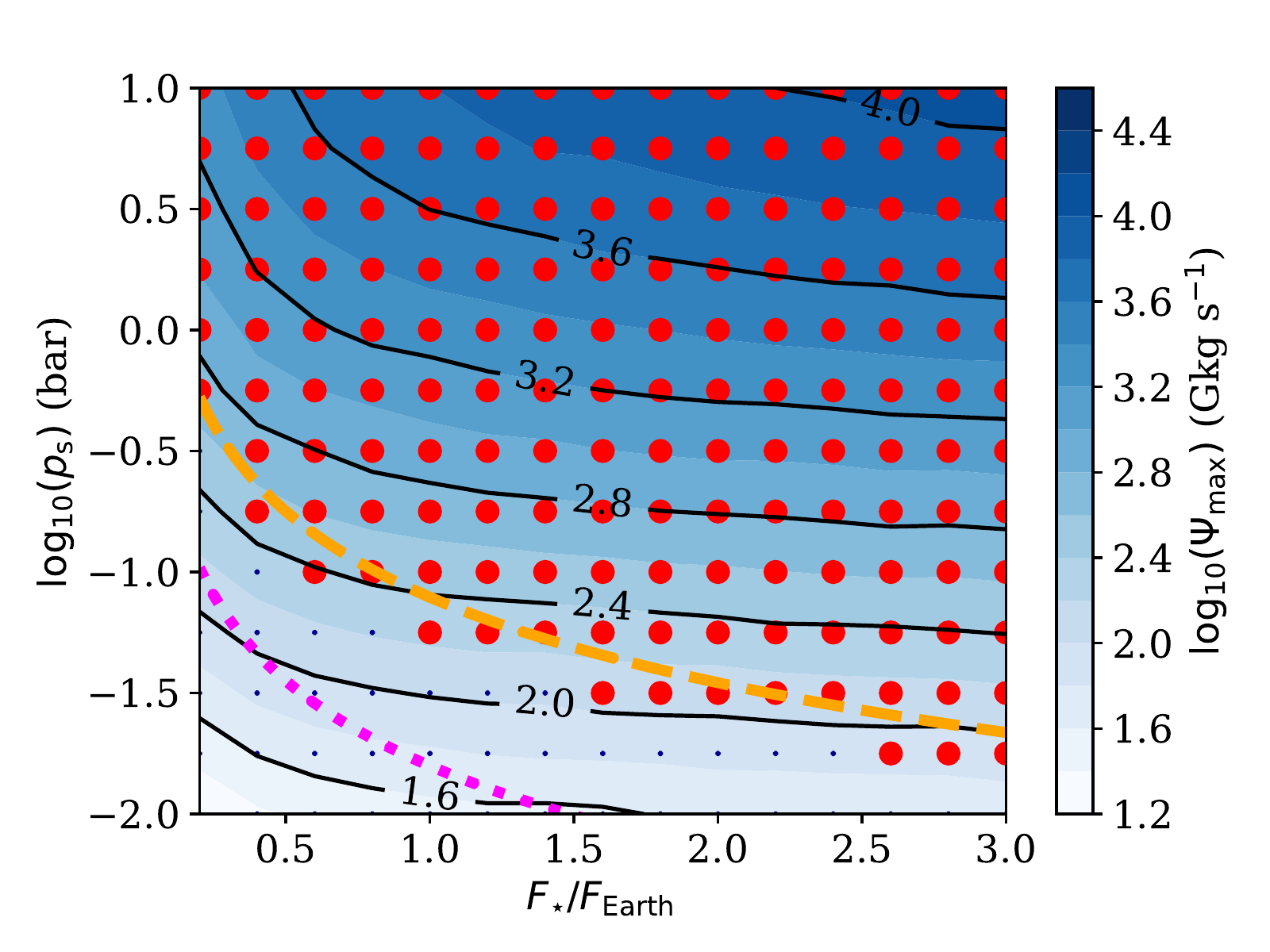}
   \includegraphics[height=\wpanel,trim = 1.0cm 1.0cm 2.8cm 0.75cm,clip]{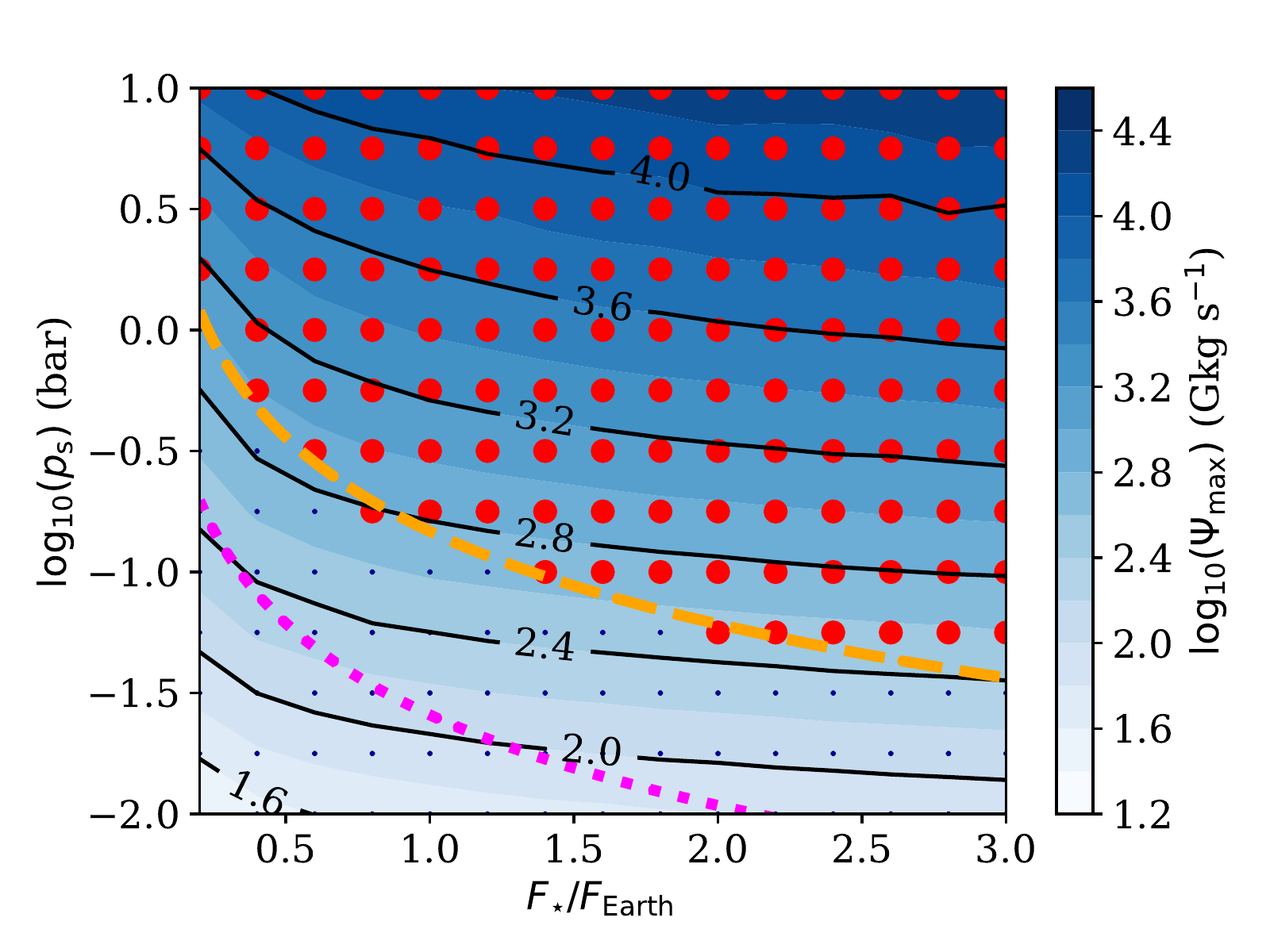} 
   \includegraphics[height=\wpanel,trim = 13.4cm 1.0cm 1.2cm 0.75cm,clip]{auclair-desrotour_fig5b.pdf} 
  \raisebox{\hraiseboxps}[1cm][0pt]{\rotatebox[origin=c]{90}{\tiny{$\log_{10} \left( \sfmax \right)$~(${\rm Gkg \ s^{-1}}$)}}} \\
  \hspace{\wepsi} 
     \raisebox{\hraisebox}[1cm][0pt]{%
   \begin{minipage}{\wbox}%
   \textsc{$\log_{10} \left( \sfmax / \sfmaxana \right)$}
\end{minipage}}
\raisebox{\hraiseboxps}[1cm][0pt]{\rotatebox[origin=c]{90}{\tiny{$\log_{10} \left( \psurf \right)$~(bar)}}}
 \hspace{\wepsi}  \includegraphics[height=\wpanelm,trim = 1.0cm 1.0cm 3.2cm 0.75cm,clip]{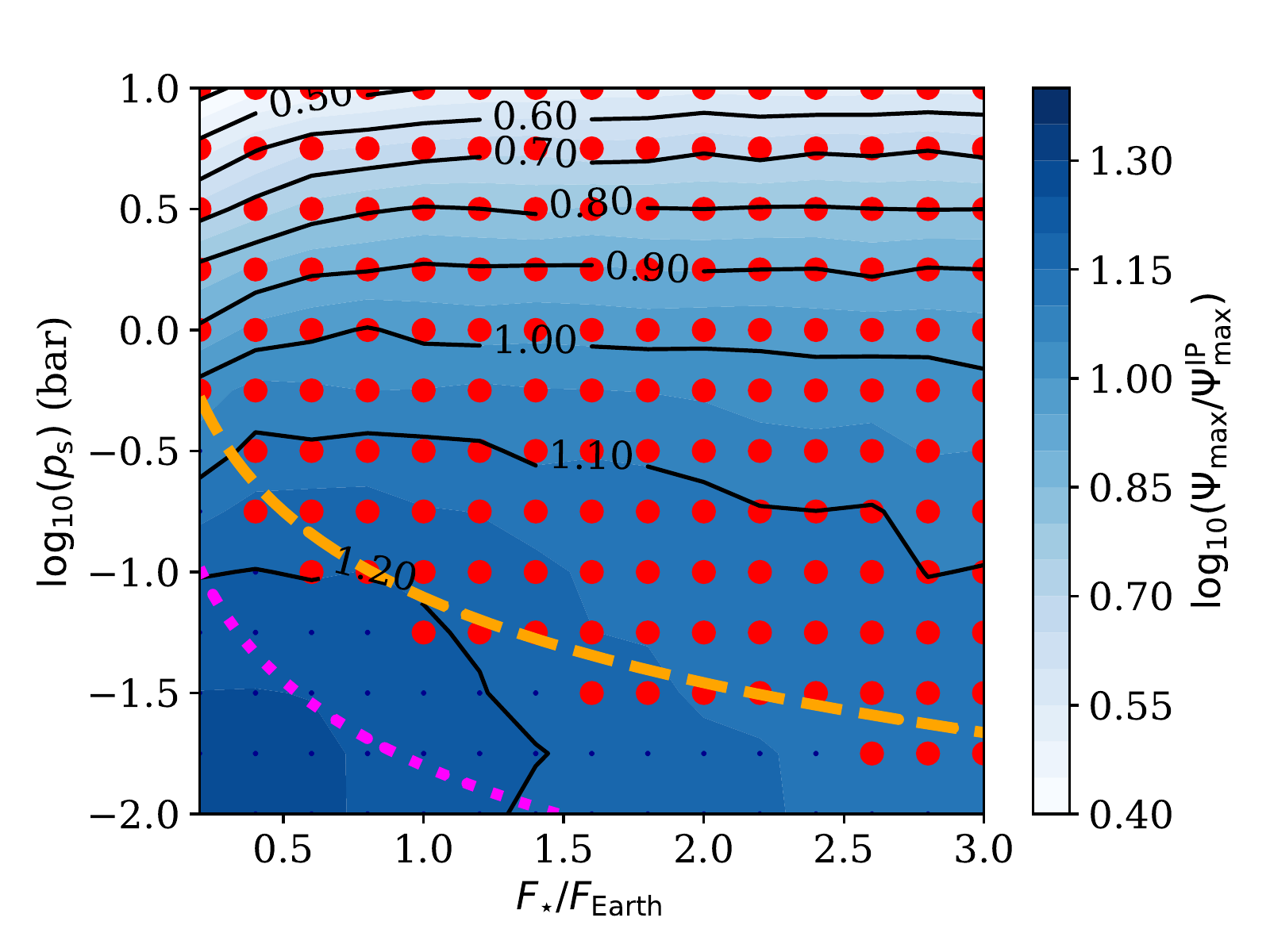}
\hspace{\wepsi}    \includegraphics[height=\wpanelm,trim = 1.0cm 1.0cm 3.2cm 0.75cm,clip]{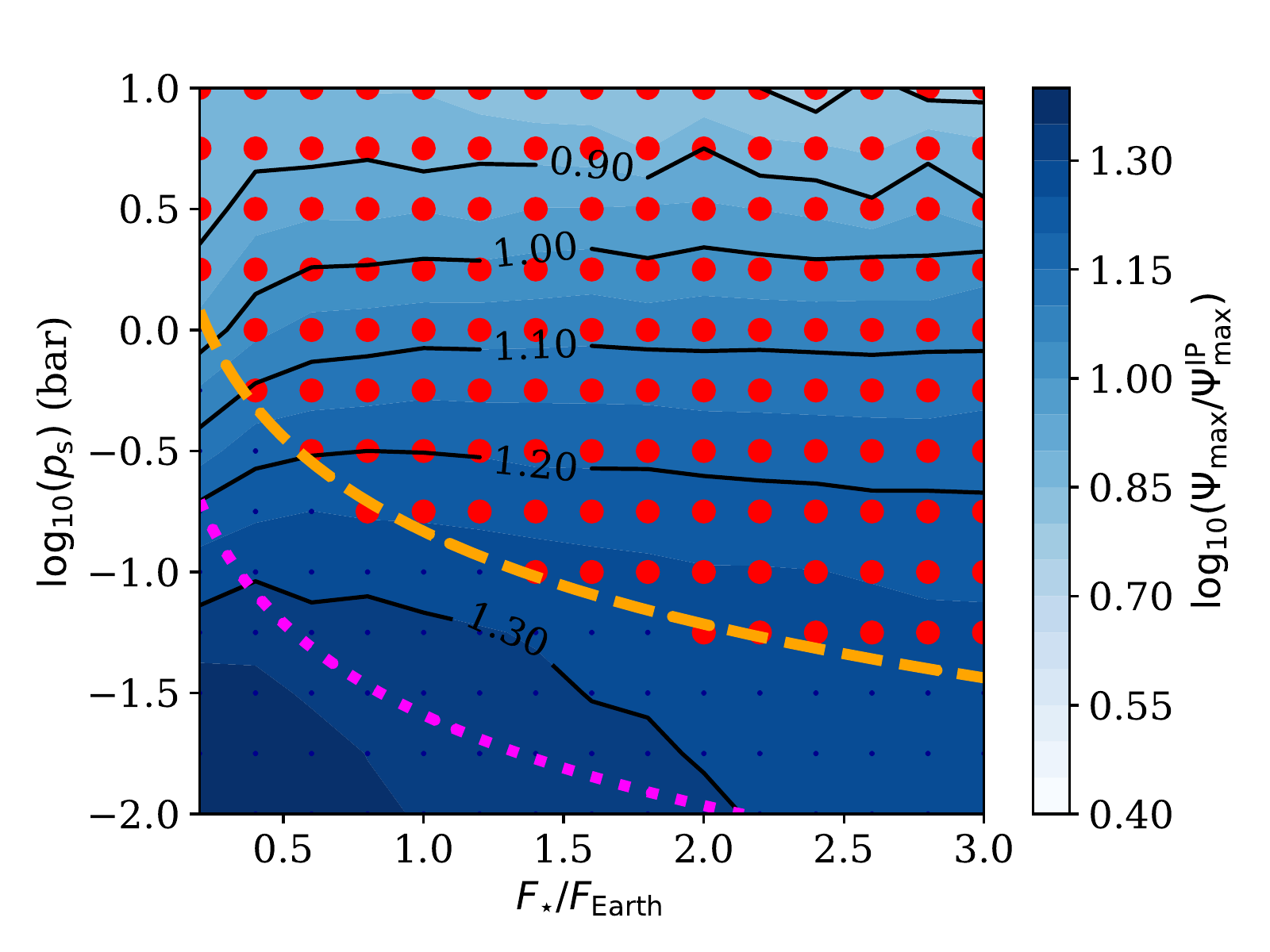} 
\hspace{\wepsi}    \includegraphics[height=\wpanelm,trim = 13.0cm 1.0cm 1.2cm 0.75cm,clip]{auclair-desrotour_fig5d.pdf}
   \raisebox{\hraiseboxps}[1cm][0pt]{\rotatebox[origin=c]{90}{\tiny{$\log_{10} \left( \sfmax / \sfmaxana \right)$}}} \\
   \hspace{\wbox} \tiny{$\Fstar / \Fearth $} \hspace{4cm}  \tiny{$\Fstar / \Fearth $} 
      \caption{Maximum of Eulerian mean streamfunction (top) and its ratio over the scaling law (bottom) for Earth-sized planets hosting Earth-like (left panels) and pure $\carbondiox$ (right panels) atmospheres. The two estimates of the Eulerian mean streamfunction are given by \eqs{maxsf}{maxsfana}, respectively. \RV{The curves and symbols are the same as in \fig{fig:stability_diagrams}.}}
       \label{fig:streamfunction}%
\end{figure*}

We now consider the evolution of the advection timescale itself, which is plotted in the case including turbulent diffusion in the PBL (\fig{fig:pbl_effect}, bottom panels). Similarly as the nightside surface temperature (\fig{fig:stability_diagrams}), $\tadv$ exhibits a relatively smooth dependence on the incident stellar flux and surface pressure, with values spanning over two orders of magnitude from ${\sim} 8$~days in the low pressure-high stellar flux regime to ${\sim} 500$~days in the high pressure-low stellar flux regime. These values are larger than those given by the analytic scaling law of \cite{KA2016} by one order of magnitude owing to the difference in the used definitions: since it is computed from the horizontal wind speed in the substellar region, the analytic advection timescale $\tadvKA$ is necessarily smaller than the advection timescale defined by \eq{tadv}, which is computed from mass flows going through the terminator annulus. Notwithstanding this scaling factor, $\tadvtd$ matches $\tadvKA$ relatively well for both Earth-like and pure $\carbondiox$ atmospheres in the optically thin regime, where the ratio varies by a factor of two. The two quantities diverge from each other as the stellar flux decreases and the surface pressure increases, the model of the present study predicting larger advection timescales in this regime. However, we remark that this discrepancy is less significant for pure $\carbondiox$ atmospheres than for Earth-like atmospheres, which suggests that the thermodynamic and absorption properties of the gas affect the large scale overturning circulation in a non-negligible way when the atmosphere is optically thick. 

\begin{figure*}[htb]
   \centering
   \includegraphics[width=0.98\textwidth,trim = 0cm 0cm 0cm 0cm,clip]{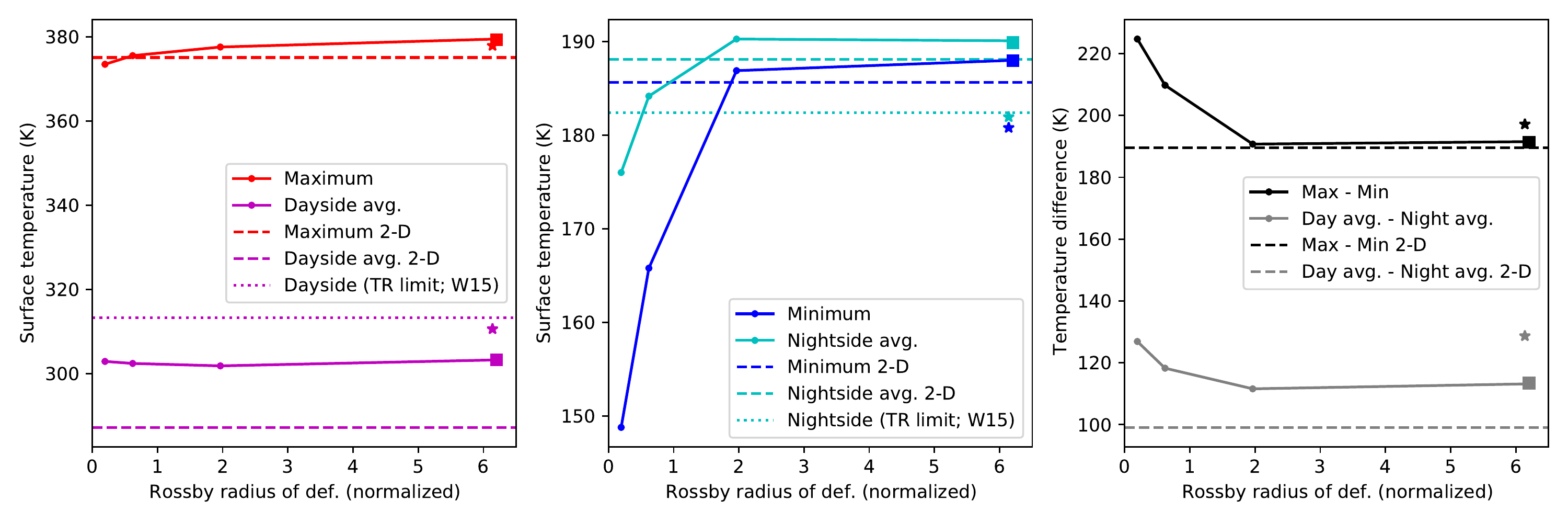} \\
   \raisebox{0.2cm}[0cm][0pt]{ \includegraphics[width=0.315\textwidth,trim = 0cm 0cm 3.2cm 0.3cm,clip]{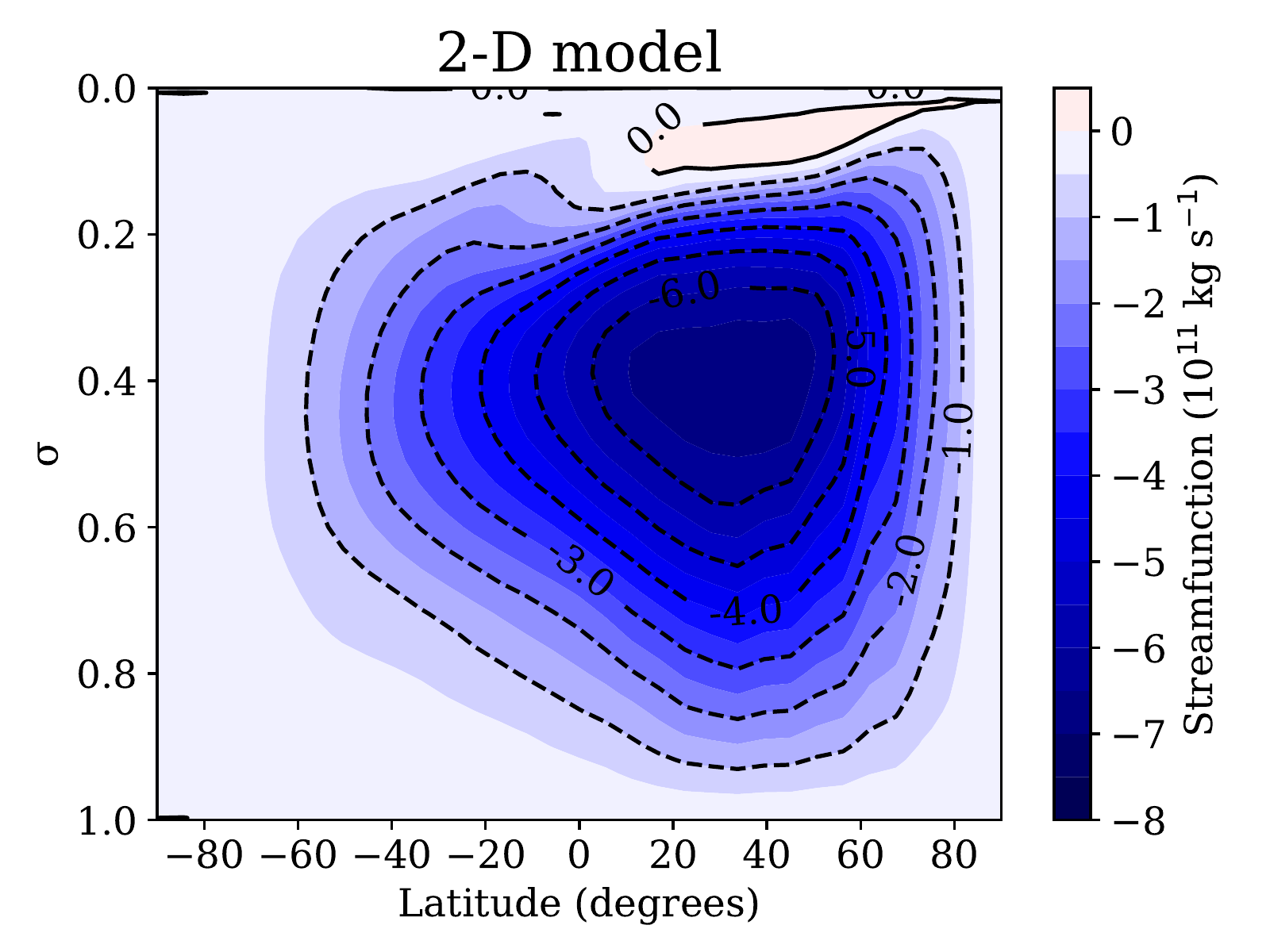}}
   \includegraphics[width=0.67\textwidth,trim = 0cm 0cm 0cm 0cm,clip]{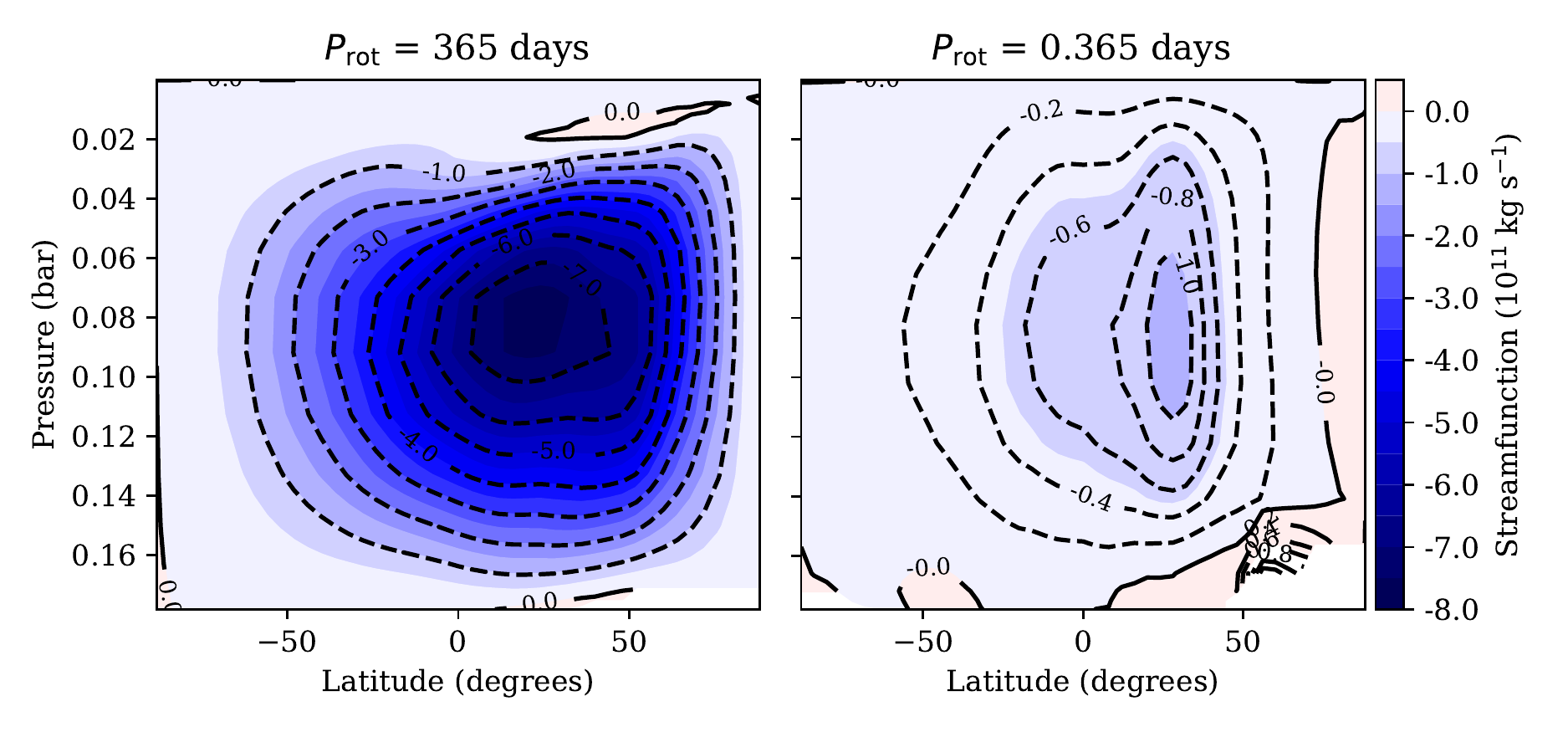}
      \caption{Dayside and nightside surface temperatures vs. normalised equatorial Rossby deformation radius ($\Rossbyradnorm$) in the gray 3-D GCM simulations performed with \texttt{THOR}. {\it Top left:} maximum (red solid line) and hemisphere-averaged (purple solid line) dayside surface temperatures. {\it Top middle:} minimum (blue solid line) and hemisphere-averaged (green solid line) nightside surface temperature. {\it Top right:} Day-night averaged (grey solid line) and extremal (black solid line) surface temperature differences. \RV{Dashed lines, dotted lines, star symbols, and square points indicate (i) the asymptotic surface temperatures computed using the 2-D model, (ii) the hemisphere-averaged surface temperatures predicted by Wordsworth's greenhouse model \citep[W15,][]{Wordsworth2015}, (iii) the surface temperatures obtained in the 3-D simulation without turbulent diffusion in the slow rotator case, and (iv) those obtained in the simulation performed with the same asymptotic scale lengths in the PBL as in the GCMM, respectively. }{\it Bottom, from left to right:} Averaged snapshots of the Eulerian mean streamfunction \citep[e.g.][]{Pauluis2008} obtained from the 2-D simulation (zero-spin rate limit), the 3-D simulation for $\period = 365$~days (slow rotator), and the 3-D simulation for $\period = 0.365$~days (fast rotator).}  
       \label{fig:Ro_temp}%
\end{figure*}

The strength of the overturning circulation can be characterised by examining the behaviour of the Eulerian mean streamfunction, which is defined in tidally locked coordinates as \citep[e.g.][]{Pauluis2008}
\begin{equation}
\euleriansf \define \frac{2 \pi \Rpla}{\ggravi} \integ{\Vtheta \sin \col }{\press}{\press}{\psurf}.
\label{euleriansf}
\end{equation}
The Eulerian mean streamfunction measures here the strength of longitude-averaged cells in the TLC. It accounts for the vorticity of the flow in a plane containing the planet-star axis. In the slow rotation regime, the large-scale cell of the predominant overturning circulation corresponds to a large region centred on a maximum of $\euleriansf$ in absolute value, the latter taking negative values owing to the flow direction (see e.g. \fig{fig:Ro_temp} in next section). The maximum value of $\euleriansf$ over the atmospheric domain, defined as 
\begin{equation}
\sfmax \define \max \left\{ -  \euleriansf \right\} ,
\label{maxsf}
\end{equation}
can be scaled analytically, as shown by \cite{IP2021} who established for sub-Neptunes the scaling law $\sfmax \scale \Fstar^{3/4}$. In the case of rocky planets, we need to account for the presence of the planet's surface, which sizes the thickness of the atmosphere. The mass flow thus depends on the atmospheric mass in addition with the stellar flux. As the atmospheric mass is directly proportional to the surface pressure, we introduce for rocky planets a scaling law of the form
\begin{equation}
\sfmaxana = \frac{\Rpla^2 \Fearth}{\Cp \Tearth} \left( \frac{\Fstar}{\Fearth} \right)^\alpha \left( \frac{\psurf}{\prefteta} \right)^\beta,
\label{maxsfana}
\end{equation}
where $\Tearth$ is the equilibrium temperature of Earth given by \eq{Teq}, $\prefteta = 1$~bar a reference pressure, and $\alpha$ and $\beta$ two real exponents to be defined. We remark that $\left( \alpha , \beta \right) = \left( 3/4,0 \right) $ corresponds to the scaling law obtained by \cite{IP2021} for sub-Neptunes. However, for the Earth-sized rocky planets of the present study, we adopt the empirical values $\left( \alpha , \beta \right) = \left( 1/2, 1 \right)$, which seem to be more appropriate as discussed further.

Figure~\ref{fig:streamfunction} shows the evolution of $\sfmax$ and of the ratio $\sfmax / \sfmaxana$ as functions of instellation and surface pressure for the Earth-like and pure $\carbondiox$ atmospheres defined in Table~\ref{tab:param_reference_case}. In both cases the maximum of the Eulerian mean streamfunction varies over more than three orders of magnitude ($\sfmax \sim 10^{10} {-} 10^{13} \ {\rm kg \ s^{-1}}$), by increasing with $\Fstar$ and $\psurf$ monotonically. The scaling law given by \eq{maxsfana} (with $\alpha = 1/2$ and $\beta = 1$) approximately captures this behaviour although it can only be considered as a rough estimate of $\sfmax$. We observe that the dependence of $\sfmax$ upon the instellation and surface pressure evolves between the optically thin and optically thick regimes with transition zones located around $\psurf \sim 0.3$~bar for Earth-like atmospheres and $0.1$~bar for pure $\carbondiox$ atmospheres. The scaling law $\sfmax \scale \Fstar^{1/2}$ is appropriate to describe the dependence of $\sfmax$ on $\Fstar$ in the regime of optically thick atmospheres in the infrared, but not in the regime of optically thin atmospheres where $\sfmax$ increases faster with $\Fstar$. Conversely, the slow evolution of $\sfmax /\sfmaxana$ with $\psurf$ at low surface pressures suggests that the scaling law $\sfmax \scale \psurf$ captures well the dependence of $\sfmax$ on the surface pressure in the regime of optically thin atmospheres, while this dependence is weaker in the regime of thick atmospheres.

\section{From slow to fast rotation}
\label{sec:slow_fast_rotation}
 Since the GCMM is designed to study the asymptotic regime of slowly rotating planets, where mean flows are predominantly driven by the day-night temperature gradient, we have ignored the effects of rotation on the general circulation until now. As discussed in \sect{sec:preliminary_scalings}, the impact of these effects is mainly controlled by the nondimensional equatorial Rossby deformation radius $\Rossbyradnorm$ introduced in \eq{Ronorm}. The zero-spin rate limit treated by the model corresponds to $\Rossbyradnorm = + \infty$. In reality Coriolis acceleration tends to deviate the divergent winds blowing between the substellar and the anti-stellar points. This alters the large-scale circulation regime, which switches from the bi-dimensional Hadley cell described by the 2-D model to the three-dimensional structure characterising fast rotators, where super-rotation develops, as $\Rossbyradnorm$ decays. In this section, we investigate the limitations of the bi-dimensional approach by benchmarking the results obtained with the 2-D instance of the GCMM against the simulations performed with \texttt{THOR} GCM \citep[][]{Mendonca2016,Deitrick2020}, which solves the three-dimensional nonhydrostatic Euler equations on an icosahedral grid. Particularly, \texttt{THOR} fully accounts for the horizontal vortical component of mean flows that is ignored in the 2-D model. 

Four simulations were performed with \texttt{THOR} for the pure $\carbondiox$ atmospheres defined by Table~\ref{tab:param_reference_case}. They each correspond to a given spin period $\period$ taken in a range spanning from fast to slow rotators all things being equal ($\period = 10^{-3}, 10^{-2}, 10^{-1}, 10^{0} \ \period_0$ with $\period_0 = 365$~days). For comparison, one additional simulation was run with the 3-D GCM in the slow rotator case ($\period = \period_0$) by assuming no turbulent diffusion or sensible surface-atmosphere heat exchanges. In these simulations, the values of the surface pressure ($\psurf = 0.18$~bar) and stellar flux ($\Fstar = \Fearth$) were chosen so that the studied cases are in the vicinity of the threshold between the stability and collapse regions predicted by the 2-D instance of the meta-model (see \fig{fig:stability_diagrams}, top right panel). The horizontal resolution of the icosahedral grid used in the model was set to $\sim 4$ degrees, and the atmosphere was divided into 40 vertical intervals logarithmically refined in the vicinity of the surface with a top altitude of 37000~m and a 2~m-thick lowest layer. For the sake of consistency, the simulations were run using the double-gray approximation for radiative transfer and a physical setup as close as possible as that implemented in the GCMM. After the circulation and radiative transfer have reached a state of equilibrium, the physical quantities were averaged over the longitude of the tidally locked coordinates to transform the three-dimensional fields into two-dimensional fields similar to those displayed in \fig{fig:snapshots_HB1993_diff}. The bulk dimensionless equatorial Rossby deformation length ($\Rossbyradnorm$) was computed from the resulting mass-averaged temperature using the expression given by \eq{Roiso}. In parallel of the simulations performed with \texttt{THOR}, a simulation corresponding to the zero-spin rate limit was run with the 2-D instance of the meta-model. 

Notwithstanding the vertical coordinates used in dynamical cores -- altitude-based in \texttt{THOR}, and mass-based in the GCMM -- the main differences between the two models essentially lie in the description of the surface thermal response, planetary boundary layer, and numerical energy dissipation. In \texttt{THOR}, the surface temperature evolution is integrated using a 0-D thermodynamic equation parametrised by an effective surface heat capacity ($\Csgr = 10^7 \ {\rm J \ K^{-1} \ m^2}$), while a 1-D soil heat transfer scheme is used in the GCMM (see \append{app:soil_heat_transfer}). In the treatment of the planetary boundary layer, the two models closely follow the method proposed by \cite{HB1993}, except for some parameters, which are set to different values. For instance, the asymptotic length scale characterising the evolution of the mixing length with altitude (\eq{mixing_length}) for the heat equation is set to three times the function used for the momentum equation (\eq{mixlength_function}) in \texttt{THOR}, while the same function is used in both cases in the GCMM. As regards numerical energy dissipation, all \texttt{THOR} simulations utilized 4th order horizontal hyperdiffusion and 3-D divergence damping with a nondimensional coefficient $D_{\rm hyp} = D_{\rm div} = 0.002$ \citep[see][for descriptions of the hyperdiffusion scheme]{Mendonca2016,Deitrick2020}. We additionally used 6th order vertical hyperdiffusion with a nondimensional coefficient of $D_{\rm ver} = 5\times10^{-4}$. Finally, a sponge layer was applied to the upper $25\%$ of the model domain to damp spuriously reflected waves off the top boundary, with a damping time scale of 5000 s \citep[see][for the sponge layer description]{Mendonca18,Deitrick2020}.

Figure~\ref{fig:Ro_temp} shows the simulation results. The dayside and nightside surface temperatures, as well as the day-night surface temperature difference, are plotted as a function of the bulk equatorial Rossby deformation radius (top panels). Solid lines designate the temperatures obtained in \texttt{THOR} 3-D simulations, and the horizontal dashed lines the temperatures obtained in the 2-D simulation performed with the meta-model. The dotted lines indicate the corresponding averaged dayside and nightside surface temperatures predicted by Wordsworth's purely radiative greenhouse model \citep[][]{Wordsworth2015}. The temperatures obtained from the 3-D simulation performed in the absence of turbulent diffusion are designated by the star symbol $\star$. In addition, the Eulerian mean streamfunction defined in \eq{euleriansf} is plotted for the 2-D model and the two extrema of the 3-D model ($\period = 365$~days and $\period = 0.365$~days) as functions of pressure (or sigma-coordinate) and the latitude of the TLC, where the North and South poles correspond to the substellar and anti-stellar points, respectively (bottom panels). 

We first consider the evolution of surface temperatures with $\Rossbyradnorm$ (\fig{fig:Ro_temp}, top panels). Whereas the dayside surface temperature predicted by 3-D GCM simulations is hardly affected by the planet's rotation, the nightside surface temperature increases monotonically with the equatorial Rossby deformation radius until it converges towards the asymptotic limit of slow rotators described by the 2-D model. The effect of rotation is particularly significant for the minimum surface temperature, which varies by ${\sim} 35$~K between the two extremal cases. This difference results from the decay of day-night advection provoked by the breaking of the day-night overturning circulation in the fast rotator regime ($\Rossbyradnorm \lesssim 1$). The super-rotating equatorial jets induced by Coriolis effects do not compensate the overturning circulation in terms of mass flux crossing the terminator annulus, which makes the nightside surface temperature decrease. However we remark that the nightside temperature stays close from the asymptotic limit of slow rotators from the moment that $\Rossbyradnorm >2$, the difference of the cases $\period = 36.5$~days and $\period = 365$~days to this limit being less than 3~K. This suggests that the 2-D model is relevant to describe the climate and large-scale circulation regime of the planet for equatorial Rossby deformation radii exceeding the critical value $\Rossbyradnorm \approx 2$. Below this value, the circulation and heat redistribution are strongly affected by the planet's rotation. 

Besides we remark that the 2-D model and \texttt{THOR} GCM are in agreement with each other for the nightside surface temperature, the latter being greater by $\sim 2$~K only in 3-D simulations. The discrepancy between the two models is more significant for the dayside average temperature, where the effects of the PBL in surface-atmosphere heat exchanges are predominant. As observed in the 2-D model, the frictional interaction of mean flows with the planet's surface taking place in the PBL tends to increase the atmospheric stability against collapse by warming up the nightside surface hemisphere. To understand whether the differences in PBL parameters between the 2-D and 3-D models could be the source of the discrepant dayside temperatures, one additional \texttt{THOR} simulation was run at the slowest rotation rate. In this simulation, the asymptotic scale lengths (for heat and momentum mixing in the boundary layer) were set equal to those used in the 2-D model. The resulting values are shown as the square points in top panels of Fig. \ref{fig:Ro_temp}. Interestingly, these results are nearly indistinguishable from the simulation with different scale lengths used for the heat and momentum equations, indicating that the scenario is insensitive to the asymptotic scale lengths used in Equation \ref{mixing_length}. 

We now consider the snapshots of the Eulerian mean streamfunction (\fig{fig:Ro_temp}, bottom panels). We remark that the vertical coordinates used for the plots in the 2-D and 3-D cases differ from each other, which leads to slight distortions of the distributions. The sigma-coordinate is such that $\sig = 1$ everywhere at planet's surface, while the surface pressure  is less than the globally averaged surface pressure on the dayside, and greater on the nightside due to the day-night thermal forcing gradient. As a consequence, isobars in sigma-coordinate are slightly shifted downwards on dayside and upwards on nightside. The 3-D simulation performed with \texttt{THOR} GCM in the slow rotation regime is in good agreement with the 2-D simulation performed with the GCMM in the zero-spin rate limit (\fig{fig:Ro_temp}, bottom left and middle panels). The two snapshots both exhibit large day-night cells of comparable strengths albeit slightly weaker in the 2-D model. The snapshot corresponding to the fast rotation regime (bottom right panel) exhibits more complex features due to Coriolis effects although a weak day-night cell still remains visible.

\section{Concluding remarks}
\label{sec:conclusions}
In this work we have developed a General Circulation Meta-Model (GCMM) to bridge the gap between the analytic solutions provided by simplified greenhouse models for synchronous planets and the numerical simulations obtained from 3-D GCMs in the asymptotic regime of slow rotators.\RV{ This model hierarchy is based on a systematic bottom-up approach in the spirit of \cite{Held2005}, wherein the number of degrees of freedom determines the key sources of complexity that are added or subtracted.} The solver of the GCMM integrates the hydrostatical primitive equations using the finite-volume method for arbitrary numbers of horizontal and vertical intervals, each configuration being an instance of the meta-model. Consistently with a previous analytical study \citep[][]{ADH2020}, the physics implemented in the meta-model includes double-gray radiative transfer, turbulent diffusion in the planetary boundary layer, and soil heat conduction. Particularly, the meta-model was designed so that the solutions obtained with the 0-D instance exactly correspond to the analytic solutions of the purely radiative box model detailed in Sect.~3.3 of \cite{ADH2020}. 

As a first step, we proceeded to a vertical model inter-comparison by running grid simulations for four instances of the meta-model (0-D, 1-D, 1.5-D, and 2-D) in the cases of dry Earth-like and pure $\carbondiox$ atmospheres. In each case, we computed from simulations the nightside surface temperature of the planet as a function of the stellar flux and surface pressure, and the resulting stability diagrams of the atmosphere against collapse. These diagrams were compared to the scaling laws predicted by the analytic theory. With the 0-D and 1-D instances of the meta-model, we recovered the stability diagrams predicted by simplified radiative models in the optically thin regime, which shows that the globally isothermal approximation used in these models is relevant is this regime. The 1.5-D instance tends to underestimate the atmospheric stability, by predicting a collapse pressure $25 \% $ to $ 80\%$ larger than that given by radiative box models. Conversely, the collapse pressure computed using the 2-D instance of the meta-model is $10\%$ to $40\%$ smaller than the analytic estimate owing to the warming effect of the PBL, which is less than the theoretical $75\%$ maximum decrease predicted by radiative models albeit still significant. 

As a second step, we investigated the role played by the planetary boundary layer in the thermal state of equilibrium and atmospheric circulation of the planet by examining with the 2-D instance of the GCMM how the turbulent diffusion taking place in the PBL alters the nightside temperature, the day-night advection timescale, and the collapse pressure. We compared the advection timescale obtained from simulations with the analytic scaling law proposed by \cite{KA2016}. We observed that the turbulent diffusion taking place in the PBL increases the nightside surface temperature by $4  {-} 14$~K around the threshold of the stability region. However, we found that the PBL can also contribute to cool down the nightside surface of the planet by acting on the day-night advection timescale in the transition zone between optically thin and optically thick atmospheres. This result was corroborated a posteriori by 3-D GCM simulations. The effect of the PBL on the large-scale circulation is complex and depends upon the interplay between the advection and radiative timescales. The day-night advection timescale estimated with the 2-D model varies over two orders of magnitude in the studied domain of stellar fluxes and surface pressures, with values ranging between 8 and 500 days. In the optically thin regime, its evolution matches relatively well the scaling law derived by \cite{KA2016} but it diverges from it at high pressures and low stellar fluxes. We noticed that this behaviour also depends on the thermodynamic and absorption properties of the atmosphere in the optically thick regime. In addition we empirically obtained, for the circulation of slowly rotating rocky planets, a scaling law analogous to that established analytically by \cite{IP2021} for sub-Neptunes.
 
 As a third and final step, we characterised the limitations of the slow rotator approximation that forms the foundations of the meta-model by performing simulations with \texttt{THOR} 3-D GCM, the latter fully accounting for the effects of rotation on mean flows. We computed from these simulations the evolution of the planet's dayside and nightside surface temperatures as functions of the dimensionless equatorial Rossby deformation radius, which controls the large-scale circulation regime of the planet. The results obtained with the 3-D GCM were compared with the outcomes of the 2-D instance of the meta-model. The 3-D GCM simulations highlight the transition between the slow and fast rotation regimes. We found that the 2-D model properly accounts for the climate and large-scale atmospheric circulation from the moment that the normalised equatorial Rossby deformation radius is greater than the critical value $\Rossbyradnorm \approx 2$, which corresponds to a broad region of the parameter space. In the slow rotation regime, the circulation and surface temperature predicted by \texttt{THOR} 3-D GCM and the 2-D instance of the meta-model are similar.

This study is a first attempt to fill the continuum between analytic greenhouse models and 3-D GCMs in a self-consistent way. The obtained results show that the meta-modelling approach is efficient to disentangle the mechanisms determining the climatic state of the planet, which are narrowly coupled together in GCMs. This approach allows for running in parallel series of models of various complexities albeit sharing the very same physical setup, so that each simulation can be interpreted using the others. As a consequence, the final outcome of the meta-model conveys information not only on the climate itself but also on the separated contributions of key physical ingredients such as radiative transfer, atmospheric structure, dynamics, and turbulent friction in the PBL. In addition with these diagnostic aspects, the meta-modelling approach appears as a robust method to refine the analytic theory of planetary climates given that it allows for assessing the relevance of solutions obtained at low spatial dimensionality with consistently generated GCM numerical solutions. Finally, we note that the meta-modelling approach can be extended to moist atmospheres and rapidly rotating planets, but we leave that to the content of a future study.

\begin{acknowledgements}
\RV{The authors thank the anonymous referee for helpful comments that improved the manuscript, and} Vera Matarese for inspiring discussions about the representational capacity of numerical simulations. They acknowledge financial support from the European Research Council via the Consolidator grant \texttt{EXOKLEIN} (grant number 771620). Kevin Heng also acknowledges partial financial support from the National Swiss Foundation, the PlanetS National Center of Competence in Research, the Center for Space \& Habitability and the MERAC Foundation. This research has made use of NASA's Astrophysics Data System. The authors gratefully acknowledge the open source softwares that made this work possible: \texttt{numpy} \citep[][]{Harris2020}, \texttt{matplotlib} \citep[][]{Hunter2007}, \texttt{scipy} \citep[][]{Virtanen2020}. \texttt{THOR} calculations were performed on UBELIX (\texttt{\url{http://www.id.unibe.ch/hpc}}), the HPC cluster at the University of Bern. 
\end{acknowledgements}


\bibliographystyle{aa}  
\bibliography{references} 

\appendix

\section{Nondimensional primitive equations}
\label{app:nondimensional_hpes}
The model solves the HPEs given by \eqsto{hpe1}{hpe4} in their nondimensional form. Although it is not used in the present study, the moisture conservation equation is included in the solver and we shall give it here for the sake of generality. In its conservative form, this equation reads \citep[e.g.][]{YS1987}
\begin{equation}
\dd{}{\time}\left(\pisurf  \qmoist  \right)+ \divx{\sig} \left( \pisurf \qmoist \Vvecth  \right) + \dd{}{\sig} \left( \pisurf \qmoist \Vsig \right) = \pisurf \dtqmoist,
\label{hpe_moisture}
\end{equation}
where $\qmoist$ designates the specific humidity of any tracer, and $\dtqmoist$ the corresponding net evaporation or condensation rate per unit mass. Since the elemental unit of time of the discretised equations is the dynamical time step $\dyndt$ used in the time-differencing scheme, it is convenient to normalise the time by $\dyndt$ so that the time lapse between two dynamical time steps is always unity.

We then introduce the reference pressure $\pressr$, temperature $\tempr$, and specific humidity $\qmoistr$, from which we can define the reference velocity $\velr$, energy per unit mass $\energyr$, density $\rhor$, force per unit mass $\forcer$, heat power per unit mass $\Qheatr$, net evaporation per unit mass $\dtqmoistr$, Exner function $\Exnerr$, and potential temperature $\tetar$,
\begin{equation}
\begin{array}{lll}
\velr \define \sqrt{\Cp \tempr}, & \energyr \define \Cp \tempr, & \rhor \define \dfrac{\pressr}{\Cp \tempr}, \\ 
\forcer \define \dfrac{\sqrt{\Cp \tempr}}{\dyndt},    & \Qheatr \define \dfrac{\Cp \tempr}{\dyndt}, & \dtqmoistr \define \dfrac{\qmoistr}{\dyndt}, \\[0.3cm]
\Exnerr \define \Cp \left( \dfrac{\pressr}{\prefteta} \right)^{\rcp}, & \tetar \define \dfrac{\Cp \tempr}{\Exnerr}.
\end{array}
\end{equation}
Besides, we make the horizontal coordinate vary within the range $\left[0, 1 \right]$ like the vertical coordinate by renormalising the colatitude ($\col$). The normalised variables and colatitude are therefore defined as  
\begin{equation}
\begin{array}{llll}
\psurfn \define \dfrac{\psurf}{\pressr}, & \pressn \define \dfrac{\press}{\pressr},& \pisurfn \define \dfrac{\pisurf}{\pressr},  & \Vthetan \define \dfrac{\Vtheta}{\velr} , \\[0.3cm]
 \Vsign \define \dyndt \Vsig, & \tempn \define \dfrac{\temp}{\tempr}, & \tetan \define \dfrac{\teta}{\tetar}, & \rhon \define \dfrac{\density}{\rhor}, \\[0.3cm]
 \gpotn \define \dfrac{\geopot}{\energyr}, & \forcehn \define \dfrac{\forceh}{\forcer}, &  \Qheatn \define \dfrac{\Qheat}{\Qheatr}, & \qmoistn \define \dfrac{\qmoist}{\qmoistr}, \\[0.3cm]
 \dtqmoistn \define \dfrac{\dtqmoist}{\dtqmoistr} , & \coln \define \dfrac{\col}{\pi}. &  &  
\end{array}
\label{normalisation}
\end{equation}

\begin{figure}[t]
   \centering
   \includegraphics[width=0.48\textwidth,trim = 0cm 0cm 0cm 0cm,clip]{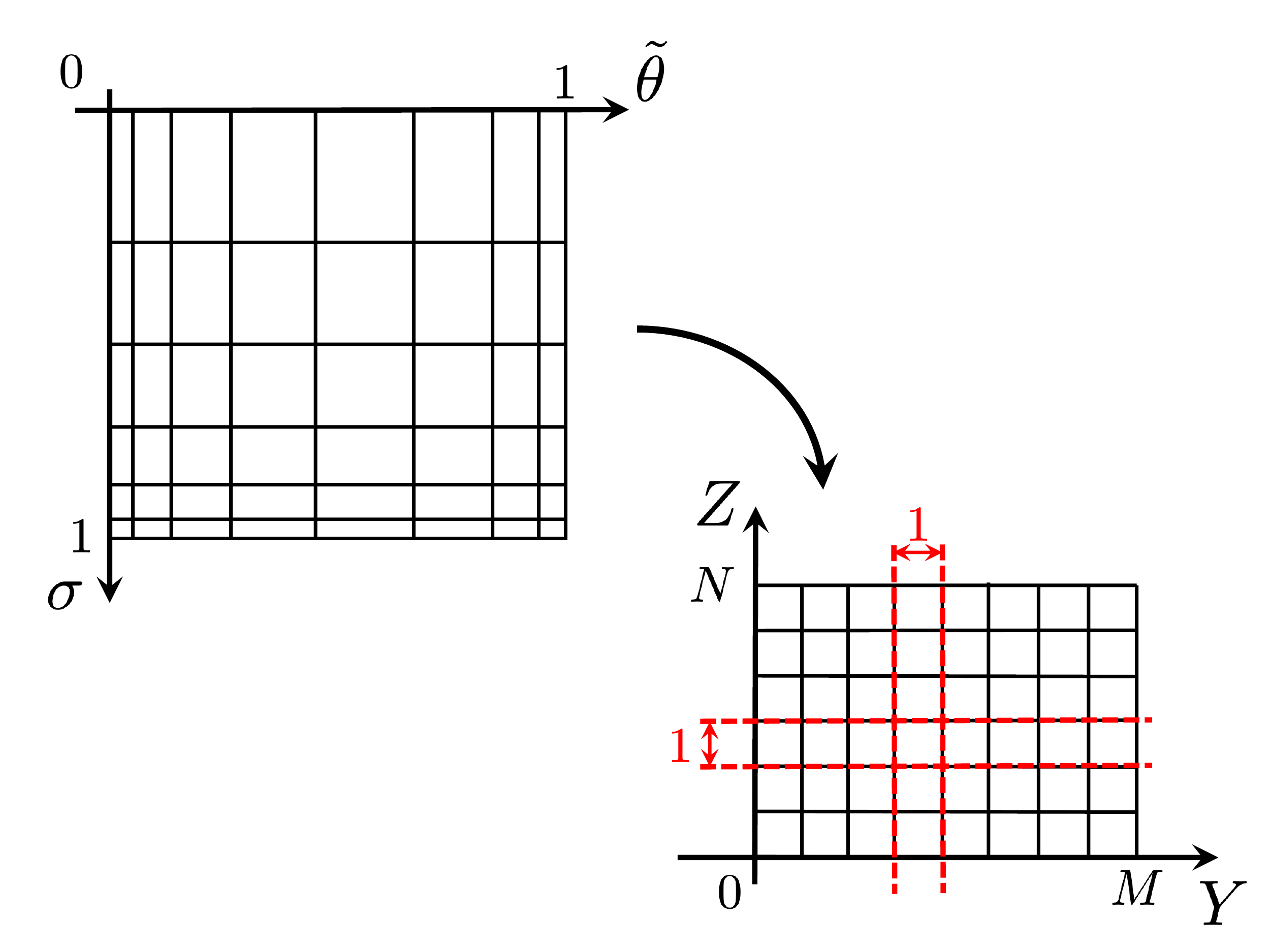}
      \caption{Change of coordinate $\left( \coln, \sig \right) \rightarrow \left( \ygrid, \zgrid \right)$, from the physical spatial coordinates to the grid spatial coordinates. The grid coordinates vary in the range $0 \leq \ygrid \leq M$ and $0 \leq \zgrid \leq N$, where $M$ and $N$ are two integers that do not need to be defined at this stage but will correspond to the numbers of horizontal and vertical grid levels in the finite-volume discretisation of the primitive equations. The metric is illustrated by the spacing variation between two isolines. The interval between two isolines in grid coordinates is of size $1$ to mark the cells of the finite-volume method employed in the following.}
       \label{fig:grid_coord}%
\end{figure}

As a second step, the spatial coordinates $\left( \coln , \sig \right)$ are converted to grid coordinates\footnote{The notation $\ygrid$ is chosen in place of $X$ for the horizontal coordinate in order to be consistent with notations conventionally used in GCMs ($X$ and $Y$ for the longitudinal and latitudinal directions, and $Z$ for the vertical direction).}, $\left( \ygrid, \zgrid \right)$. In this transformation, the normalised colatitude $\coln = \coln \left( \ygrid \right) $ and vertical coordinate $\sig = \sig \left( \zgrid \right) $ are assumed to be monotonic functions of $\ygrid$ and $\zgrid$, respectively. These functions are defined further so that the interval between the two boundaries of a finite volume is equal to 1 both in the horizontal and in the vertical directions, as showed by \fig{fig:grid_coord}. The transformation $\left( \coln, \sig \right) \rightarrow \left( \ygrid, \zgrid \right)$ is defined at any point by the horizontal and vertical metric coefficients, $\cy$ and $\cz$, defined as
\begin{align}
&  \cy \define \dd{\coln}{\ygrid} , & \cz \define \dd{\sig}{\zgrid}.
\end{align}
which yields the change relations of partial derivatives 
\begin{align}
& \dd{}{\coln} = \frac{1}{\cy} \dd{}{\ygrid}, & \dd{}{\sig} = \frac{1}{\cz} \dd{}{\zgrid}, \\
& \dd{}{\ygrid} = \cy \dd{}{\coln}, &  \dd{}{\zgrid} = \cz \dd{}{\sig}.
\end{align}
We remark that the change of coordinate is just a dilation if the adopted horizontal and vertical spacings are uniform since $\cy$ and $\cz$ are constants in this case. We also notice that $\cz < 0 $ with the chosen sigma coordinate given that $\sigma$ decreases monotonically as the altitude increases.

To include the metric in the primitive equations, we introduce the respective covariant and contravariant horizontal velocities,
\begin{align}
&& \vcov \define \cy \Vthetan, && \vcon \define \frac{\Vthetan}{\cy},
\end{align}
the normalised area density (area per unit length),
\begin{equation}
\arean \define \cy \sin \left( \col \right)
\end{equation}
mass density\footnote{The minus sign in the expression of $\massn$ is here to ensure $\massn>0$, which is the convention used in the model.},
\begin{equation}
\massn \define - \pisurfn \arean \cz
\end{equation}
and horizontal and vertical mass flux densities,
\begin{align}
&& \Vflux \define \massn \vcon , && \Wflux \define \massn \frac{\Vsign}{\cz}. 
\end{align}

After the above manipulations, the system of HPEs given by \eqsto{hpe1}{hpe4} and \eq{hpe_moisture} become the normalised primitive equations 
\begin{align}
\label{prim1}
\dd{\massn}{\timen} + \bcourant \dd{\Vflux}{\ygrid} + \dd{\Wflux}{\zgrid} & =  0 ,  \\
\dd{}{\timen} \left(  \massn \Vthetan \right) + \bcourant \dd{}{\ygrid} \left( \Vflux \Vthetan \right) + \dd{}{\zgrid} \left( \Wflux \Vthetan \right) & \\
+ \frac{\bcourant}{\cy} \massn \left( \dd{\gpotn}{\ygrid} + \tetan \dd{\Exnern}{\ygrid}  \right)  & =  \massn \forcehn , \nonumber \\
\dd{}{\timen} \left( \massn \tetan \right) + \bcourant \dd{}{\ygrid} \left( \Vflux \tetan \right) + \dd{}{\zgrid} \left( \Wflux \tetan \right) & = \frac{ \massn \Qheatn}{\Exnern}, \\
\dd{\gpotn}{\zgrid} + \rcp \tetan \dd{\Exnern}{\zgrid}  & = 0, \\
\label{prim5}
\dd{}{\timen} \left( \massn \qmoistn \right) + \bcourant \dd{}{\ygrid} \left( \Vflux \qmoistn \right) +  \dd{}{\zgrid} \left( \Wflux \qmoistn \right)  & = \massn  \dtqmoistn,
\end{align}
where the normalised Exner function, derived from \eq{teta_exner}, is expressed as a function of the normalised pressure $\pressn$,
\begin{equation}
\Exnern \define \frac{\Exner}{\Exnerr} = \pressn^{\rcp}.
\end{equation}

As may be noticed, with the normalisation adopted in \eq{normalisation}, the dynamics of the nondimensional HPEs is controlled by one unique dimensionless parameter,
\begin{equation}
\bcourant \define \dfrac{\velr \dyndt}{\pi \Rpla} = \dfrac{\sqrt{\Cp \tempr} \dyndt}{\pi \Rpla},
\label{bcourant}
\end{equation}
\noindent which is a global Courant number weighting the contribution of horizontal advection\footnote{The global Courant number given by \eq{bcourant} should be multiplied by the number of horizontal cells to obtain the Courant number used in the CFL (Courant, Friedrichs, and Lewy) numerical stability condition \citep[][]{CFL1928}.}. The density is not a variable of the primitive equations but can be evaluated using the perfect gas equation,
\begin{equation}
\pressn = \rcp \rhon \tempn,
\end{equation}
while the vertical velocity can be calculated from the conversion formula
\begin{equation}
 \Vz =  \frac{\Vzr}{\massn} \left[ \massn \dd{\gpotn}{\timen} + \bcourant \Vflux \dd{\gpotn}{\ygrid} + \Wflux \dd{\gpotn}{\zgrid} \right] ,
\end{equation}
where we have introduced the characteristic vertical velocity 
\begin{equation}
\Vzr \define \frac{\energyr}{\ggravi \dyndt}.
\end{equation}

\section{Dynamical core}
\label{app:dynamical_core}
The system of prognostic equations given by \eqsto{prim1}{prim5} is integrated numerically using a standard finite-volume method. This appendix details the main features of the dynamical core of our model. 
\subsection{Grid}
\begin{figure*}[t]
   \centering
   \includegraphics[width=0.6\textwidth,trim = 0cm 0cm 0cm 0cm,clip]{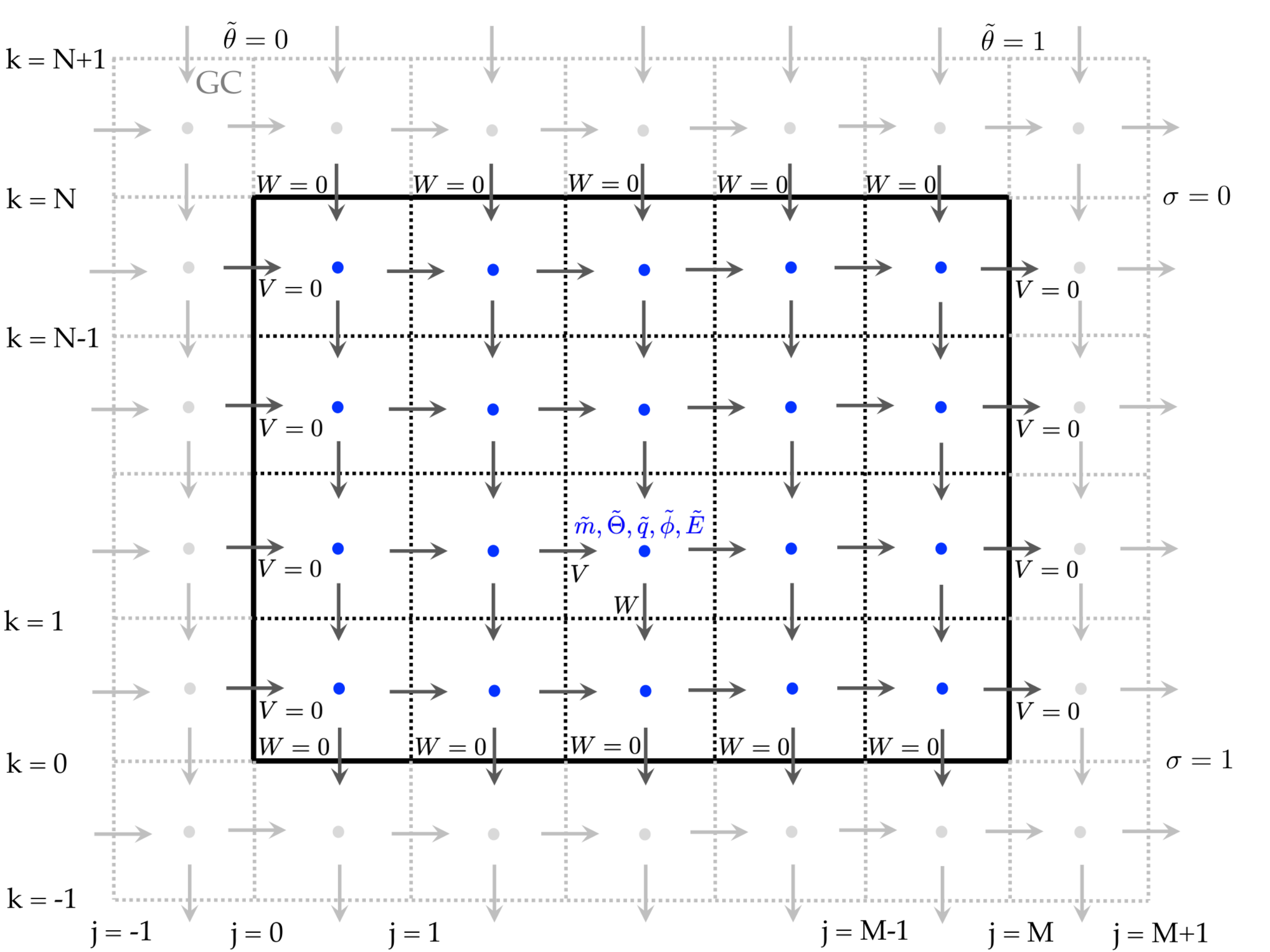}
      \caption{Staggered grid. In the adopted finite-volume method, the horizontal axis is divided into $\hlev$ intervals and the vertical axis into $\vlev$ intervals, which represents $\hlev \times \vlev$ cells (here, $\hlev = 5$, and $\vlev = 4$). Horizontal levels are indexed by $\jhor$ and vertical levels by $\kver$. The left and right boundaries of the physical domain correspond to $\coln =  0$ (sub-stellar point) and $\coln = 1$ (anti-stellar point), respectively. The top and bottom boundaries correspond to $\sig = 0$ (space) and $\sig = 1$ (ground), respectively. The horizontal and vertical mass flows, $\Vflux$ and $\Wflux$, are evaluated at vertical and horizontal cell interfaces, respectively, while the mass, temperature, geopotential, specific humidity, and Exner function (blue) are evaluated at cell centers. The physical domain is surrounded by ghost cells (grey), which are employed to facilitate vectorisation. }
       \label{fig:grid}%
\end{figure*}

The primitive equations are discretised and solved on a staggered Arakawa~C grid \citep[][]{AL1977}. The atmosphere is divided into elemental cells. Mass fluxes and velocities are evaluated at cell interfaces, and volume quantities ($\massn$, $\tetan$, $\qmoistn$, $\gpotn$, $\Exnern$) at cells centers, except the pressure, which is evaluated at horizontal cells interfaces. A diagram representing the grid is shown by \fig{fig:grid}.

As a first step, the spatial domain defined by $\left(\coln , \sig \right) \in \left[ 0 , 1 \right]^2 $ is divided into non-uniform intervals both along the horizontal and vertical directions. There are $\hlev$ intervals for the colatitude, and $\vlev$ intervals for the vertical coordinate, which represents $\hlev \times \vlev$ cells. The vertical spacing can be specified arbitrarily. It determines the coordinates of cell boundaries. We note that the $\sig$ levels of cell centers can be explicitly computed from the $\sig$ levels of cell boundaries if $\ptop = 0$ (standard sigma-coordinate) as discussed in \sect{ssec:exner}. 

\begin{figure}[htb]
   \centering
   \includegraphics[width=0.48\textwidth,trim = 0cm 9.3cm 4.8cm 0cm,clip]{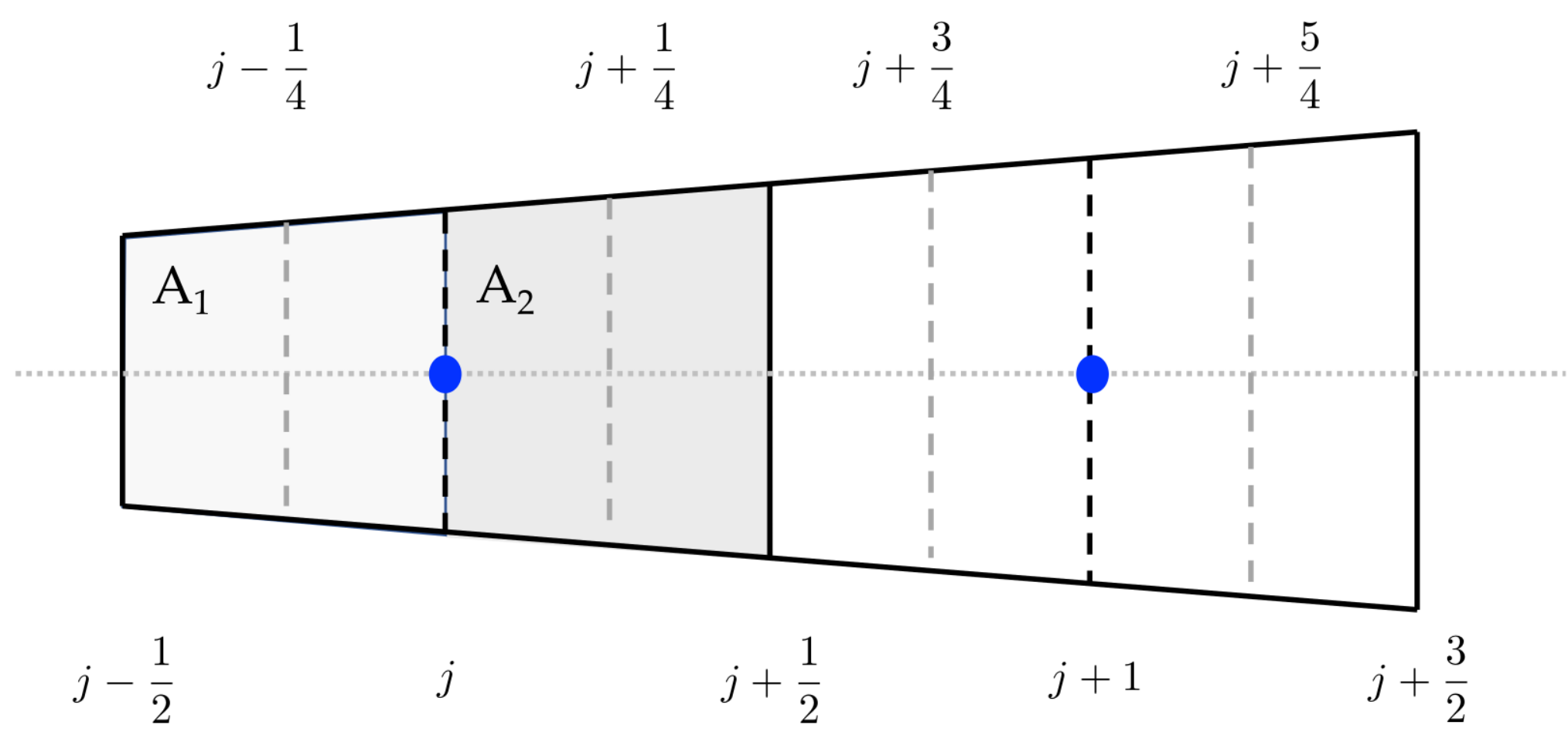}
      \caption{Subdivision of a surface area into two subsurface areas, $A_1$ and $A_2$ (light grey and grey regions). The big blue dots indicate the centers of horizontal intervals. The areas are calculated analytically by taking into account their dependence on colatitude. The configuration of the diagram corresponds to small horizontal intervals, where this dependence is approximately linear.}
       \label{fig:surface_area}%
\end{figure}

As as second step, the change of coordinates $\left( \coln , \sig \right) \rightarrow \left( \ygrid, \zgrid \right) \in \left[ 0 , \hlev \right] \times \left[ 0 , \vlev \right] $ is applied, and the corresponding metric coefficients $\cy$ and $\cz$ are calculated. Typically, for  a uniform grid, $\coln = \ygrid / \hlev$, and thus $\cy = 1/\hlev$ everywhere. The case of non-uniform horizontal intervals is more complicated, since this implies $\coln$-dependent coefficients. To treat the general case, we adapt the method employed for the horizontal grid in the \texttt{LMDZ} GCM to the one-dimensional case: each surface area is divided into two subsurface areas, as shown by \fig{fig:surface_area}, and the $\cy$ are evaluated at subareas centers and boundaries \citep[][]{Sadourny1975a,Sadourny1975b}. The $\cz$ are just the difference between two vertical levels. We introduce the difference and average operators, 
\begin{align}
\label{diff_operator}
& \diff{\XX}{\psi} \left( \XX \right) = \psi \left( \XX+\frac{1}{2} \right) - \psi \left( \XX - \frac{1}{2} \right), \\
\label{average_operator}
& \moy{\XX}{\psi} \left( \XX \right) = \frac{1}{2} \left[ \psi \left( \XX + \frac{1}{2} \right) + \psi \left( \XX - \frac{1}{2} \right)  \right],
\end{align}
where $\XX$ is a placeholder for $\ygrid$ or $\zgrid$ and $\psi$ any variable. In this formalism, the $\cz$ are simply expressed as $\cz = \diffz{\sig}$.

In the present work, we use a vertical grid refined near the surface and the upper boundary. The sigma-coordinates of vertical levels are given by the function
\begin{equation}
\sig \left( \xx \right) = \frac{1}{2} \left[ 1 + \cos \left( \pi \xx^a \right) \right],
\end{equation}
where $\xx \define \zgrid / \vlev$ takes its values between 0 and 1. Here, the exponent $a$ is a dilation coefficient that controls the growth or decay rate of vertical intervals in the vicinity of the lower and upper bounds. Introducing the vertical coordinate difference between the lowest model level and the surface, $\deltasigfirst$, this parameter is defined as 
\begin{equation}
a \define \frac{\ln \left[ \pi^{-1} \arccos \left( 1 - 2 \deltasigfirst \right)  \right]}{\ln \left( 1 / \vlev \right)}.
\end{equation}
For an isothermal temperature profile near planet's surface, $\deltasigfirst \approx \zzfirst / \Hpress $, where $\Hpress$ is the local pressure height. In practice, we set $\deltasigfirst  = 3.0 \times 10^{-3}$.

\subsection{Exner function}
\label{ssec:exner}
The calculation of the Exner function follows the method described by \cite{AL1977} and \cite{Hourdin1994note}, which is summarised here. Instead of computing $\Exner$ at the middle of the layer by extrapolating the pressure at the middle of the layer from the level pressures at the interfaces, which is computationally expensive, one rather uses a supplementary relationship between the interface levels and the Exner function. This relationship is directly derived from considerations about energy conservation principles within the atmospheric air column. Particularly, in the hydrostatic approximation, the internal and potential energy are proportional, which implies \citep[][]{AL1977,Hourdin1994note}
\begin{equation}
\integ{\geopot}{\mass}{0}{\masscol} = \integ{\Rspec \temp }{\mass}{0}{\masscol},
\label{conservation_exner}
\end{equation} 
where $\mass = \density \ddroit \zz$ is the infinitesimal parcel of mass per unit surface, and $\masscol$ the mass of the air column per unit surface. The expression of $\geopot$ as a function of $\teta$ and $\Exner$ is simply obtained by making use of the hydrostatic balance equation given, in grid vertical coordinates, by
\begin{equation}
\dd{\geopot}{\zgrid} + \teta \dd{\Exner}{\zgrid} = 0,
\end{equation}
and reads
\begin{equation}
\geopot = \integ{\dd{\geopot}{\zgrid'}}{\zgrid'}{0}{\zgrid} = - \integ{\rcp \teta \dd{\Exner}{\zgrid'}}{\zgrid'}{0}{\zgrid},
\end{equation}
which allows us, noticing that $ \Rspec \temp = \rcp \teta \Exner$, to rewrite \eq{conservation_exner} as
\begin{equation}
\integ{ \left[  \integ{\teta \dd{\Exner}{\zgrid'} }{\zgrid'}{0}{\zgrid} + \rcp \teta \Exner  \right] }{\mass}{0}{\masscol} = 0.
\end{equation}
The discretised form of the first term is given by 
\begin{equation}
\integ{\teta \dd{\Exner}{\zgrid}}{\zgrid'}{0}{\zgridl} = \sum_{\kver = 1}^{\kver} \left( \frac{\tetai{\kver}+ \tetai{\kver-1}}{2} \right) \left( \Exneri{\kver} - \Exneri{\kver-1} \right) + \tetai{0} \left( \Exneri{0} - \Exneri{\isurf} \right),
\end{equation}
where $\tetal$, $\Exnerl$ are the potential temperature and Exner function evaluated at the middle of the layer, indexed by $\lver = 0, \ldots, \vlev-1$ (for $\vlev$ vertical intervals), and $\Exneri{\isurf} = \Cp \left( \press / \prefteta \right)^{\rcp} $ is the Exner function evaluated at the planet's surface. By making use of the difference and average operators introduced in \eqs{diff_operator}{average_operator}, the preceding equation can be rewritten in the compact form
\begin{equation}
\integ{\teta \dd{\Exner}{\zgrid}}{\zgrid'}{0}{\zgridl} = \sum_{\lver = 0}^{\vlev-1} \left[  \moyz{\teta} \diffz{\Exner} \right]_{\kver},
\end{equation}
with
\begin{align}
& \left[  \moyz{\teta} \diffz{\Exner} \right]_{\kver} = \! \left( \frac{\tetai{\kver}+ \tetai{\kver-1}}{2} \right) \left( \Exneri{\kver} - \Exneri{\kver-1} \right), &  \kver = 1, \ldots, \vlev-1,\nonumber\\
& \left[  \moyz{\teta} \diffz{\Exner} \right]_{0}  = \tetai{0} \left( \Exneri{0} - \Exneri{\isurf} \right).  & 
\end{align}
Then, introducing the mass $\massl$ per unit area of layer $\lver$ and interchanging the sums in the double integral, we obtain
\begin{align}
\sum_{\lver = 0}^{\vlev - 1} \massl  \integ{\teta \dd{\Exner}{\zgrid}}{\zgrid'}{0}{\zgridl}  & =  \sum_{\lver =0}^{\vlev-1} \massl \sum_{\kver = 0}^{\lver} \left[ \moyz{\teta} \diffz{\Exner} \right]_{\kver}, \\
& =  \sum_{\kver = 0}^{\vlev-1} \left[ \moyz{\teta} \diffz{\Exner} \right]_{\kver} \sum_{\lver = \kver}^{\vlev-1} \massl.
\end{align}
In the above equation, we recognise the atmospheric mass per unit surface above the $\kver$ interfaces ($\kver = 0, \ldots , \vlev$), which is expressed in the framework of the hydrostatic approximation as 
\begin{equation}
\sum_{\lver = \kver}^{\vlev} \massl = \frac{\pressi{\kver}}{\ggravi}.
\end{equation}
It follows 
\begin{equation}
\sum_{\lver = 0}^{\vlev - 1} \massl \integ{\teta \dd{\Exner}{\zgrid}}{\zgrid'}{0}{\zgridl}   = \frac{1}{\ggravi} \sum_{\kver = 0}^{\vlev-1} \left[ \moyz{\teta} \diffz{\Exner} \right]_{\kver} \pressi{\kver}.
\end{equation}
Finally, expanding the coefficients of the sum, we remark that this later may be rewritten 
\begin{equation}
\sum_{\kver = 0}^{\vlev-1} \left[ \moyz{\teta} \diffz{\Exner} \right]_{\kver} \pressi{\kver} = \sum_{\kver = 0}^{\vlev-1} \tetai{\kver} \left[ \moyz{ \press \diffz{\Exner}} \right]_{\kver} ,
\end{equation}
with the conventions
\begin{align}
& \left[  \moyz{ \press \diffz{\Exner}} \right]_{0} = \frac{1}{2}  \left( \Exneri{1} - \Exneri{0} \right) \pressi{1} +  \left( \Exneri{0} - \Exneri{\isurf} \right) \pressi{0} \\ 
& \left[  \moyz{ \press \diffz{\Exner}} \right]_{\vlev-1}  = \frac{1}{2} \left( \Exneri{\vlev-1} + \Exneri{\vlev-2} \right) \pressi{\vlev-1} + \left( \Exneri{\itop} - \Exneri{\vlev-1} \right) \ptop. 
\end{align}
Thus, in the general case, the Exner function can be evaluated at cell centers by applying, at all vertical levels, the relation
\begin{equation}
\moyz{\press \diffz{\Exner}} = - \rcp \Exner \diffz{\press}. 
\end{equation}
The use of standard sigma coordinates appreciably reduces the complexity of the problem. Assuming $\ptop = 0$ and introducing the notation $\scord = \sig^{\rcp}$, the preceding equation simplifies to 
\begin{equation}
\moyz{\sig \diffz{\scord}} = \rcp \scord \diffz{\sig},
\end{equation}
which allows to pre-calculate the sigma coordinates of cell centers levels. Indexing these levels by $\kver = 0 , \ldots , \vlev-1$ and the intermediate (or interface) $\sig$ levels by $\kver = 0 , \ldots , \vlev$, we have
\begin{align}
 \left[ \moyz{\sig \diffz{\scord}}  \right]_{\kver} =  & \rcp \scord_\kver \left[ \diffz{\sig} \right]_\kver \ {\rm for}  \ \kver = 1, \ldots , \vlev-2, \\
 \left[ \moyz{\sig \diffz{\scord}} \right]_0 = & \frac{1}{2} \left( \scord_1 - \scord_0 \right) \sig_1 + \left( \scord_0 - \scord_{\isurf} \right) \sig_0 = \rcp \scord_0  \left( \sig_1 -\sig_0\right), \nonumber \\
 \left[ \moyz{\sig \diffz{\scord}} \right]_{\vlev-1} = & \frac{1}{2} \left( \scord_{\vlev-1}  - \scord_{\vlev-2} \right) \sig_{\vlev-1} = \rcp \scord_{\vlev-1} \left( \sig_{\vlev} - \sig_{\vlev-1} \right) ,\nonumber
\end{align}
where $\scord_\isurf = 1$ and $\scord_\itop = 0$ are the value of $\scord$ at planet's surface and atmospheric upper boundary, respectively. 

In practice, the Exner function is evaluated at grid centers levels by solving an algebraic equation of the form $\Amat \Xvect = \Bmat$. In the case of standard sigma coordinates ($\ptop = 0$), this calculation is performed only once, when the grid is constructed. The algebraic system to solve then writes
\begin{small}
\begin{equation}
\begingroup 
\setlength\arraycolsep{2pt}
\begin{bmatrix}
 A_{0,0} & A_{0,1} &  & &  \\
A_{1,0}  & A_{1,1} & A_{1,2} &  & \\ 
 & A_{\kver,\kver-1} & A_{\kver,\kver} & A_{\kver,\kver+1} &  \\
 & & A_{\vlev-2,\vlev-3} & A_{\vlev-2,\vlev-2} & A_{\vlev-2,\vlev-1} \\
 & & & A_{\vlev-1,\vlev-2} & A_{\vlev-1,\vlev-1}
\end{bmatrix} \! \! 
\endgroup
\begin{bmatrix}
\scord_0 \\
\scord_1 \\
\scord_\kver \\
\scord_{\vlev-2} \\
\scord_{\vlev-1}
\end{bmatrix}
\! \! = \! \!
\begin{bmatrix}
B_0 \\ B_1 \\ B_\kver \\ B_{\vlev-2} \\ B_{\vlev - 1}
\end{bmatrix} ,
\end{equation}
\end{small}
where the $\scord_\kver$ is the coordinate $\scord = \sig^{\rcp}$ evaluated at the mid-level of the $\kver$-layer, $\Amat$ the tridiagonal matrix of coefficients 
\begin{align}
& A_{\kver, \kver} = \left( \frac{1}{2} + \rcp \right) \left( \sig_\kver - \sig_{\kver +1} \right) \ {\rm for} \  \kver = 1, \ldots , \vlev-2 ; \\
& A_{0,0} = \left( 1 + \rcp \right) \sig_0 - \left( \frac{1}{2} + \rcp \right) \sig_1,  \\
& A_{\vlev-1,\vlev-1} = \left( \frac{1}{2} + \rcp \right) \sig_{\vlev-1} - \left( 1 + \rcp \right) \sig_{\vlev}, \\
& A_{\kver,\kver+1} = - A_{\kver +1,\kver} = \frac{1}{2} \sig_{\kver+1},
\end{align}
and $\Bmat$ the vector of coefficients 
\begin{align}
& B_\kver = 0  \ {\rm for} \ \kver = 1, \ldots, \vlev-2; \\
& B_0 = \sig_0 \scord_{\isurf}, \\
& B_{\vlev-1} = - \sig_{\vlev} \scord_{\itop}.
\end{align}
In these equations, $\scord_\isurf = 1$ and $\scord_\itop = 0$ are the values of $\scord$ at planet's surface and atmospheric top, respectively. This procedure can be applied from the moment that $\vlev>1$. If there is only one layer, then the mid-level coordinate of the unique layer is set to $\scord_0 = \left( \sig_0 + \sig_1 \right) /2 $.  

\subsection{Discretisation of the primitive equations}
The normalised prognostic equations given by \eqsto{prim1}{prim5} are discretised by making use of the difference and average operators defined by \eqs{diff_operator}{average_operator}, and become
\begin{align}
\dd{\massn}{\timen} + \bcourant \diffy{\Vflux} + \diffz{\Wflux}  & =  0, \\
\dd{}{\timen} \left( \moyy{\massn} \Vthetan \right) + \bcourant \diffy{} \left( \moyy{\Vflux} \moyy{\Vthetan}  \right) + \diffz{\left( \moyy{\Wflux} \moyz{\Vthetan} \right)}    & \nonumber \\
 + \frac{\bcourant}{\cy} \left[ \moyy{\massn} \diffy{\gpotn} + \moyy{ \massn \tetan} \left(\diffy{\Exnern} \right) \right] & =  \moyy{\massn}  \forcehn , \\
\dd{}{\timen} \left( \massn \tetan \right) + \bcourant \diffy{} \left( \Vflux \moyy{\tetan} \right) + \diffz{} \left( \Wflux \moyz{\tetan} \right) & =  \frac{\massn \Qheatn}{ \Exnern}, \\
\dd{}{\timen} \left( \massn \qmoistn \right) + \bcourant \diffy{} \left( \Vflux \moyy{\qmoistn} \right) + \diffz{} \left( \Wflux \moyz{\qmoistn} \right)  & = \massn  \dtqmoistn, \\
\label{prim5_dis}
\diffz{\gpotn} + \moyz{\tetan} \diffz{\Exnern}  & = 0.
\end{align}
In the right-hand members of these equations, the horizontal force is specified at horizontal interface levels, where the horizontal mass fluxes and velocities are also evaluated, and $\Qheatn$ and $\qmoistn$ at cell centers. The vertical velocity can be calculated afterwards, when it is necessary, using the formula
\begin{align}
\Vz = \frac{\Vzr}{\moyz{\massn}} \left[ \moyz{ \massn \dd{\gpotn}{\timen}} + \bcourant \moyz{\moyy{\Vflux \diffy{\gpotn}}} + \Wflux \diffz{\gpotn} \right].
\end{align}

\subsection{Time-differencing scheme}

\begin{figure}[htb]
   \centering
   \includegraphics[width=0.48\textwidth,trim = 0cm 6.2cm 4cm 0cm,clip]{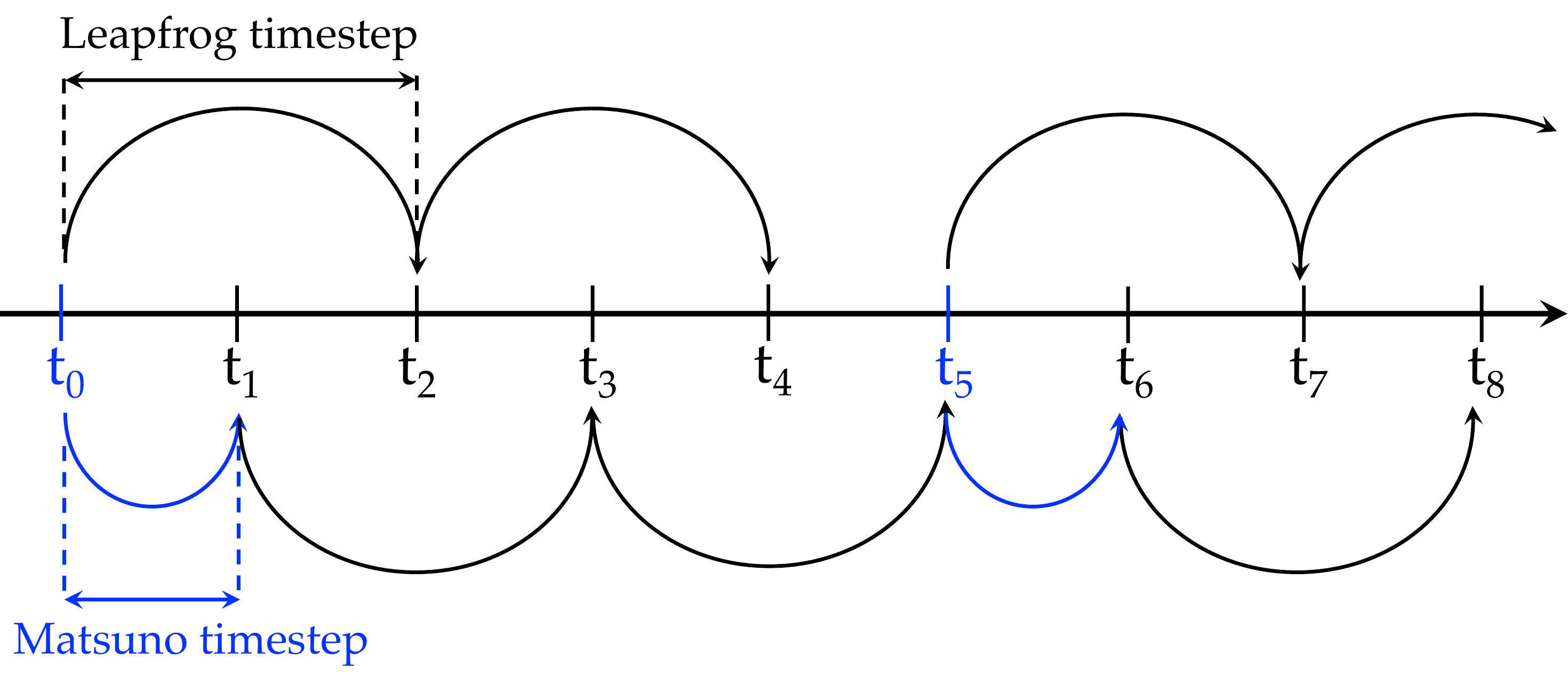}
      \caption{Time-differencing scheme implemented in the solver. The scheme is based on a leapfrog time step \citep[e.g.][Sect.~5.5.2]{Lauritzen2011}, which is replaced by a Matsuno time step \citep[][]{Matsuno1966a} every $\nmat =  5$ steps.}
       \label{fig:time_scheme}%
\end{figure}

Following the method used in many general circulation models \citep[][]{Hansen1983,YS1987,Hourdin2006}, the temporal integration of the primitive equations is based on the so-called leapfrog scheme, which is a centred explicit scheme \citep[e.g.][]{press2007numerical}. Since the leapfrog scheme tends to generate a spurious growth of numerical instabilities over time \citep[e.g.][Sect.~5.5]{Lauritzen2011}, a Matsuno time step \citep[][]{Matsuno1966a} is introduced every $n_{\rm MT}$ steps (with $\nmat = 5$) to stabilise the integration, as illustrated by \fig{fig:time_scheme}. Mathematically, denoting the time derivative operator by $\timederop$, any dynamical variable by $\physvar$, and indexing time steps by $\nn$, a leapfrog step is expressed as
\begin{equation}
\physvari{\ntime} = \physvari{\ntime-2} + 2 \dtimen \timederop \left( \physvari{\ntime-1} \right), 
\end{equation}
while a Matsuno step \citep[][]{Matsuno1966a} consists in the succession of two steps, 
\begin{align}
& \physvari{\ntime}^* = \physvari{\ntime-1} + \dtimen \timederop \left( \physvari{\ntime - 1} \right), \\
& \physvari{\ntime} = \physvari{\ntime-1} + \dtimen \timederop \left( \physvari{\ntime}^* \right),
\end{align}
where the superscript $*$ is used to designate the intermediate virtual time step. In \fig{fig:time_scheme}, the leapfrog time step is used to compute variables at all dates except $\time_0$ and $\time_5$, where it is replaced by a Matsuno step. Sink terms associated with numerical dissipation (hyper-diffusion, sponge layer, etc.) are evaluated periodically every $\nmat$ dynamical time step, before Matsuno time steps. Physical tendencies are computed every $\nphys$ dynamical time step (with $\nphys = 10$). 

\section{Numerical dissipation}

\def\wpanel{0.33\textwidth}
\def\wlegend{0.32\textwidth}
\def\hraisebox{0.20\textwidth}
\begin{figure*}[t]
   \centering
  \hspace{0cm} \textsc{$\diffcoeff = 10^{-5}$} \hspace{4.5cm} \textsc{$\diffcoeff = 10^{-4}$} \hspace{4.5cm} \textsc{$\diffcoeff = 10^{-3}$} \\[0.3cm]
   \includegraphics[width=\wpanel,trim = 2.5cm 0.0cm 0.8cm 0.9cm,clip]{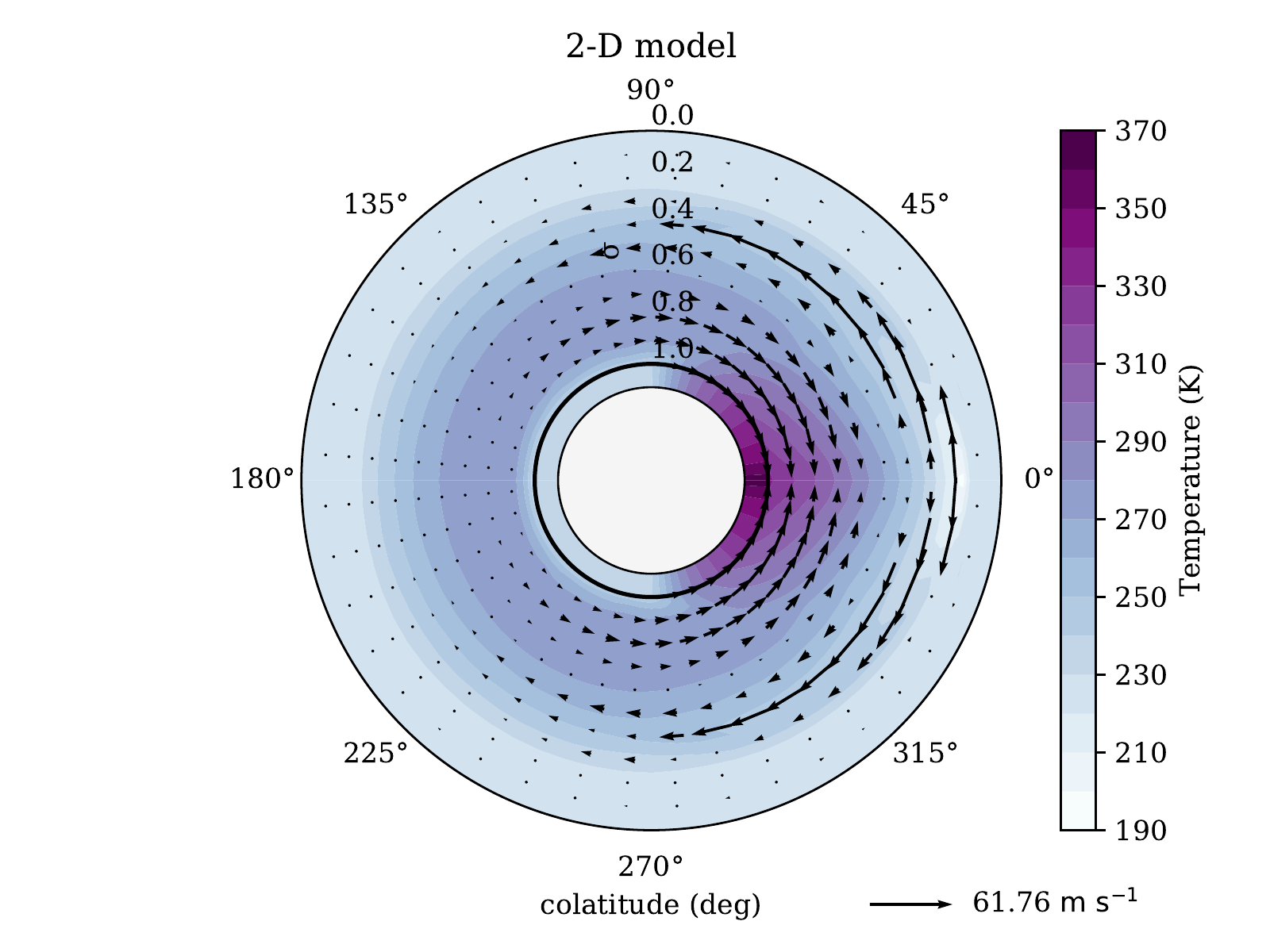}
   \includegraphics[width=\wpanel,trim = 2.5cm 0.0cm 0.8cm 0.9cm,clip]{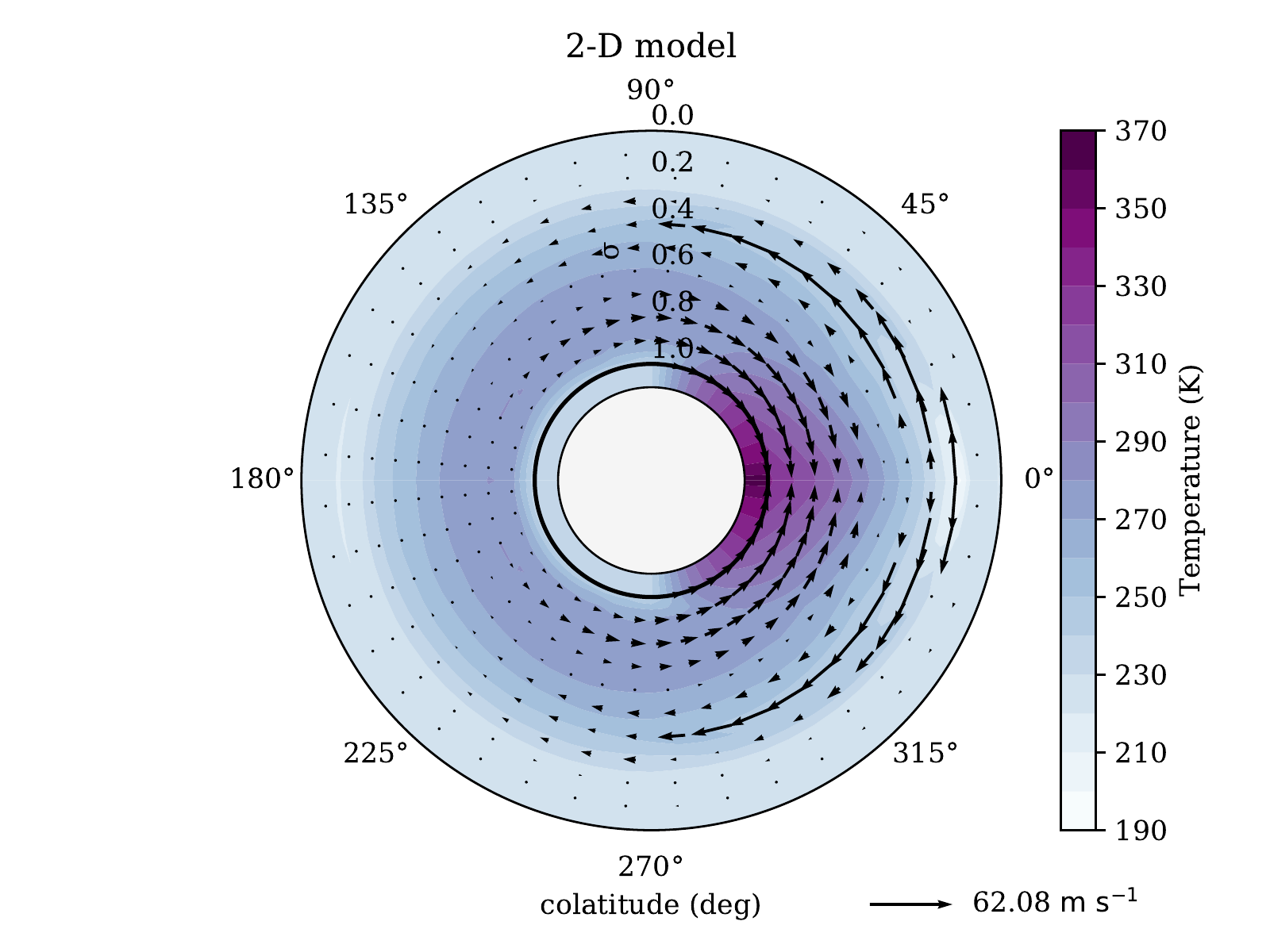} 
   \includegraphics[width=\wpanel,trim = 2.5cm 0.0cm 0.8cm 0.9cm,clip]{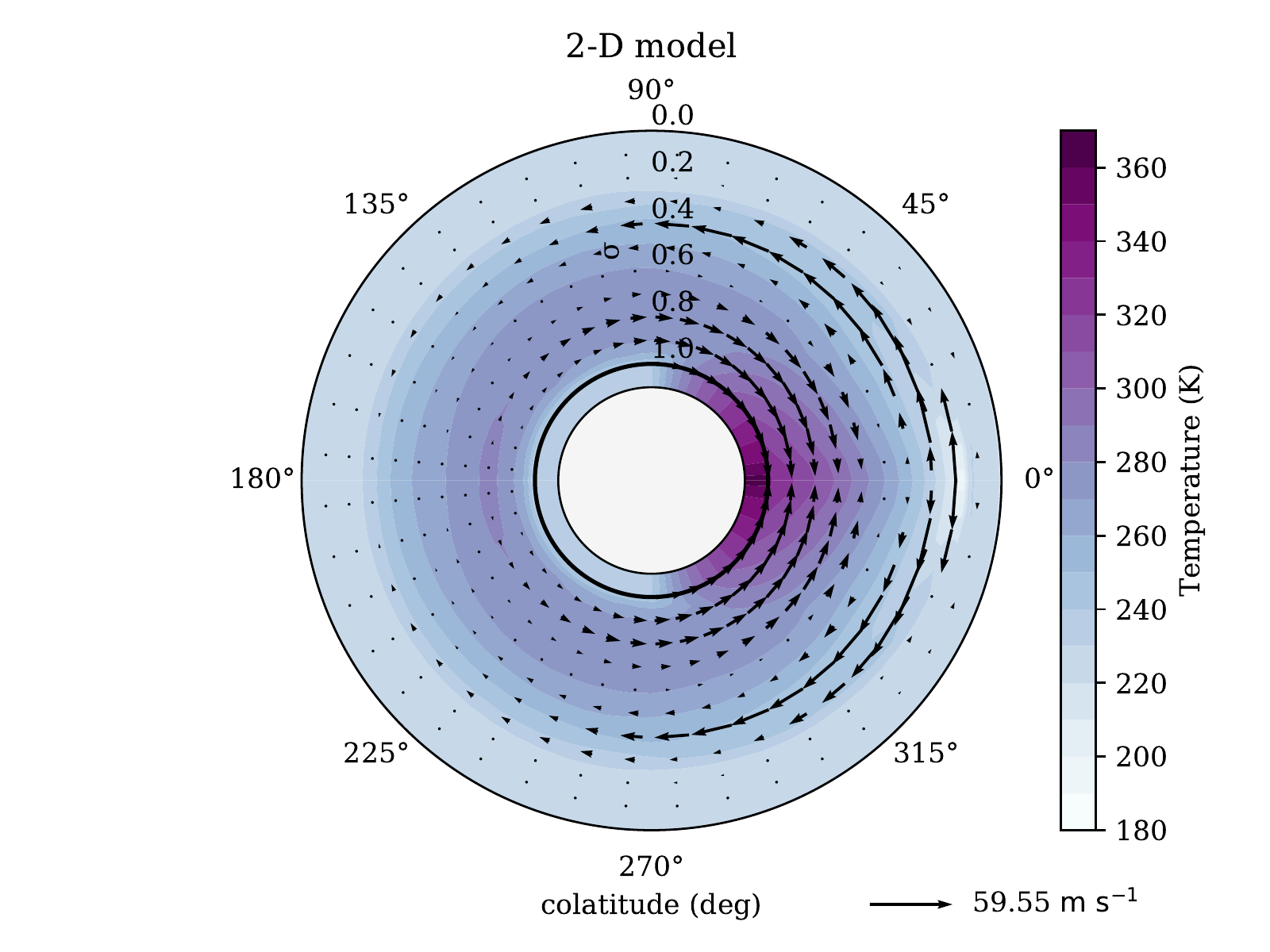} 
      \caption{\RV{Two-day averaged temperature snapshots for various values of hyper-diffusion parameter (see \eq{diffcoeff}). {\it Left:} $\diffcoeff = 10^{-5}$. {\it Middle:} $\diffcoeff = 10^{-4}$. {\it Right:} $\diffcoeff = 10^{-3}$. Simulations were performed for the Earth-like case of Table~\ref{tab:param_reference_case} with a stellar irradiation of 1366~${\rm W \ m^{-2}}$ and a 1 bar surface pressure, similarly as in \fig{fig:snapshots_HB1993_diff}.} }
       \label{fig:validation_hydiff}%
\end{figure*}

\subsection{Horizontal hyper-diffusion}
\label{app:hyper_diff}
To dissipate energy at grid scale, we introduce a bi-harmonic diffusion \citep[][]{Lauritzen2011}, which is a fourth-order diffusion. This is obtained by applying twice the Lapacian operator to the temperature and horizontal velocity. For any scalar quantity $\physvar$, the horizontal Laplacian operator reads, in our two-dimensional system of coordinates, 
\begin{equation}
\laph \physvar =  \Rpla^{-2} \frac{1}{\sin \col} \dd{}{\col} \left( \sin \col \dd{\physvar}{\col} \right) .
\end{equation} 
In grid coordinates, it is expressed as
\begin{align}
\laph \physvar & =  \left( \pi \Rpla \right)^{-2} \left( \cy \sin \col \right)^{-1} \dd{}{\ygrid} \left( \frac{\sin \col }{\cy} \dd{\physvar}{\ygrid} \right) \\
 & = \left( \pi \Rpla \right)^{-2} \laphn \physvar, \nonumber
\end{align}
where $\laphn$ designates the normalised horizontal Laplacian operator,
\begin{equation}
\laphn \physvar \define \left( \cy \sin \col \right)^{-1} \dd{}{\ygrid} \left( \frac{\sin \col }{\cy} \dd{\physvar}{\ygrid} \right).
\end{equation}
We introduce the hyper-Laplacian operator, which is the Laplacian of order $\qdiff$, defined as $\nabn{2 \qdiff} \define \laph \ldots \laph$. Basically, for the fourth-order hyper-diffusion, $\qdiff = 2$, and $\nabn{4} \define \laph \laph$. In the general case, the $2 \qdiff$-order hyper-diffusion term of any variable $\physvar$ is defined by 
\begin{equation}
\Fdiff \define \left( - 1 \right)^{\qdiff+1} \Kdiffq \nabn{2 \qdiff} \physvar, 
\label{fdiff}
\end{equation}
where $\Kdiffq$ is the hyper-diffusivity. This parameter can be written as a function of the diffusion timescale $\taudiff$ and mean horizontal grid spacing $\meandeltacol \define \pi / \hlev $ \citep[][Sect.~13.3]{Lauritzen2011},
\begin{equation}
\Kdiffq = \frac{1}{2 \taudiff} \left( \frac{\Rpla \meandeltacol}{2} \right)^{2 \qdiff}.
\label{Kdiffq}
\end{equation}
The normalised hyper-diffusivity $\Kdiffqn$ is thus given by 
\begin{equation}
\Kdiffqn \define \frac{\Kdiffq}{\left( \pi \Rpla \right)^{2 \qdiff}} = \frac{1}{2 \taudiff \left( 2 \hlev \right)^{2 \qdiff}}.
\end{equation}
In order to adapt the hyper-diffusivity to the horizontal grid resolution, it is convenient to introduce the nondimensional diffusion parameter \citep[][]{TS2004,Mendonca2016}
\begin{equation}
\diffcoeff \define \frac{\dyndt}{2^{2 \qdiff+1} \taudiff},
\label{diffcoeff}
\end{equation}
which allows us to rewrite diffusivity parameters as 
\begin{align}
\Kdiffq = \diffcoeff \frac{\meandeltacol^{2 \qdiff}}{\dyndt} , && \Kdiffqn = \diffcoeff \frac{1}{\dyndt \hlev^{2 \qdiff}}.
\end{align}
Bi-harmonic diffusion is used in the model, with a diffusion parameter set to $\diffcoeff = 6.25 \times 10^{-4} $ in agreement with the order of magnitude of commonly used values \citep[][Sect.~13.3]{Lauritzen2011}. 

In the finite-volume approach, hyper-diffusion can lead to numerical instabilities near the poles because of the singularity $\sin^{-1} \left( \col \right) $ of the Laplacian operator. Typically, assuming that $\physvar$ is an oscillatory function of the form $\physvar \left( \col \right) = \sin \left( \nosc \pi \col \right)$ yields

\begin{equation}
\laph \physvar = \frac{\left( \nosc \pi \right)^2}{\sin \col} \left[ \left( \nosc \pi \right)^{-1} \cos \left( \col \right) \cos \left( \nosc \pi \col \right) - \sin \left( \col \right) \sin \left( \nosc \pi \col \right)   \right] ,
\end{equation}
and thus $\laph \physvar \scale \sin^{-1} \left( \col \right) $ in polar regions. To remediate to this problem, it is often chosen to compensate the singularity introduced by the Laplacian by annihilating the diffusivity parameter at the poles \citep[][Sect.~13.3]{Lauritzen2011}. Basically, the isotropic diffusivity introduced in \eq{fdiff} is multiplied by $\sin^{\canis \qdiff} \left( \col \right)$, where $\canis$ is a coefficient of anisotropy ($\canis = 0.5-1$ typically). This anisotropic diffusion solves the stability problem but can also induce unphysical boundary effects near the poles. Following \cite{Majewski2002}, we apply a bi-harmonic diffusion ($\qdiff = 2$) to the temperature and horizontal velocity $\Vtheta$. We use the same diffusion coefficient for both terms. Thus, 
\begin{align}
& \Fdiffv = - \Kdiffbin \sin^{2 \canis} \left( \col \right) \nabnhn{4} \Vtheta, \\
& \Fdifftemp = - \Kdiffbin \sin^{2 \canis} \left( \col \right) \nabnhn{4} \temp.
\end{align}
In addition, we apply to the top layer of the model an harmonic diffusion of the form
\begin{align}
& \Fdiffv =  - \Kdiffhan \sin \left( \col \right) \laphn \Vtheta, \\
& \Fdifftemp = - \Kdiffhan \sin \left( \col \right) \laphn \temp. 
\end{align}

\RV{In order to verify that the mean flow and temperature distribution are insentitive to the hyper-diffusion scheme, simulations were run for various values of the nondimensional diffusion parameter $\diffcoeff$ introduced in \eq{diffcoeff}, namely $\diffcoeff = 10^{-5}, 10^{-4},10^{-3}$. These validation tests were performed for the Earth-like case of Table~\ref{tab:param_reference_case} with the same surface pressure and stellar flux as in \fig{fig:snapshots_HB1993_diff}. Figure~\ref{fig:validation_hydiff} shows the two-day averaged temperature snapshots obtained for each value of $\diffcoeff$, as well as the associated mean flows. We observe that varying the value of $\diffcoeff$ over two orders of magnitude hardly alters the temperatures and wind speeds. The minimum nightside temperature after convergence is $\Tnight = 231.5$~K, $\Tnight = 231.7$~K, and $\Tnight = 231.8$~K, for $\diffcoeff = 10^{-5}, 10^{-4},10^{-3}$, respectively. Similarly, the maximum wind speed varies between $59.55 \ {\rm m \ s^{-1}}$ and $62.08 \ {\rm m \ s^{-1}}$ (see \fig{fig:validation_hydiff}). These variations correspond to a 0.1$\%$ difference for the minimum surface temperature, and to a 4.2$\%$ difference for the maximum wind speed. Winds are thus more sensitive to the hyper-diffusion scheme than the nightside surface temperature, although the observed dependence is relatively weak in both cases. Nevertheless, this dependence is expected to be more important for extreme values of $\diffcoeff$ because such values would lead either to underdissipated or to overdissipated flow fields, as shown by \cite{TC2011}. Typically, with a value of $\diffcoeff$ smaller than the adopted one by several orders of magnitude, the flow rapidly becomes numerically unstable at grid scale, which induces spurious fluctuations and may cause the run to abort. }

\subsection{Sponge layer}
\label{app:sponge_layer}
In extreme cases (low surface pressure and high stellar flux), the strong thermal forcing of the atmosphere on the dayside generates instabilities that can lead to negative pressures near the top of the model. Particularly, it generates internal gravity waves that propagate upwards with amplitudes becoming very large as the atmospheric density tends to zero\footnote{Although they exist, these waves are poorly resolved in the hydrostatic approximation since the buoyancy  term in the vertical momentum equation is missing.}. These waves are reflected downwards by the upper boundary, which acts as a wall (no vertical mass flow). Owing to the weak density, such extreme fluctuations have dramatic repercussions on the computation of mass flows and are thereby a source of unphysical values that make runs abort. Thus, in addition with the hyperdiffusion scheme, the use of a sponge layer is necessary in cases where the surface pressure is low and the stellar irradiation is strong. 

The sponge layer is a numerical dissipation process that strongly damps wind flows diverging from a prescribed equilibrium profile in the upper regions of the atmosphere, with an efficiency increasing with the altitude. In the present model, we use a Rayleigh friction sponge, which is based on a linear relaxation term of generic form \citep[e.g.][Sect.~13.4]{Lauritzen2011}
\begin{equation}
\dd{\Vtheta}{\time} = - \kspon \left( \Vtheta  - \Vthetaspon \right),
\end{equation}
where $\kspon$ designates the Rayleigh damping coefficient of the sponge layer and $\Vthetaspon$ the equilibrium velocity profile near the upper boundary. This profile is set to the standard value $\Vthetaspon = 0$. Following \cite{PK2002}, we opt for a vertical profile of the Rayleigh coefficient of the form
\begin{equation}
\kspon \left( \sig \right) = 
\left\{
\begin{array}{ll}
0 & \mbox{if} \ \sig\geq \sigspon, \\ 
\ksponmax \left( 1 - \frac{\sig}{\sigspon} \right)^2 & \mbox{if} \ \sig < \sigspon. 
\end{array}
\right.
\end{equation}
In the above piecewise function, $\sigspon$ corresponds to the critical normalised pressure below which the sponge layer is applied while $\ksponmax$ is the maximum value of the Rayleigh friction coefficient. This parameter has dimensions of a frequency and is the inverse of the minimum damping timescale of the sponge layer $\tsponmin$. The smaller $\tsponmin$, the stronger the damping in the sponge layer. In the model, the maximum Rayleigh coefficient is set to $\ksponmax = 0.5 \ {\rm day^{-1}}$, while the thickness of the sponge layer is defined as a function of the stellar flux and surface pressure.

\section{Convective adjustment scheme}
\label{app:convadj}
The turbulent diffusion scheme implemented described by \sect{ssec:pbl} is not sophisticated enough to prevent superadiabatic vertical temperature gradients,
\begin{equation}
\dd{\teta}{\zz} <0. 
\end{equation}
If such an unstable profile is produced by the model, it may generate numerical errors in the solution and lead the run to crash. In order to prevent this behaviour, a convective adjustment scheme can be activated in simulations. This scheme tends to regularise the potential temperature profile every physical timestep by correcting the tendencies in heat fluxes while conserving the entropy over the air column. The convective adjustment scheme used in the present work is similar to that implemented in the \texttt{LMDZ} and \texttt{THOR} GCMs \citep{Hourdin1993,MB2020}. 
\begin{figure}[htb]
   \centering
   \includegraphics[width=0.35\textwidth,trim = 0.cm 0cm 12.cm 9.cm,clip]{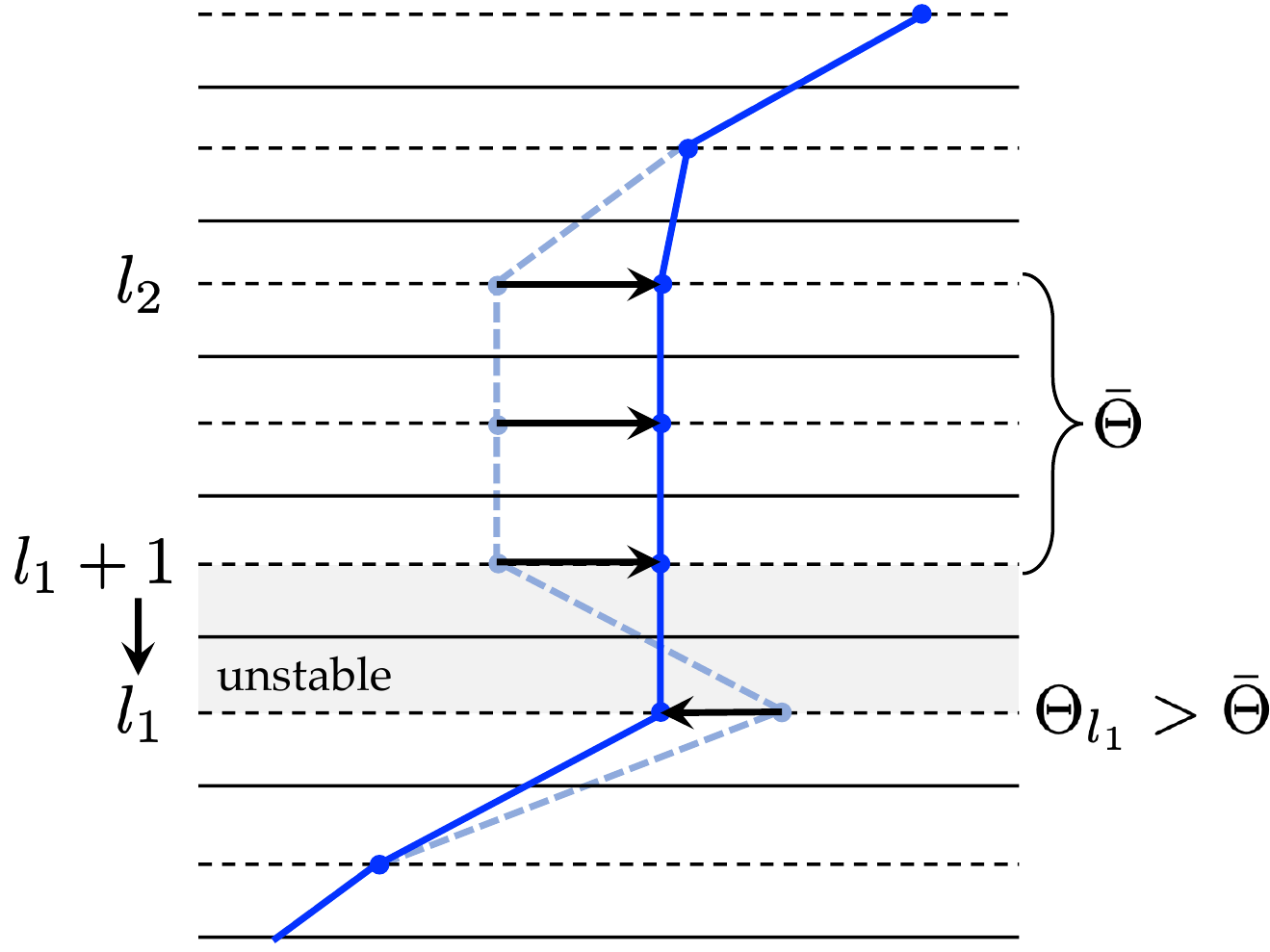}
      \caption{Construction of the stable temperature profile in the convective adjustment scheme. An initially unstable region of the air column spreading from layer $l_1$ to layer $l_2$ is extended both downwards and upwards until the temperature profile is stable. Every step, the mass-averaged potential temperature of the region is adjusted to include the contribution of the current unstable layer.}
       \label{fig:convadj}%
\end{figure}

The principle of the scheme is contained in two steps. During the first step, the unstable intervals of the vertical temperature profiles extrapolated from tendencies are detected and a stable profile is constructed by averaging the potential temperature over unstable layers. Starting from the ground, the interval bounded by layers $l_1$ and $l_2$ with $l_1 \leq l_2$ is extended both downwards and upwards incrementally until $l_1 $ and $l_2$ correspond to the bottom and top layers of the atmosphere, respectively. In an unstable region, the potential temperature is set to the mass-averaged adiabatic potential temperature,
\begin{equation}
\tetaadiab \define \frac{\integ{\teta  }{\mass}{\mbot}{\mtop} }{\integ{}{\mass}{\mbot}{\mtop}},
\end{equation}
where $\infvar{\mass} = \density \infvar{\zz}$ is an infinitesimal parcel of mass of the air column, $\mbot$ the mass of the air column below the lower boundary of the mixed region, and $\mtop$ the mass below the upper boundary of the mixed region. If $\tetai{\kver} > \tetaadiab$ for a layer $\kver$ underneath the mixed region, the mass-averaged adiabatic potential temperature is adjusted by including the contribution of the layer and $\tetai{\kver}$ is set to $\tetaadiab$. Figure~\ref{fig:convadj} illustrates how the adiabatic profile is adjusted to stabilise the layer $l_1$, where the vertical gradient of potential temperature is negative.

The second step of the scheme consists in evaluating the new tendencies resulting from the stable temperature profile. The tendency for the entropy equation is straightforwardly obtained from the adiabatic temperature profile constructed at the first step. For the momentum equation, an estimate of the instability of the atmosphere is computed from the relative enthalpy exchange that is necessary to restore the adiabatic profile from the original profile,
\begin{equation}
\alphaconv \define \frac{\integ{\abs{\teta - \tetaadiab} }{\mass}{\mbot}{\mtop} }{\integ{\teta}{\mass}{\mbot}{\mtop}}.
\end{equation}
This parameter corresponds to the fraction of the mesh on which the angular momentum is mixed \citep[in practice, the condition $\alphaconv<1$ is always verified in simulations; e.g.][]{Hourdin2006}. The extrapolated horizontal velocity used to compute the tendency for the horizontal momentum is corrected by a factor $\alphaconv \left( \Vthetaadiab - \Vtheta \right) $, where $\Vthetaadiab$ is the mass-averaged velocity defined as
\begin{equation}
\Vthetaadiab \define \frac{\integ{\Vtheta}{\mass}{\mbot}{\mtop}}{\integ{}{\mass}{\mbot}{\mtop}}.
\end{equation}

\section{Radiative transfer scheme}
\label{app:rad_transfer}
\begin{figure}[t]
   \centering
   \includegraphics[width=0.40\textwidth,trim = 0cm 3.4cm 9.5cm 0cm,clip]{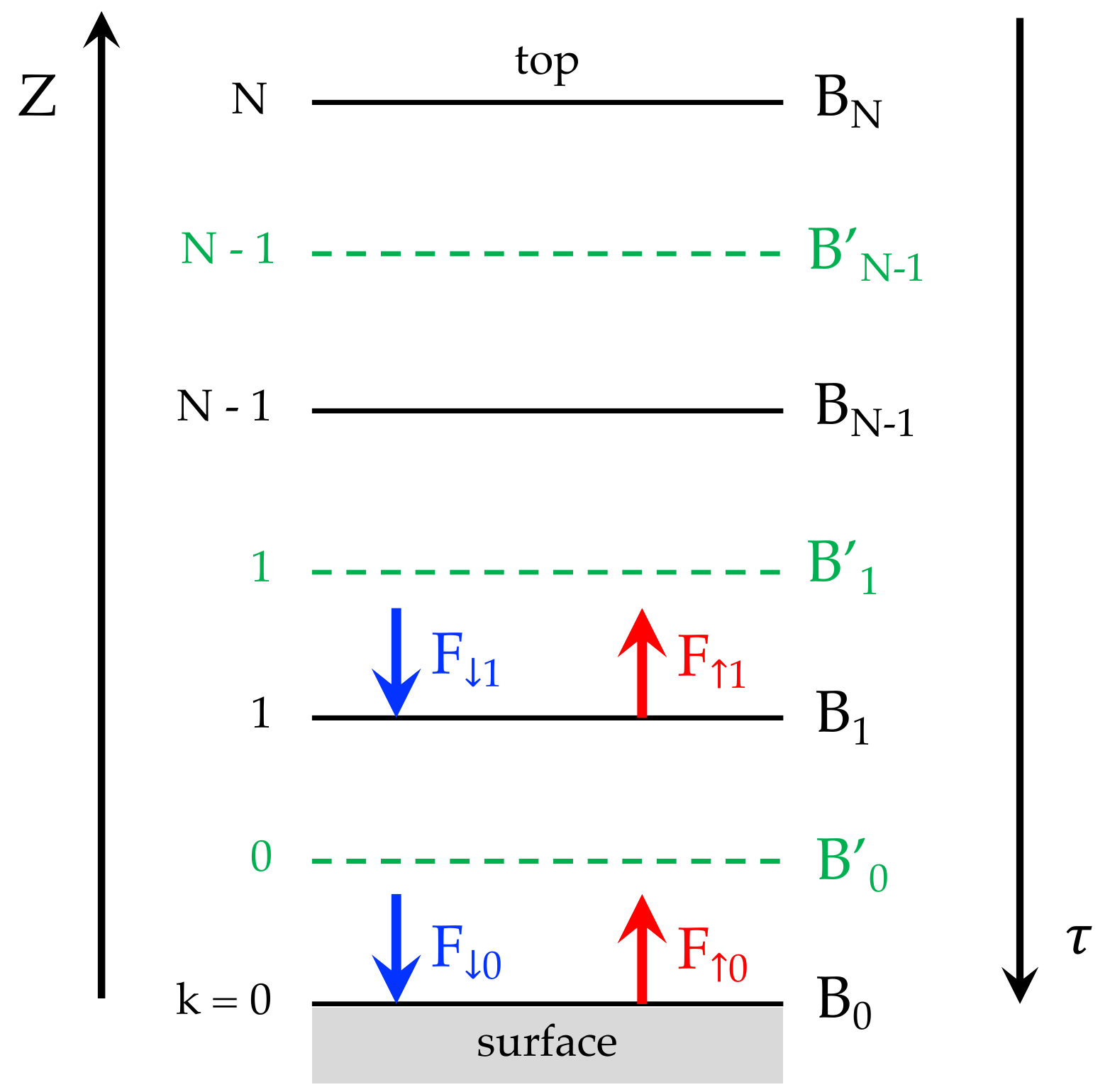}
      \caption{Radiative transfer in double-gray approximation, as described in \append{app:rad_transfer}. The equations for the shortwave and longwave radiative fluxes are solved separately. In each band, the coupled upward and downward fluxes are calculated by making use of Thomas algorithm (see \append{app:thomas_algorithm}). The superscript $\prime$ designates the vertical gradient of the black body emission flux $\Bbbi{\kver}^\prime =  \atx{\DD{\Bblackbody}{\optdepth}}{\kver}$ introduced in \eqs{radiative_trans_1}{radiative_trans_2}.}
       \label{fig:rad_transfer}%
\end{figure}

The two radiative transfer equations given by \eqs{eqrt1}{eqrt2} are solved periodically every $\nrad$~physical time step (with $\nrad = 6 $) as illustrated by \fig{fig:rad_transfer}. First we integrate them analytically between two intermediate vertical levels, indexed by $\kver$ and $\kver+1$, with $\kver = 0, \ldots , \vlev-1$. This leads to 
\begin{align}
\label{radiative_trans_1}
\Fdowni{\kver} =  \etafluxi{\kver}^{-1} & \left\{  \lambi{\kver} \Fdowni{\kver+1} +\mui{\kver} \Fupi{\kver}  - \lambi{\kver} \Bbbi{\kver+1} \nonumber  + \left[ \etafluxi{\kver} - \mui{\kver} \right] \Bbbi{\kver} \right.  \nonumber \\
 & \left. - \left( \gamplus^2 - \gamminus^2 \right) \left( 1 - \ftransi{\kver} \right) \left( \gamplus + \gamminus \ftransi{\kver} \right) \atx{\DD{\Bblackbody}{\optdepth}}{\kver}  \right\},
\end{align}
\begin{align}
\label{radiative_trans_2}
\Fupi{\kver+1} =   \etafluxi{\kver}^{-1} & \left\{  \lambi{\kver} \Fupi{\kver} + \mui{\kver} \Fdowni{\kver+1}  - \lambi{\kver} \Bbbi{\kver} \nonumber  + \left[  \etafluxi{\kver} + \mui{\kver} \right] \Bbbi{\kver+1} \right. \nonumber \\
 & \left. + \left( \gamplus^2 - \gamminus^2 \right) \left( 1 - \ftransi{\kver} \right) \left( \gamplus + \gamminus \ftransi{\kver} \right) \atx{\DD{\Bblackbody}{\optdepth}}{\kver}  \right\},
\end{align}
where the $\Bbbi{\kver}$ designate the values of the blackbody emission $\Bblackbody$ introduced in \eq{eqrt2} ($\Bblackbody= 0$ for the shortwave) interpolated at intermediate vertical levels, $\atx{\DD{\Bblackbody}{\optdepth}}{\kver}$ the values of its derivatives evaluated at internal vertical levels, $\gampm$ the usual coupling coefficients, 
\begin{equation}
\gampm = \frac{1}{2} \left( 1 \pm \betascat \right),
\end{equation}
and where we have introduced the transmission functions 
\begin{equation}
\ftransi{\kver} \define \expo{\optdepth_{\kver+1} - \optdepth_{\kver}},
\end{equation} 
and the coefficients 
\begin{equation}
\begin{array}{ll}
\displaystyle \etafluxi{\kver} \define \gamplus^2 - \left( \gamminus \ftransi{\kver} \right)^2, & \displaystyle \lambi{\kver} \define \left(  \gamplus^2 - \gamminus^2  \right) \ftransi{\kver}, \\
\displaystyle \mui{\kver} \define \gamminus \gamplus \left( 1 - \ftransi{\kver}^2 \right). 
\end{array}
\end{equation}

These relations can be put in the matrix form 
\begin{align}
\begin{bmatrix}
- \mui{\kver} & \etafluxi{\kver} & 0  & - \lambi{\kver} \\ 
- \lambi{\kver} &  0 & \etafluxi{\kver} & - \mui{\kver}
\end{bmatrix}
\begin{bmatrix}
\Fupi{\kver} \\ \Fdowni{\kver} \\ \Fupi{\kver+1} \\ \Fdowni{\kver+1}
\end{bmatrix}
=
\begin{bmatrix}
\bcompij{\kver}{1} \\
\bcompij{\kver}{2}
\end{bmatrix}
,
\end{align}
where $\bcompij{\kver}{1}$ and $\bcompij{\kver}{2} $ are given by
\begin{align}
\bcompij{\kver}{1} = &  - \lambi{\kver} \Bbbi{\kver+1} + \left( \etafluxi{\kver} - \mui{\kver} \right) \Bbbi{\kver} \\
 &  - \left( \gamplus^2 - \gamminus^2 \right) \left( 1 - \ftransi{\kver} \right) \left( \gamplus + \gamminus \ftransi{\kver} \right) \atx{\DD{\Bblackbody}{\optdepth}}{\kver}, \nonumber
\end{align}
\begin{align}
\bcompij{\kver}{2} = &  - \lambi{\kver} \Bbbi{\kver} + \left( \etafluxi{\kver} + \mui{\kver} \right) \Bbbi{\kver+1} \\
 &  + \left( \gamplus^2 - \gamminus^2 \right) \left( 1 - \ftransi{\kver} \right) \left( \gamplus + \gamminus \ftransi{\kver} \right) \atx{\DD{\Bblackbody}{\optdepth}}{\kver}. \nonumber
\end{align}
They are completed by relations derived from the lower and upper boundary conditions, which, in the general case, are of the form
\begin{align}
\label{bc_bottom}
\bcsurfi{0} \Fupi{0} + \bcsurfi{1} \Fdowni{0} + \bcsurfi{2} \Fupi{1} + \bcsurfi{3} \Fdowni{1} = \bcsurfb, \\
\label{bc_top}
\bctopi{0} \Fupi{\vlev-1} + \bctopi{1} \Fdowni{\vlev-1} + \bctopi{2} \Fupi{\vlev} + \bctopi{3} \Fdowni{\vlev} = \bctopb,
\end{align}
where the subscripts $\isurf$ and $\itop$ denote the coefficients associated with the planet's surface or with the top of the atmosphere, respectively. Therefore, introducing the vectors $\Fvecti{\kver} \define \transp{\left( \Fupi{\kver} , \Fdowni{\kver} \right)}$, we can write the discretised equations as a linear algebraic system of the form

\begin{equation}
\begin{bmatrix}
\Bmati{0} & \Cmati{0} &  & & \\
\Amati{1} & \Bmati{1} & \Cmati{1} & & \\
	       & \Amati{\kver} & \Bmati{\kver} & \Cmati{\kver} &  \\
	       &		       & \Amati{\vlev-1} & \Bmati{\vlev-1} & \Cmati{\vlev-1} \\
	       &		       &			 & \Amati{\vlev} & \Bmati{\vlev}
\end{bmatrix}
\begin{bmatrix}
\Fvecti{0} \\
\Fvecti{1} \\
\Fvecti{\kver} \\
\Fvecti{\vlev-1} \\
\Fvecti{\vlev}
\end{bmatrix}
=
\begin{bmatrix}
\bvecti{0} \\ \bvecti{1} \\ \bvecti{\kver} \\ \bvecti{\vlev-1} \\ \bvecti{\vlev}
\end{bmatrix}
.
\end{equation}
In this system, the sub-matrices $\Amati{\kver}$, $\Bmati{\kver}$, and $\Cmati{\kver}$ are expressed as
\begin{align}
& \Amati{\kver} = 
\begin{bmatrix}
- \lambi{\kver-1} & 0 \\
0 & 0
\end{bmatrix}
, & \Bmati{\kver} = 
\begin{bmatrix}
\etafluxi{\kver-1}  & - \mui{\kver-1} \\
- \mui{\kver} & \etafluxi{\kver}
\end{bmatrix}
,& 
\end{align}
\begin{align}
& \Cmati{\kver} = 
\begin{bmatrix}
0 & 0 \\
0 & - \lambi{\kver}
\end{bmatrix}
,&  \kver = 1, \ldots , \vlev-1,
\end{align}
\begin{align}
& \Bmati{0} = 
\begin{bmatrix}
\bcsurfi{0} & \bcsurfi{1} \\
- \mui{0} & \etafluxi{0}
\end{bmatrix}
, & \Cmati{0} = 
\begin{bmatrix}
\bcsurfi{2} & \bcsurfi{3} \\ 
0 & - \lambi{0}
\end{bmatrix}
, 
\end{align}
\begin{align}
 & \Amati{\vlev} = 
\begin{bmatrix}
- \lambi{\vlev-1} & 0 \\
\bctopi{0} & \bctopi{1} 
\end{bmatrix}
, & \Bmati{\vlev} = 
\begin{bmatrix}
\etafluxi{\vlev-1} & - \mui{\vlev-1} \\
\bctopi{2} & \bctopi{3} 
\end{bmatrix}
,
\end{align}
and the corresponding vectors are expressed as
\begin{align}
 & \bvecti{\kver} = 
\begin{bmatrix}
\bcompij{\kver-1}{2} \\ \bcompij{\kver}{1}
\end{bmatrix}
, & \kver = 1, \ldots , \vlev-1, 
\end{align}
\begin{align}
 & \bvecti{0} = 
\begin{bmatrix}
\bcsurfb \\ \bcompij{0}{1} 
\end{bmatrix}
, &  \bvecti{\vlev} = 
\begin{bmatrix}
\bcompij{\vlev-1}{2} \\ \bctopb 
\end{bmatrix}
. 
\end{align}
As the matrix of the system is a block tridiagonal matrix, the system can be solved by making use of Thomas algorithm (see \append{app:thomas_algorithm}). We note that the shortwave and longwave fluxes can be integrated in parallel since there are decoupled. In practice, the coefficients of boundaries conditions introduced in \eqs{bc_bottom}{bc_top} are, for the shortwave, 

\begin{equation}
\begin{array}{lllll}
\bcsurfi{0} = 1, & \bcsurfi{1}=-\Asurfsw, & \bcsurfi{2} = 0 ,& \bcsurfi{3} = 0,  & \bcsurfb = 0 , \\
\bctopi{0} = 0, & \bctopi{1} = 0, & \bctopi{2} = 0 , & \bctopi{3} = 1, & \bctopb = \Finc,
\end{array}
\end{equation}
and, for the  longwave,
\begin{equation}
\begin{array}{lllll}
\bcsurfi{0} = 1, & \bcsurfi{1}=0, & \bcsurfi{2} = 0 ,& \bcsurfi{3} = 0,  & \bcsurfb = \emissurf \sigmaSB \Tsurf^4 , \\
\bctopi{0} = 0, & \bctopi{1} = 0, & \bctopi{2} = 0 , & \bctopi{3} = 1, & \bctopb = 0,
\end{array}
\end{equation}
where $\Finc$ designates the incident stellar flux, given by
\begin{equation}
\Finc \left( \col \right) = \left\{
\begin{array}{ll}
 \Fstar \cos \col, & \mbox{if} \ 0^\degree \leq \col \leq 90^\degree, \\
 0, & \mbox{if} \ 90^\degree < \col \leq 180^\degree.
\end{array}
\right.
\end{equation}

\section{Turbulent diffusion scheme}
\label{app:turbdiff_scheme}
\subsection{Stability and asymptotic length scale functions}
\label{app:stability_functions}

The functions $\fmom $ and $\fheat$ introduced in \eqs{flux_mom}{flux_heat} are piecewise functions of the surface-layer bulk Richardson number given by \eq{Ribulk}. They are defined in the model following the formulation proposed by \cite{HB1993}, which was established experimentally in the Earth case. In the unstable regime ($\Ribulk <0$), they are given by 
\begin{align}
& \fmom \left( \Ribulk \right) = 1 - \frac{10 \Ribulk}{1 + 75 \Cneutral \sqrt{ \left( 1 + \zzfirst / \zrough \right) \abs{\Ribulk}} }, \\
& \fheat \left( \Ribulk \right) =  1 - \frac{15 \Ribulk}{1 + 75 \Cneutral \sqrt{\left( 1 + \zzfirst / \zrough \right) \abs{\Ribulk} } },
\end{align}
and, in the stable regime ($\Ribulk \geq 0$), by
\begin{equation}
\fmom \left( \Ribulk \right) = \fheat \left(\Ribulk \right) = \frac{1}{1+10 \Ribulk \left( 1 + 8 \Ribulk \right)}. 
\end{equation} 
The functions $\fdiffx$ introduced in \eq{Kdiffx} to characterise the dependence of eddy diffusivities upon the gradient Richardson number are the same for both momentum and heat diffusion. This function is defined, in the unstable regime ($\Riz < 0$), as
\begin{equation}
\fdiffx \left( \Riz \right) = \sqrt{ 1 - 18 \Ribulk},
\end{equation}
and, in the stable regime ($\Riz \geq  0$), as
\begin{equation}
\fdiffx \left( \Riz \right) = \fmom \left( \Riz \right) = \frac{1}{1 + 10 \Riz \left( 1 + 8 \Riz \right)}. 
\end{equation}
Finally, following \cite{HB1993}, the asymptotic length scale is defined, for both momentum and heat diffusivities, as the piecewise function
\begin{equation}
\mixingL \left( \zz \right) = \left\{
\begin{array}{ll} 
 300  & \mbox{if} \ \zz \leq 1 \ \mbox{km} \\ 
 30 + 270 \, \exp \left( 1 - \zz / 1000 \right) & \mbox{if} \ \zz > 1 \ \mbox{km},
\end{array}
 \right.
 \label{mixlength_function}
\end{equation}
where $\zz$ and $\mixingL$ are expressed in meters. This function enforces a mixing length of 300~m from the surface to $\zz = 1$~km, and a smooth interpolation to the free atmospheric value, which is set to 30~m. 

\subsection{Discretisation of diffusion equations}
\label{app:discretisation_diff}
\begin{figure}[t]
   \centering
   \includegraphics[width=0.40\textwidth,trim = 0cm 0cm 9.5cm 0cm,clip]{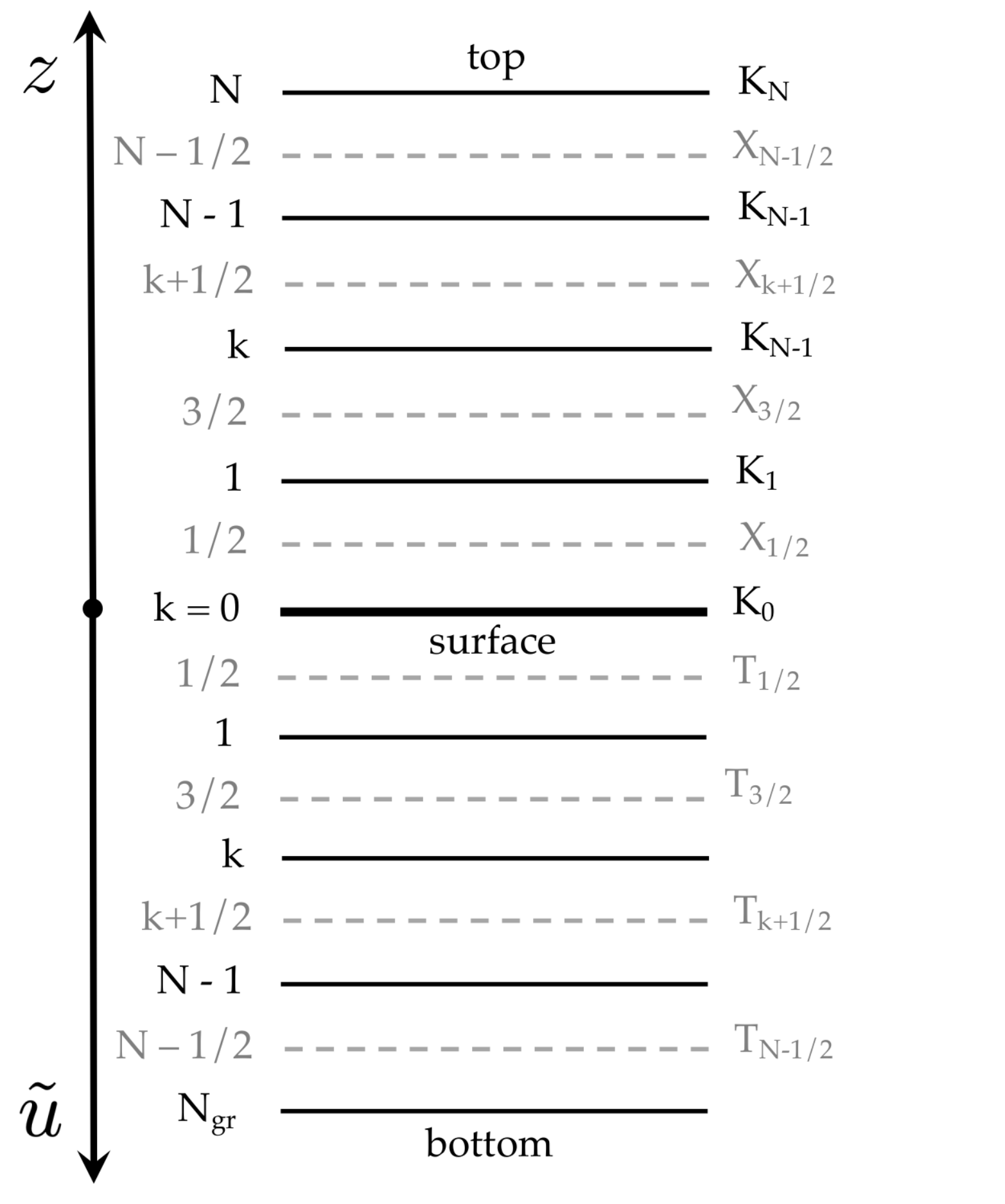}
      \caption{Discretisation of the atmosphere and ground into vertical levels.}
       \label{fig:diffusion}%
\end{figure}

The contribution of turbulent diffusion to the physical tendencies is computed every physical time step. Between the surface and the top of the atmosphere (internal levels corresponding to $\kver = 1, \ldots , \vlev-2$), the discretised equations derived from \eq{diffusive_term} are given by

\begin{equation}
\frac{\xvkn{\kver+1/2}{\ntime} - \xvkn{\kver+1/2}{\ntime-1}}{\dtphys} =   \frac{\area}{\masskn{\kver+1/2}{\ntime}  }   \! \left[ \rhokn{\kver+1}{\ntime} \Kdiffxkn{\kver+1}{\ntime} \frac{\diff{\zz}{\xvkn{\kver+1}{\ntime}}}{\diff{}{\zkn{\kver+1}{\ntime}}}  - \rhokn{\kver}{\ntime} \Kdiffxkn{\kver}{\ntime} \frac{\diff{\zz}{\xvkn{\kver}{\ntime}}}{\diff{}{\zkn{\kver}{\ntime}}}  \right] \! \! ,
\end{equation}
with $\area$ the area of the surface parcels defined by horizontal intervals. We note that integer indices in subscripts indicate levels separating vertical intervals, and non-integer indices centers of vertical intervals (see \fig{fig:diffusion}). For $\kver = 0$ (lower boundary condition),
\begin{align}
\frac{\xvkn{1/2}{\ntime} - \xvkn{1/2}{\ntime-1}}{\dtphys} = \frac{\area}{\masskn{1/2}{\ntime}} \left[  \rhokn{1}{\ntime} \Kdiffxkn{1}{\ntime} \frac{\diff{\zz}{\xvkn{1}{\ntime}}}{\diff{}{\zkn{1}{\ntime}}} - \Fturb  \right],
\end{align}
and, for $\kver = \vlev-1$ (upper boundary condition), 
\begin{align}
\frac{\xvkn{\vlev-1/2}{\ntime} - \xvkn{\vlev-1/2}{\ntime-1}}{\dtphys} = -  \frac{\area}{\masskn{\vlev-1/2}{\ntime} } \left[   \rhokn{\vlev-1}{\ntime} \Kdiffxkn{\vlev-1}{\ntime} \frac{\diff{\zz}{\xvkn{\vlev-1}{\ntime}}}{\diff{}{\zkn{\vlev-1}{\ntime}}}  \right],
\end{align}
where $\Fturb$ is the downward turbulent flux at the surface-atmosphere interface. Similarly as the equations of vertical diffusion, these equations are put into the form 
\begin{align}
& \ck{\kver+1/2} \left( \xvkn{\kver+1/2}{\ntime} - \xvkn{\kver+1/2}{\ntime-1} \right)    = \dk{\kver+1} \diff{\zz}{\xvkn{\kver+1}{\ntime}} - \dk{\kver} \diff{\zz}{\xvkn{\kver}{\ntime}}, \\
& \ck{1/2} \left( \xvkn{1/2}{\ntime} - \xvkn{1/2}{\ntime-1} \right)  = \dk{1} \diff{\zz}{\xvkn{1}{\ntime}}  - \Fturb  , \nonumber \\
& \ck{\vlev-1/2} \left( \xvkn{\vlev-1/2}{\ntime} - \xvkn{\vlev-1/2}{\ntime-1} \right)  = - \dk{\vlev-1} \diff{\zz}{\xvkn{\vlev-1}{\ntime}}, \nonumber 
\end{align}
where we have introduced the coefficients
\begin{align}
& \ck{\kver +1/2} \define \frac{ \masskn{\kver+1/2}{\ntime} }{ \area \dtphys}, & \dk{\kver} \define \frac{\rhokn{\kver}{\ntime} \Kdiffxkn{\kver}{\ntime} }{\diff{}{\zkn{\kver}{\ntime}}}.
\end{align}

This algebraic system is solved using the tridiagonal matrix algorithm (\append{app:thomas_algorithm}). We introduce, for $\kver  =1, \ldots, \vlev-1$, the recursion relation 
\begin{equation}
\xvkn{\kver+1/2}{} = \alphak{\kver} \xvkn{\kver-1/2}{} + \betak{\kver},
\label{xk_rec}
\end{equation}
where the coefficients $\alphak{\kver}$ and $\betak{\kver}$ are given by
\begin{align}
& \alphak{\kver} = \frac{\dk{\kver}}{\Deltak{\kver}}, & \betak{\kver} = \frac{1}{\Deltak{\kver}} \left( \ck{\kver+1/2} \xvkn{\kver+1/2}{\ntime-1} + \dk{\kver+1} \betak{\kver+1}  \right),
\end{align}
with $\Deltak{\kver} = \ck{\kver+1/2} + \dk{\kver} +  \left( 1 - \alphak{\kver+1} \right) \dk{\kver+1}$. The $\alphak{\kver}$ and $\betak{\kver}$ are first computed downwards starting from the upper boundary, where 
\begin{align}
& \alphak{\vlev-1} = \frac{\dk{\vlev-1}}{\ck{\vlev-1/2} + \dk{\vlev-1}}, & \betak{\vlev-1} = \frac{\ck{\vlev-1/2} \xvkn{\vlev-1/2}{\ntime-1} }{\ck{\vlev-1/2} + \dk{\vlev-1}}. 
\end{align}
Then, the $\xvkn{\kver+1/2}{\ntime}$ are computed upwards, from $\kver=0$ to $\kver = \vlev - 1$, using the recursion relation given by \eq{xk_rec} and starting from
\begin{equation}
\xvkn{1/2}{\ntime} = \frac{ \dk{1} \betak{1} - \Fturb + \ck{1/2} \xvkn{1/2}{\vlev-1} }{\ck{1/2} + \dk{1} \left( 1 - \alphak{1} \right)}. 
\end{equation}
This allows for calculating the source terms of the momentum, thermodynamic, and moisture conservation equations. Basically,
\begin{equation}
\begin{array}{lll}
\displaystyle \forceh = \DD{\Vtheta}{\time} , & \displaystyle \Qheat = \left( \frac{\press}{\prefteta} \right)^{\rcp} \DD{}{\time} \left( \Cp \teta \right), & \displaystyle \dtqmoist = \DD{\qmoist}{\time}. 
\end{array}
\end{equation}
 
\section{Soil heat transfer scheme}
\label{app:soil_heat_transfer}
The one-dimensional heat conduction equation given by \eq{heat_conduction} is solved by means of a finite difference method adapted from Appendix~B.1 of \cite{WCD2016}. The domain is discretised into $\vlevgr$ vertical intervals of $\zzn$ following a geometric law of scale factor~$\alpha$ (\fig{fig:diffusion}). Denoting by $\kver = 0, \ldots , \vlevgr$ the vertical grid levels ($\kver = 0$ corresponding to the surface, and $\kver = \vlevgr$ to the inner boundary of the domain), we express all of the $\zznk{\kver}$ as functions of the thickness of the first layer $\zznk{1}$. Literally, the depths of the full and intermediate levels are respectively given by 
\begin{align}
& \zznk{\kver} = \frac{\alpha^\kver - 1}{\alpha-1} \zznk{1}, & \kver = 0, \ldots , \vlevgr \\
& \zznk{\kver + 1/2} = \frac{\alpha^{\kver+1/2}-1}{\alpha-1} \zznk{1}, & \kver = 0, \ldots, \vlevgr-1. 
\end{align}
\begin{figure*}[htb]
   \centering
   \includegraphics[width=0.7\textwidth,trim = 5.5cm 0cm 0cm 0cm,clip]{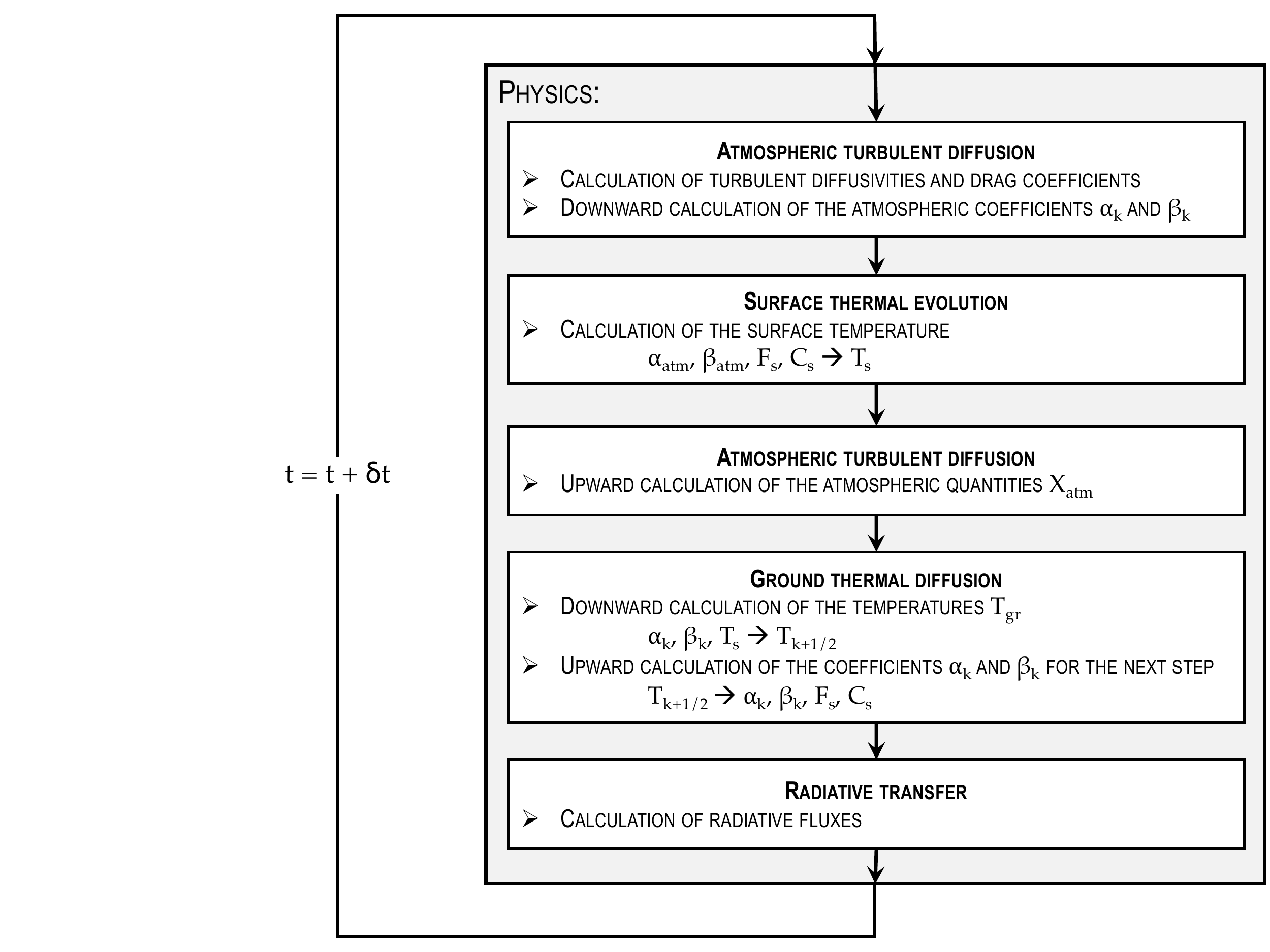}
      \caption{Temporal integration scheme for the calculation of physical tendencies resulting from turbulent diffusion and soil heat conduction. The notation $\delta t$ designates a physical timestep ($\delta t = \nphys \Delta t$ with $\nphys = 10$).}
       \label{fig:physics_loop}%
\end{figure*}
We take $\vlevgr=6$, $\alpha = 2$ and $\zznk{1}  =  0.1\ {\rm s^{1/2}}$. To solve the heat equation, we use an implicit scheme. The set of discretised equations is written, for $1 \leq \kver \leq \vlevgr-1$, as 
\begin{align}
\label{eqrt_dis1}
& \frac{\tempkn{\kver+1/2}{\ntime} - \tempkn{\kver+1/2}{\ntime-1}}{\dtphys} = \frac{1}{\dzkn{\kver+1/2}{}}  \left[ \frac{\diff{\zzn}{\tempkn{\kver+1}{\ntime}} }{\dzkn{\kver+1}{}} - \frac{\diff{\zzn}{\tempkn{\kver}{\ntime}} }{\dzkn{\kver}{}} \right], 
\end{align}
with $\diff{\zzn}{\tempkn{\kver}{\ntime}} = \tempkn{\kver+1/2}{\ntime} - \tempkn{\kver}{\ntime}$, and the boundary conditions yield
\begin{align}
\label{eqrt_dis2}
\frac{\tempkn{1/2}{\ntime} - \tempkn{1/2}{\ntime-1}}{\dtphys} & = \frac{1}{\dzkn{1/2}{}} \left[ \frac{\diff{\zzn}{\tempkn{1}{\ntime}}}{\dzkn{1}{}} + \frac{1}{\inertiagr} \left( \sum \Fdown \left( \Tsurf \right) -\emissurf  \sigmaSB \Tsurf^4 \right)  \right] ,
\end{align}
and 
\begin{align}
\label{eqrt_dis3}
\frac{\tempkn{\vlevgr-1/2}{\ntime} - \tempkn{\vlevgr-1/2}{\ntime-1}}{\dtphys} & = - \frac{1}{\dzkn{\vlevgr-1/2}{}} \left[  \frac{\diff{\zzn}{\tempkn{\vlevgr-1}{\ntime}}}{\dzkn{\vlevgr-1}{}} \right]. 
\end{align}

Introducing the coefficients $\ck{\kver +1/2}$ and $\dk{\kver}$ defined as
\begin{align}
& \ck{\kver +1/2} \define \frac{\dzkn{\kver+1/2}{}}{\dtphys}, & \dk{\kver} \define \frac{1}{\dzkn{\kver}{}}, 
\end{align}
the above equations (\eqsto{eqrt_dis1}{eqrt_dis3}) are rewritten as 
\begin{align}
& \ck{\kver+1/2} \left( \tempkn{\kver+1/2}{\ntime} - \tempkn{\kver+1/2}{\ntime-1} \right)    = \dk{\kver+1} \diff{\zzn}{\tempkn{\kver+1}{\ntime}} - \dk{\kver} \diff{\zzn}{\tempkn{\kver}{\ntime}}, \\
& \ck{1/2} \left( \tempkn{1/2}{\ntime} - \tempkn{1/2}{\ntime-1} \right)  = \dk{1} \diff{\zzn}{\tempkn{1}{\ntime}}  + \inertiagr^{-1} \left( \sum \Fdown \left( \Tsurf \right) - \emissurf \sigmaSB \Tsurf^4 \right)  , \nonumber \\
& \ck{\vlevgr-1/2} \left( \tempkn{\vlevgr-1/2}{\ntime} - \tempkn{\vlevgr-1/2}{\ntime-1} \right)  = - \dk{\vlevgr-1} \diff{\zzn}{\tempkn{\vlevgr-1}{\ntime}}, \nonumber 
\end{align}
and the system is put into the standard algebraic form
\begin{equation}
\begin{bmatrix}
\Bcoeffi{0} & \Ccoeffi{0} &  & & \\
\Acoeffi{1} & \Bcoeffi{1} & \Ccoeffi{1} & & \\
	       & \Acoeffi{\kver} & \Bcoeffi{\kver} & \Ccoeffi{\kver} &  \\
	       &		       & \Acoeffi{\vlevgr-2} & \Bcoeffi{\vlevgr-2} & \Ccoeffi{\vlevgr-2} \\
	       &		       &			 & \Acoeffi{\vlevgr-1} & \Bcoeffi{\vlevgr-1}
\end{bmatrix}
\!
\begin{bmatrix}
\tempkn{1/2}{\ntime} \\
\tempkn{3/2}{\ntime}  \\
\tempkn{\kver+1/2}{\ntime}  \\
\tempkn{\vlevgr-3/2}{\ntime} \\
\tempkn{\vlevgr-1/2}{\ntime}
\end{bmatrix}
\! = \!
\begin{bmatrix}
\bcoeffi{0} \\ \bcoeffi{1} \\ \bcoeffi{\kver} \\ \bcoeffi{\vlevgr-2} \\ \bcoeffi{\vlevgr-1}
\end{bmatrix}
\! ,
\label{soil_heat_system}
\end{equation}
with the coefficients
\begin{equation}
\begin{array}{ll}
\Acoeffi{\kver} = - \dk{\kver} & \mbox{for} \ \kver = 1, \ldots , \vlevgr-1, \\ 
\Bcoeffi{\kver} = \dk{\kver} + \dk{\kver+1} + \ck{\kver+1/2} & \mbox{for} \ \kver = 0, \ldots , \vlevgr-1, \\
\Ccoeffi{\kver} = \dk{\kver+1} &\mbox{for} \ \kver = 0, \ldots , \vlevgr-2, \\
\bcoeffi{\kver} = \ck{\kver+1/2} \tempkn{\kver+1/2}{\ntime-1} & \mbox{for} \ \kver = 1, \ldots , \vlevgr-1, \\ 
\end{array}
\end{equation}
\begin{equation}
\begin{array}{l}
\Bcoeffi{0} = \dk{1} + \ck{1/2}, \\ 
\Bcoeffi{\vlevgr-1} = \dk{\vlevgr-1} + \ck{\vlevgr-1/2}, \\ 
\bcoeffi{0} = \ck{1/2} \tempkn{1/2}{\ntime-1} + \inertiagr^{-1} \left( \sum \Fdown \left( \Tsurf \right) - \emissurf \sigmaSB \Tsurf^4 \right).
\end{array}
\end{equation}

The algebraic system given by \eq{soil_heat_system} is solved by making use of Thomas algorithm (\append{app:thomas_algorithm}). As a first step, the temperatures of two consecutive levels are linked together by the recursion relation
\begin{equation}
\tempkn{\kver+1/2}{\ntime} = \alphak{\kver} \tempkn{\kver-1/2}{\ntime} + \betak{\kver}, 
\end{equation}
where the coefficients $\alphak{\kver}$ and $\betak{\kver}$ are defined, for $\kver = 1, \ldots, \vlevgr - 2 $, as 
\begin{equation}
\begin{array}{ll}
\displaystyle \alphak{\kver} = \frac{\dk{\kver}}{\Deltak{\kver}}, & \displaystyle  \betak{\kver} = \frac{1}{\Deltak{\kver}} \left( \ck{\kver +1/2} \tempkn{\kver+1/2}{\ntime-1} + \dk{\kver+1} \betak{\kver+1} \right),
\end{array}
\end{equation} 
with $\Deltak{\kver} = \ck{\kver+1/2} + \dk{\kver} + \left( 1 - \alphak{\kver+1} \right) \dk{\kver+1}$. At the inner boundary,  the zero-flux condition leads to 
\begin{equation}
\begin{array}{ll}
\displaystyle \alphak{\vlevgr-1} = \frac{\dk{\vlevgr-1}}{\Deltak{\vlevgr-1}} , & \displaystyle \betak{\vlevgr-1} = \frac{\ck{\vlevgr-1/2} \tempkn{\vlev-1/2}{\ntime-1}}{\Deltak{\vlevgr-1}},
\end{array}
\end{equation}
with $\Deltak{\vlevgr-1} = \dk{\vlevgr-1} + \ck{\vlevgr-1/2}$. Thus, the coefficients $\alphak{\kver}$ and $\betak{\kver}$ are integrated upwards from the inner boundary. 

As a second step, the equation of the surface boundary condition is put into the form
\begin{equation}
\Csstar \frac{\tempkn{1/2}{\ntime} - \tempkn{1/2}{\ntime-1}}{\dtphys} = \Fsstar + \sum \Fdown - \emissurf \sigmaSB \Tsurf^4 - \Fupsw.
\label{Tsurf_Cstar}
\end{equation}
where $\Csstar $ and $\Fsstar$ are expressed as 
\begin{align}
\Csstar & = \inertiagr \dtphys \left[  \ck{1/2} + \dk{1} \left( 1 - \alphak{1} \right) \right], \\
\Fsstar  & = \inertiagr \dk{1} \left[ \betak{1} + \left( \alphak{1} - 1 \right) \tempkn{1/2}{\ntime-1} \right]. 
\end{align}
In order to obtain an equation for the surface temperature, we proceed to a linear interpolation near the surface, which yields
\begin{align}
& \Tsurf = \left( 1 + \muint \right) \tempkn{1/2}{} - \muint \tempkn{3/2}{}, & \mbox{with} \ \muint = \frac{\zzkn{1/2}{}}{\zzkn{3/2}{} - \zzkn{1/2}{}}. 
\end{align}
Combining the above equation with the recursion relation $\tempkn{3/2}{\ntime} = \alphak{1} \tempkn{1/2}{\ntime} + \betak{1}$, we rearrange \eq{Tsurf_Cstar} into 
\begin{equation}
\Csgr \frac{\Tsurf^{\ntime} - \Tsurf^{\ntime-1}}{\dtphys} = \Fsgr + \sum \Fdown - \Fupsw - \emissurf \sigmaSB \left( \Tsurf^\ntime \right)^4,
\label{surface_heat_equation}
\end{equation}
where the heat capacity per unit surface $\Csgr$ and upcoming flux $\Fsgr$ are given by 

\begin{align}
\Csgr & = \frac{\Csstar}{1 + \muint \left( 1 - \alphak{1} \right)}, \\
\Fsgr & = \Fsstar + \frac{\Csstar}{\dtphys} \tempkn{1/2}{\ntime-1} - \Csstar \left( \frac{\Tsurf^{\ntime-1} + \muint \betak{1}}{\dtphys} \right). 
\end{align}
In practice, this equation is linearised and solved with an implicit scheme. The parameters $\Csgr$ and $\Fsgr$ can be set to constants if one does not wish to solve the vertical diffusion within the ground. This yields the surface temperature of the current step $\Tsurf^\ntime$, which allows for calculating the temperatures at ground levels by using the recursion relation downwards. 

\def\wpanel{0.33\textwidth}
\def\wlegend{0.32\textwidth}
\def\hraisebox{0.20\textwidth}
\begin{figure*}[t]
   \centering
  \hspace{0cm} \textsc{$\inertiagr = 10^2 \ {\rm J \ m^{-2} \ s^{-1/2} \ K^{-1}}$} \hspace{2.cm} \textsc{$\inertiagr = 10^3 \ {\rm J \ m^{-2} \ s^{-1/2} \ K^{-1}}$} \hspace{2.cm} \textsc{$\inertiagr = 10^4 \ {\rm J \ m^{-2} \ s^{-1/2} \ K^{-1}}$} \\[0.3cm]
   \includegraphics[width=\wpanel,trim = 2.5cm 0.0cm 0.8cm 0.9cm,clip]{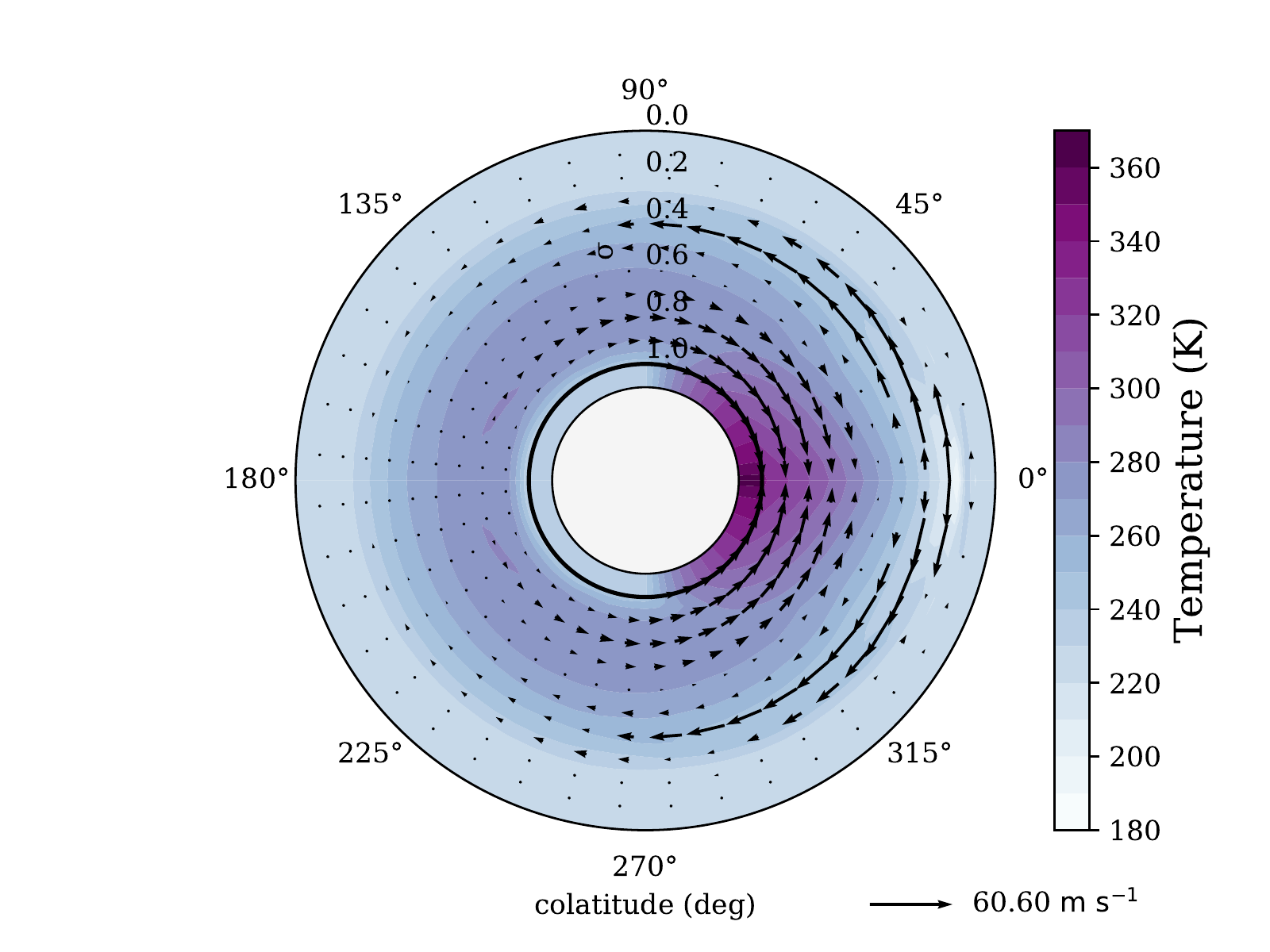}
   \includegraphics[width=\wpanel,trim = 2.5cm 0.0cm 0.8cm 0.9cm,clip]{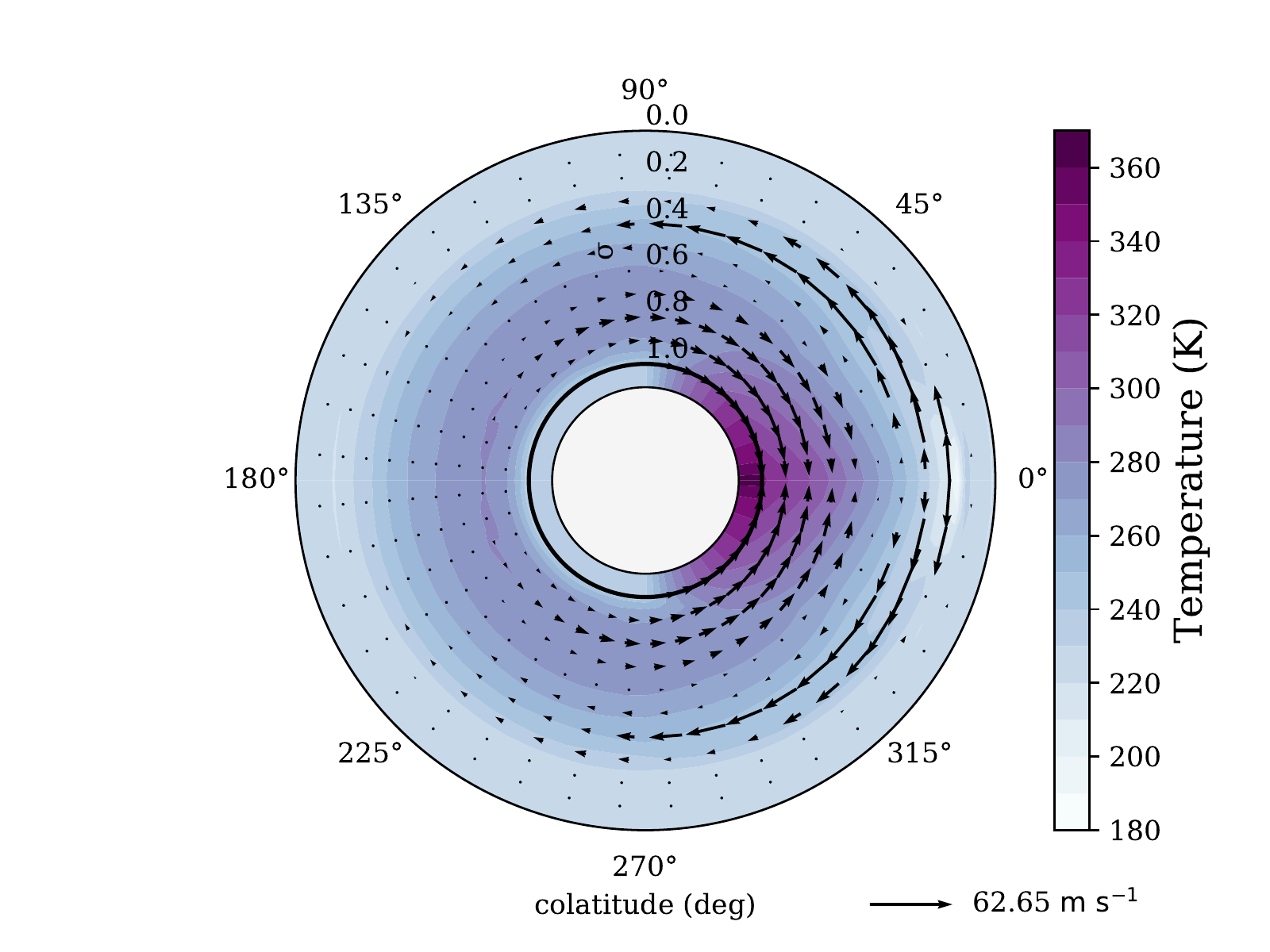} 
   \includegraphics[width=\wpanel,trim = 2.5cm 0.0cm 0.8cm 0.9cm,clip]{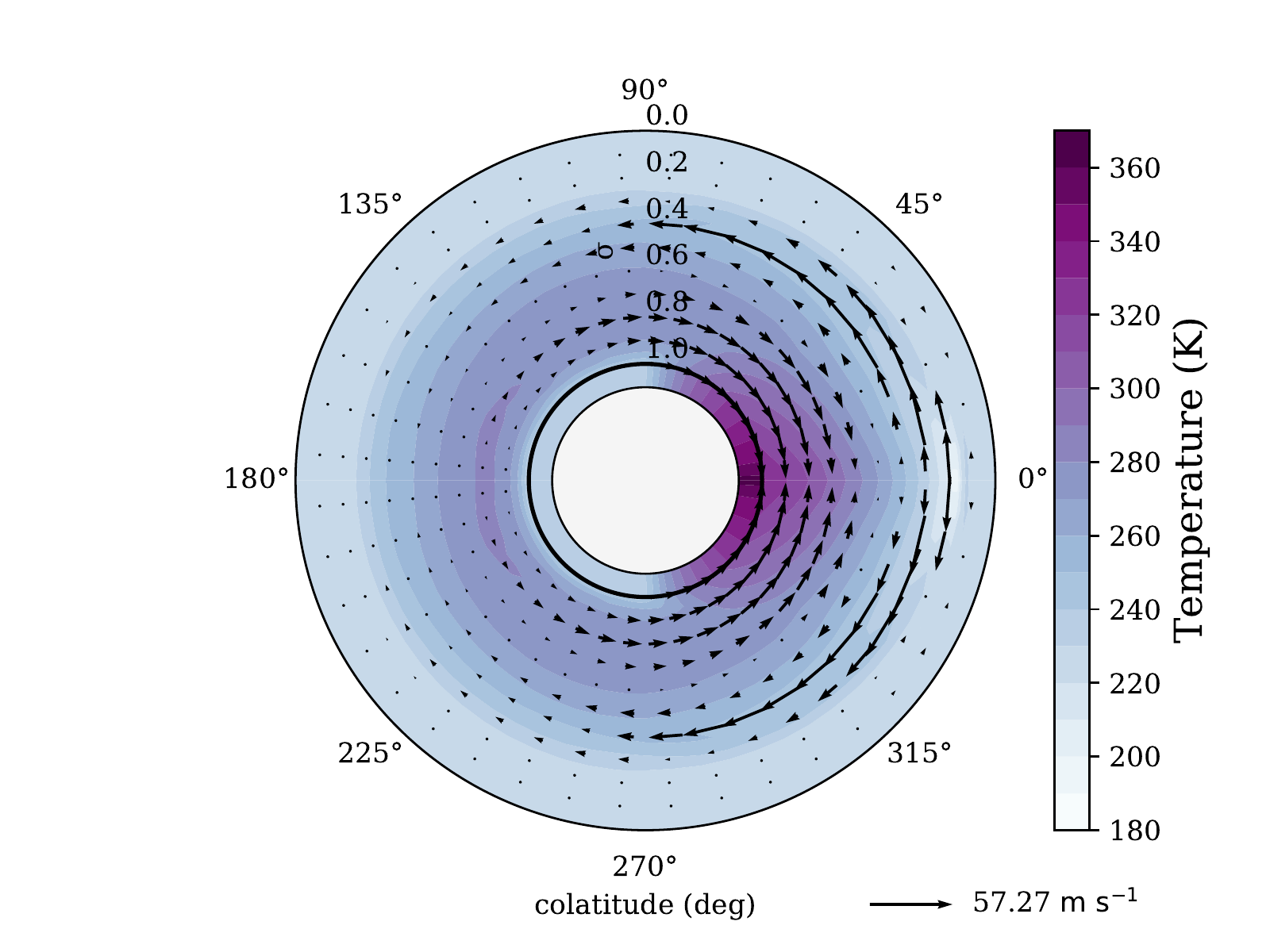} 
      \caption{\RV{Two-day averaged temperature snapshots for various values of soil thermal inertia (see \eq{inertiagr}). {\it Left:} $\inertiagr = 10^2 \ {\rm J \ m^{-2} \ s^{-1/2} \ K^{-1}}$. {\it Middle:} $\inertiagr = 10^3 \ {\rm J \ m^{-2} \ s^{-1/2} \ K^{-1}}$. {\it Right:} $\inertiagr = 10^4 \ {\rm J \ m^{-2} \ s^{-1/2} \ K^{-1}}$. Simulations were performed for the Earth-like case of Table~\ref{tab:param_reference_case} with a stellar irradiation of 1366~${\rm W \ m^{-2}}$ and a 1 bar surface pressure, similarly as in \fig{fig:snapshots_HB1993_diff}.} }
       \label{fig:validation_soil}%
\end{figure*}

Linearising and discretising the surface temperature evolution equation, given by \eq{surface_heat_equation}, we obtain
\begin{align}
\Csgr \frac{\Tsurf^{\ntime} - \Tsurf^{\ntime-1}}{\dtphys} = &  \Fsgr + \sum \Fdown - \Fupsw + \Fturb - \emissurf \sigmaSB \left( \Tsurf^{\ntime-1} \right)^4 \nonumber \\ 
 & - 4 \emissurf \sigmaSB \left( \Tsurf^{\ntime-1} \right)^3 \left( \Tsurf^\ntime - \Tsurf^{\ntime-1} \right).
\end{align}
The downward heat flux associated with turbulent diffusion is expressed in its general form as 
\begin{equation}
\Fturb = - A \Tsurf^{\ntime} + B,
\label{Fturb_A_B}
\end{equation}
which allows for writing the surface temperature of the current step as 
\begin{equation}
\Tsurf^\ntime = \frac{\frac{\Csgr}{\dtphys} \Tsurf^{\ntime-1} + 3 \emissurf \sigmaSB \left( \Tsurf^{\ntime-1} \right)^4 + \Fsgr + \sum \Fdown - \Fupsw + B}{\frac{\Csgr}{\dtphys} + 4 \emissurf \sigmaSB \left( \Tsurf^{\ntime - 1} \right)^3 + A}. 
\end{equation}
The evolution of the surface content of any tracer in liquid or solid phase can be described with a similar equation. It induces a source-sink term in the moisture conservation equation. In the case of temperature the downward turbulent flux is given by 
\begin{equation}
\Fturb = - \Fsheat = \Cheat \rhofirst \Cp  \abs{\Vvecthfirst} \left( \tetafirst - \tetasurf \right),
\end{equation}
which is a function of the mean fields near the surface (see \eq{Fsheat}). The coefficients $A$ and $B$ introduced in \eq{Fturb_A_B} are thus expressed as
\begin{align}
& A = \Cheat \rhofirst \Cp  \abs{\Vvecthfirst} \left( \frac{\psurf}{\prefteta} \right)^{- \rcp}, & B = \Cheat \rhofirst \Cp  \abs{\Vvecthfirst} \tetafirst.
\end{align}

As shown by \fig{fig:physics_loop}, the soil heat transfer scheme is coupled with the turbulent diffusion scheme (\append{app:turbdiff_scheme}) through the equation of surface thermal evolution, which may lead to consistency issues. In order to preserve the consistency of the diffusion scheme from the lowest level of the atmosphere to the the highest level of the ground conduction model, the two steps of the calculation are permuted in chronological order: temperatures are computed first, and the coefficients $\alphak{\kver}$ and $\betak{\kver}$ are computed then, and conserved for the next step. 

\RV{Simulations were run for various values of the ground thermal inertia given by \eq{inertiagr} ($\inertiagr = 10^2, 10^3, 10^4 \ {\rm J \ m^{-2} \ s^{-1/2} \ K^{-1}}$) for the Earth-like planet of \fig{fig:snapshots_HB1993_diff} in order to investigate how numerical solutions depend upon the soil thermal response. The two-day averaged temperature snapshots obtained from these simulations are shown by \fig{fig:validation_soil}. We observe that varying the ground thermal inertia over two orders of magnitude does not significantly alter the climate state of equilibrium. The change in $\inertiagr$ essentially affects the maximum wind speed, which varies by $\sim 9\%$. The minimum surface temperature on nightside $\Tnight$ hardly varies, as it switches from 231.3~K for $\inertiagr = 10^2, 10^3 \ {\rm J \ m^{-2} \ s^{-1/2} \ K^{-1}}$ to 231.8~K for $\inertiagr = 10^4 \ {\rm J \ m^{-2} \ s^{-1/2} \ K^{-1}}$. This insensitivity of mean fields to the soil vertical conduction is consistent with the fact that the circulation reaches a steady state where mean flows are essentially invariant in time. With variations, mean flows might be more substantially affected by vertical thermal diffusion in the soil.}

\section{Thomas algorithm for block tridiagonal matrices}
\label{app:thomas_algorithm}

The tridiagonal matrix algorithm \citep[TDMA, or Thomas algorithm,][]{Thomas1949} can be used to solve a system of equations that involves a block tridiagonal matrix of the form
\begin{equation}
\begin{bmatrix}
\Bmati{0} & \Cmati{0} &  & & \\
\Amati{1} & \Bmati{1} & \Cmati{1} & & \\
	       & \Amati{\kver} & \Bmati{\kver} & \Cmati{\kver} &  \\
	       &		       & \Amati{\vlev-2} & \Bmati{\vlev-2} & \Cmati{\vlev-2} \\
	       &		       &			 & \Amati{\vlev-1} & \Bmati{\vlev-1}
\end{bmatrix}
\begin{bmatrix}
\xvecti{0} \\
\xvecti{1} \\
\xvecti{\kver} \\
\xvecti{\vlev-2} \\
\xvecti{\vlev-1}
\end{bmatrix}
=
\begin{bmatrix}
\bvecti{0} \\ \bvecti{1} \\ \bvecti{\kver} \\ \bvecti{\vlev-2} \\ \bvecti{\vlev-1}
\end{bmatrix}
,
\end{equation}
where the $\Amati{\kver}$, $\Bmati{\kver}$, $\Cmati{\kver}$ are sub-matrices indexed by $\kver = 0 , \ldots, \vlev$, and the $\xvecti{\kver}$ and $\bvecti{\kver}$ vectors of appropriate dimensions. As a first step, the matrix is triangularised, meaning that the system is transformed into a system where the matrix is block triangular. The new system is written as 
\begin{equation}
\begin{bmatrix}
1 & \gamati{0} &  & & \\
  & 1 & \gamati{1} & & \\
	       & & 1 & \gamati{\kver} &  \\
	       &		       & & 1 & \gamati{\vlev-2} \\
	       &		       &			 &  & 1
\end{bmatrix}
\begin{bmatrix}
\xvecti{0} \\
\xvecti{1} \\
\xvecti{\kver} \\
\xvecti{\vlev-2} \\
\xvecti{\vlev-1}
\end{bmatrix}
=
\begin{bmatrix}
\betavi{0} \\ \betavi{1} \\ \betavi{\kver} \\ \betavi{\vlev-2} \\ \betavi{\vlev-1}
\end{bmatrix}
.
\end{equation}
The matrices $\gamati{\kver}$ and vectors $\betavi{\kver}$ are computed forwards using the recursion relations
\begin{align}
& \gamati{0} = \Bmati{0}^{-1} \Cmati{0}, & \\ 
& \gamati{\kver} = \left( \Bmati{\kver} - \Amati{\kver} \gamati{\kver-1} \right)^{-1} \Cmati{\kver}, & \kver = 1, \ldots , \vlev-2;
\end{align}
and
\begin{align}
& \betavi{0} = \Bmati{0}^{-1} \bvecti{0}, & \\
& \betavi{\kver} = \left( \Bmati{\kver} - \Amati{\kver} \gamati{\kver-1} \right)^{-1} \left( \bvecti{\kver} - \Amati{\kver} \betavi{\kver-1} \right), & \kver = 1, \ldots, \vlev-1.
\end{align}
As a second step, the solution vectors $\xvecti{\kver}$ are computed backwards (backward sweep) using the recursion relation 
\begin{align}
& \xvecti{\vlev-1} = \bvecti{\vlev-1}, & \\
& \xvecti{\kver} = \betavi{\kver} - \gamati{\kver} \xvecti{\kver+1},  & \kver = \vlev-2 , \ldots , 0. 
\end{align}

\section{Interhemispheric mass flow rate}
\label{app:mass_flow_rate}
At the terminator ($\col = 90^\degree$ in tidally locked coordinates), the total mass flow rate (i.e. mass that passes through the terminator annulus per unit of time in one direction) is given by
\begin{equation}
F_{\rm mass} = \pi \Rpla \integ{\abs{\Vtheta} \rho}{\zz}{0}{\zz_{\rm top}}. 
\end{equation}
The day-night advection timescale $\tadv$ corresponds to the mean timescale necessary for one particle to accomplish one full cycle of the day-night overturning circulation. It measures the renewal rate of the air and the strength of the cell. The smaller $\tadv$ and the faster air is advected from dayside to nigthside. Introducing the total mass of the atmosphere $M_{\rm atm} \define \left( 4 \pi \Rpla^2 \psurf \right) / \ggravi$, this timescale can be defined as
\begin{equation}
\tadv  \define \frac{M_{\rm atm}}{F_{\rm mass}}. 
\end{equation}
In sigma-coordinate, the circulation timescale is expressed as
\begin{equation}
\tadv = \frac{4 \Rpla}{\integ{\abs{\Vtheta}}{\sig}{0}{1}},
\end{equation}
the integral being performed at the terminator ($\col = 90^\degree$).

\end{document}